%% file: master.tex
\documentclass{aa}
\usepackage{txfonts}
\usepackage{natbib}
\usepackage{graphicx}
\bibpunct{(}{)}{;}{a}{}{,} 
\usepackage{subfigure}
\usepackage[margin=10pt,font=small,labelfont=bf,labelsep=endash]{caption}%
 \newcommand{\mic}{$\mu$m}
 
 \newcommand{\feii}{[\ion{Fe}{ii}]}
 \newcommand{\brg}{Br$\gamma$}

 \newcommand{\kms}{~km~s$^{-1}$}
 \newcommand{\hei}{\ion{He}{i}}
 \newcommand{\msun}{M$_\odot$}

\begin{document}

\title{\object{The nuclear gas disk of NGC 1566 dissected by SINFONI and ALMA}
	\thanks{Based on the ESO-VLT proposal ID: 090.B-0657(A)}
 }


\author{S. Smaji\'c\inst{1,}\inst{2}
	\and L. Moser\inst{1}
	\and A. Eckart\inst{1,}\inst{2}
	\and G. Busch\inst{1}
	\and F. Combes\inst{3}
	\and S. Garc\'ia-Burillo\inst{4}
	\and M. Valencia-S.\inst{1}
	\and M. Horrobin\inst{1}
	}

\institute{I. Physikalisches Institut, Universit\"at zu K\"oln, Z\"ulpicher Str. 77, 50937 K\"oln, Germany\\email: smajic@ph1.uni-koeln.de
	\and Max-Planck-Institut f\"ur Radioastronomie, Auf dem H\"ugel 69, 53121 Bonn, Germany
	\and Observatoire de Paris, LERMA, PSL, CNRS, Sorbonne Univ. UPMC and College de France, F-75014, Paris, France
	\and Observatorio Astronómico Nacional (OAN)-Observatorio de Madrid, Alfonso XII 3, 28014, Madrid, Spain
	}

\date{Received ???/ Accepted ???}

\input{abstract}
\label{abstract}


\maketitle

\input{introduction}



\input{obs_red}

\input{res_and_disc}

\input{discussion}

\input{con_sum}

\begin{acknowledgements}
The authors kindly thank the staff at the Observatory Paranal for their assistance during the observation.
Additionally, the authors thank the anonymous referees for fruitful comments and suggestions.
This paper makes
use of the following ALMA data: ADS/JAO.ALMA\#2011.0.00208.S. ALMA
is a partnership of ESO (representing its member states), NSF (USA) and
NINS (Japan), together with NRC (Canada) and NSC and ASIAA (Taiwan),
in cooperation with the Republic of Chile. The Joint ALMA Observatory is
operated by ESO, AUI/NRAO and NAOJ. The National Radio Astronomy
Observatory is a facility of the National Science Foundation operated under cooperative agreement by Associated Universities, Inc. 
We use data products from the Two Micron All Sky Survey, which is a joint project of the University of Massachusetts and the Infrared Processing and Analysis Center/California Institute of Technology, funded by the National Aeronautics and Space Administration and the National Science Foundation.
We used observations made with the NASA/ESA Hubble Space Telescope, obtained from the data archive at the Space Telescope Institute. STScI is operated by the association of Universities for Research in Astronomy, Inc. under the NASA contract NAS 5-26555.
We used observations made with the NASA/ESA Hubble Space Telescope, and obtained from the Hubble Legacy Archive, which is a collaboration between the Space Telescope Science Institute (STScI/NASA), the Space Telescope European Coordinating Facility (ST-ECF/ESA), and the Canadian Astronomy Data Centre (CADC/NRC/CSA).
This work was supported in part by the Deutsche Forschungsgemeinschaft
(DFG) via the Bonn Cologne Graduate School (BCGS),
and via grant SFB 956, as well as by
the Max Planck Society and the University of Cologne through
the International Max Planck Research School (IMPRS) for Astronomy and
Astrophysics and by the German federal department for education and research (BMBF) under the project number 50OS1101.
We had fruitful discussions
with members of the European Union funded COST Action MP0905: Black
Holes in a violent Universe and the
COST Action MP1104:
Polarization as a tool to study the Solar System and beyond.
We received funding from the
European Union Seventh Framework Programme (FP7/2007-2013)
under grant agreement No.312789.
F.C. acknowledges the European Research Council
for the Advanced Grant Program Num 267399-Momentum.
\end{acknowledgements}


\bibliographystyle{bibtex/aa}
\bibliography{bibtex/mybib,bibtex/zotero,bibtex/book}

\begin{appendix}

\section{Figures}
\begin{figure*}[htbp]
\centering
\subfigure[log(H$_2$(1-0)S(1)/Br$\gamma$)]{\includegraphics[width=0.33\textwidth]{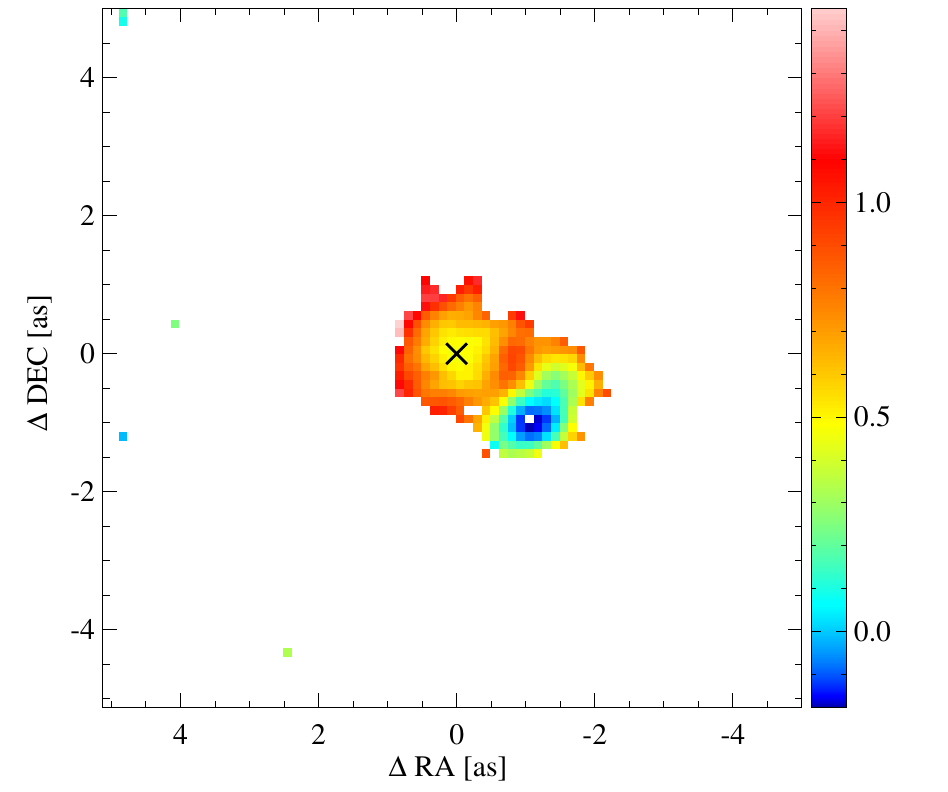}\label{fig:h212obrg}}
\subfigure[log($\mbox{[}$\ion{Fe}{ii}$\mbox{]}$/Br$\gamma$)]{\includegraphics[width=0.33\textwidth]{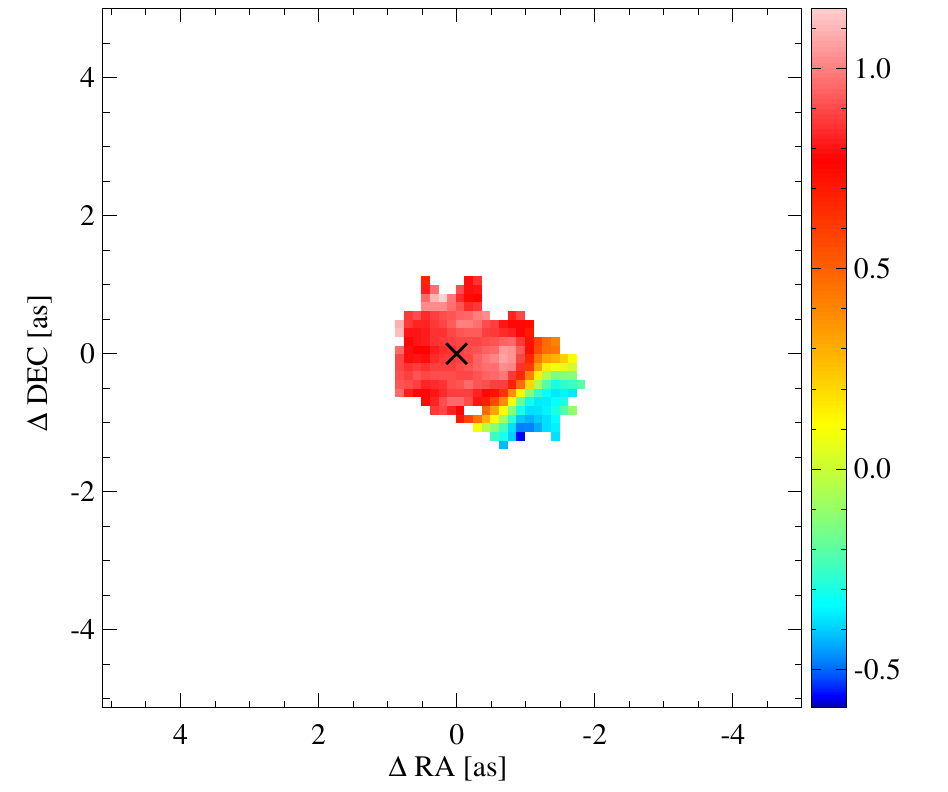}\label{fig:FeiioBrg}}
\subfigure[log(H$_2$(1-0)S(1)/\ion{He}{i})]{\includegraphics[width=0.33\textwidth]{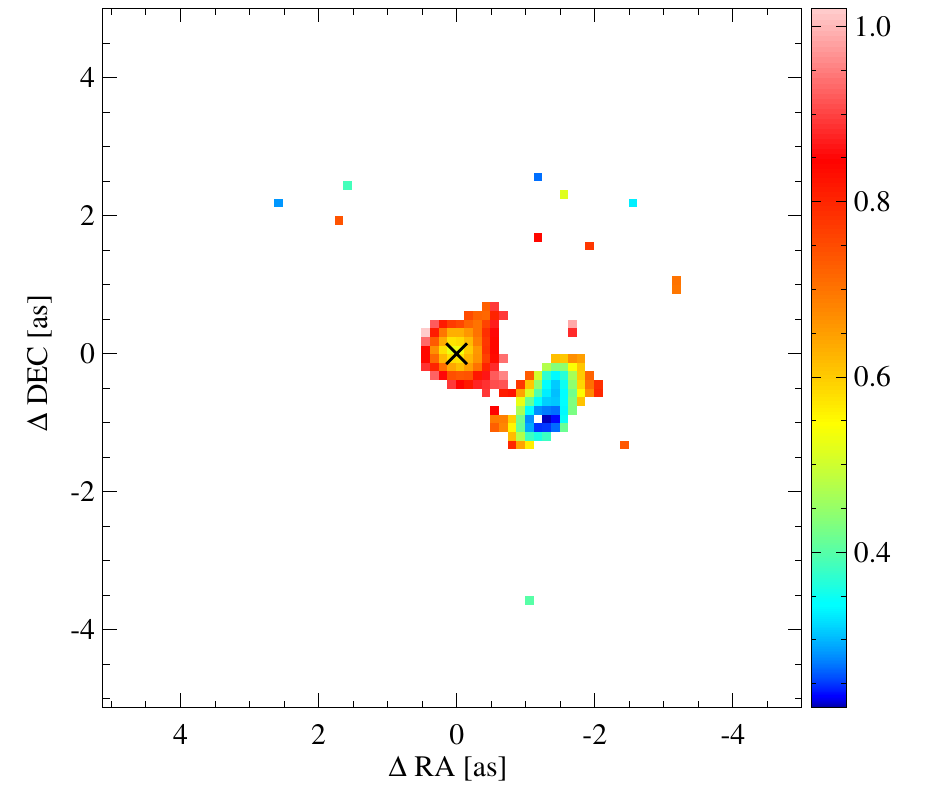}\label{fig:H212oHei}}
\subfigure[H$_2$(2-1)S(1)/H$_2$(1-0)S(1)]{\includegraphics[width=0.33\textwidth]{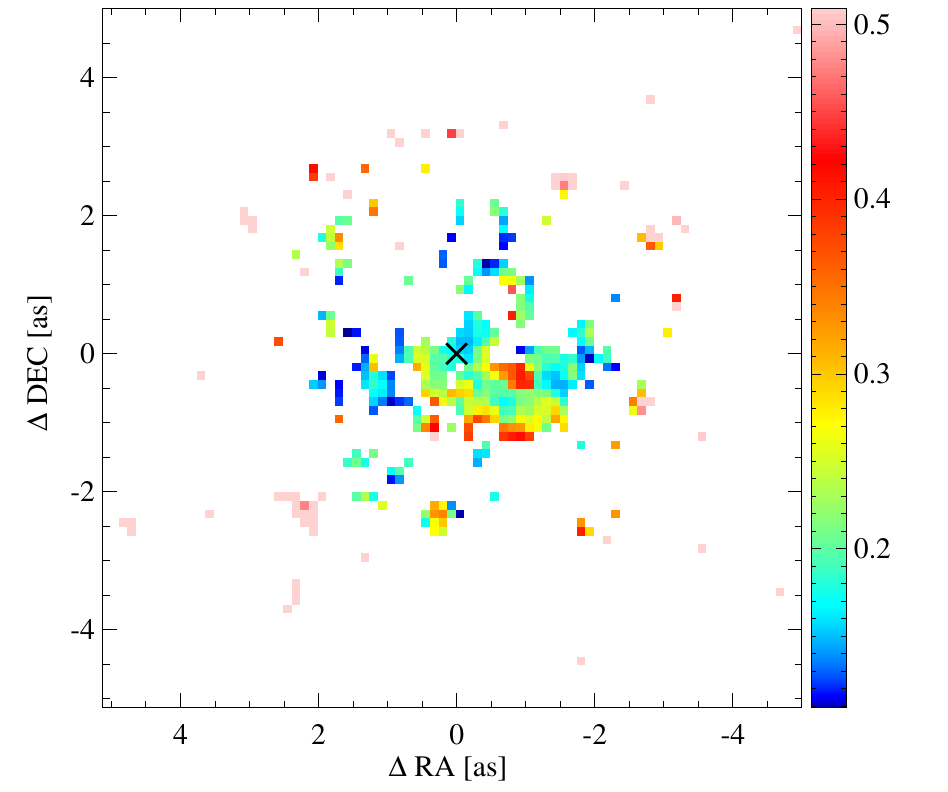}\label{fig:H2248oH212}}
\subfigure[H$_2$(1-0)S(2)/H$_2$(1-0)S(0)]{\includegraphics[width=0.33\textwidth]{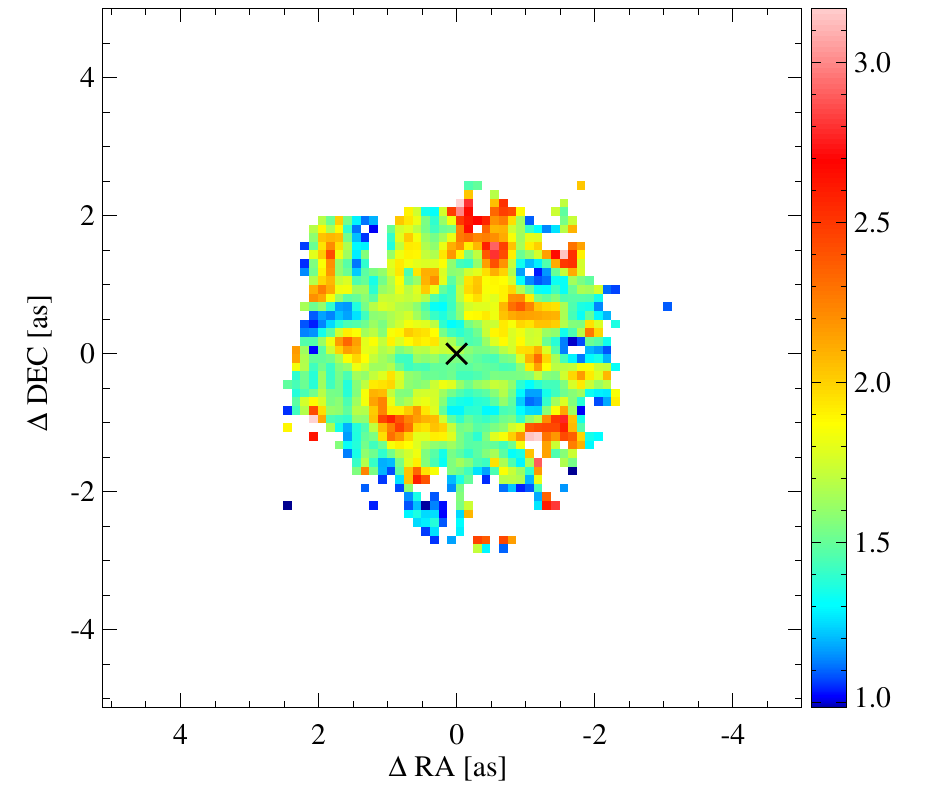}\label{fig:H203oH222}}
\subfigure[H$_2$(1-0)S(3)/H$_2$(1-0)S(1)]{\includegraphics[width=0.33\textwidth]{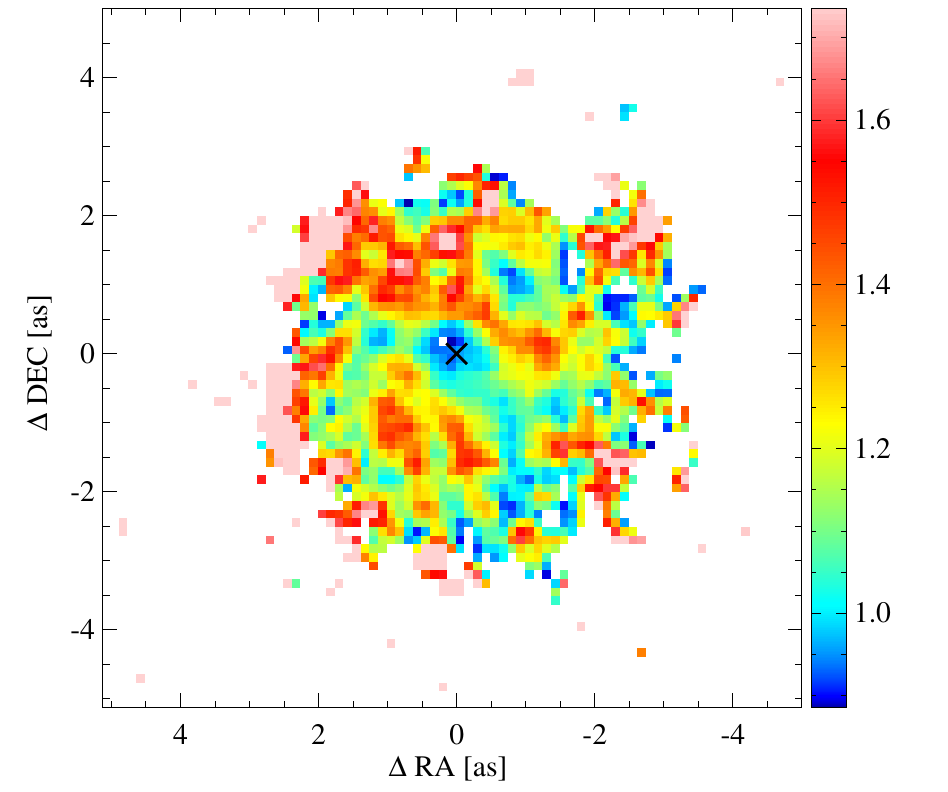}\label{fig:H2196oH212}}
\caption{Shown are the logarithmic line ratios of H$_2$(1-0)S(1) over Br$\gamma$ and [\ion{Fe}{ii}] over Br$\gamma$ in \subref{fig:h212obrg} and \subref{fig:FeiioBrg}. Panel \subref{fig:H212oHei} shows the logarithmic H$_2$(1-0)S(1) over \ion{He}{i} ratio. Panel \subref{fig:H2248oH212},\subref{fig:H203oH222},\subref{fig:H2196oH212} show H$_2$(2-1)S(1) over H$_2$(1-0)S(1), H$_2$(1-0)S(2) over H$_2$(1-0)S(0) and H$_2$(1-0)S(3) over H$_2$(1-0)S(1).}
\label{fig:linesrall}
\end{figure*}

We have attached the line maps to all important emission lines that were detected except for the H$_2$ Q-branch. The Q-branch is situated at the end of K-band and therefore the maps show noise features as well. However, the flux and FWHM of these lines is given in table \ref{tab:region} for the regions discussed here. Additionally, several line ratio maps are presented in Fig. \ref{fig:linesrall}

\begin{figure*}[htbp]
\centering
\subfigure[Flux H$_2$(1-0)S(3)]{\includegraphics[width=0.33\textwidth]{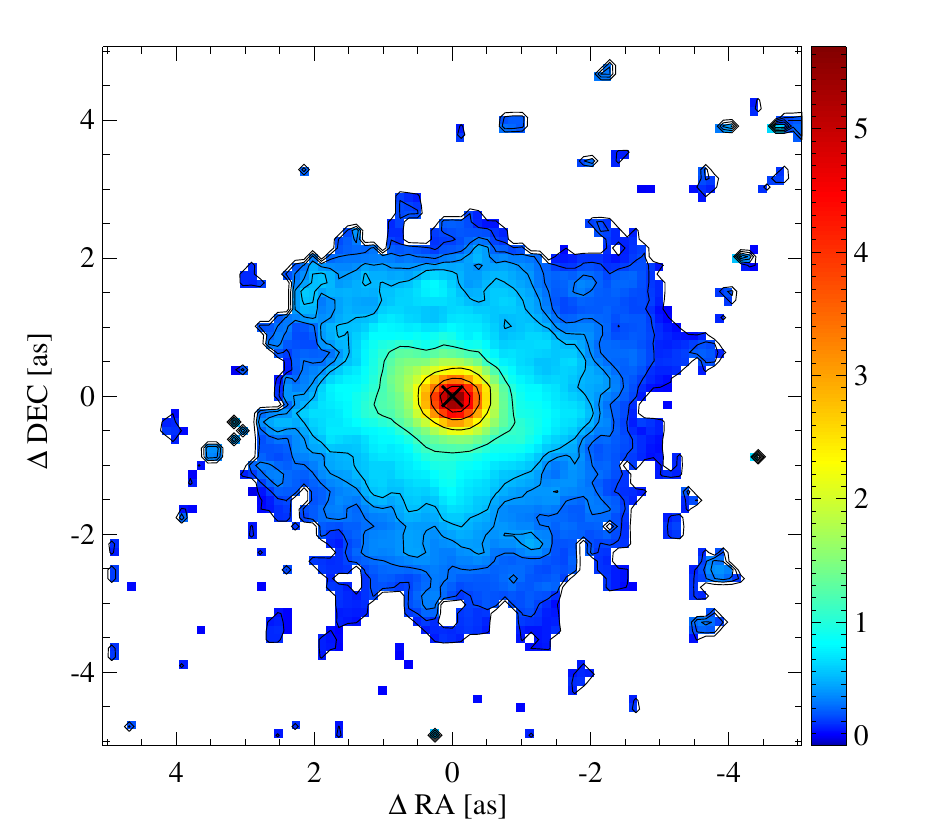}\label{fig:h21957flux}}
\subfigure[FWHM H$_2$(1-0)S(3)]{\includegraphics[width=0.33\textwidth]{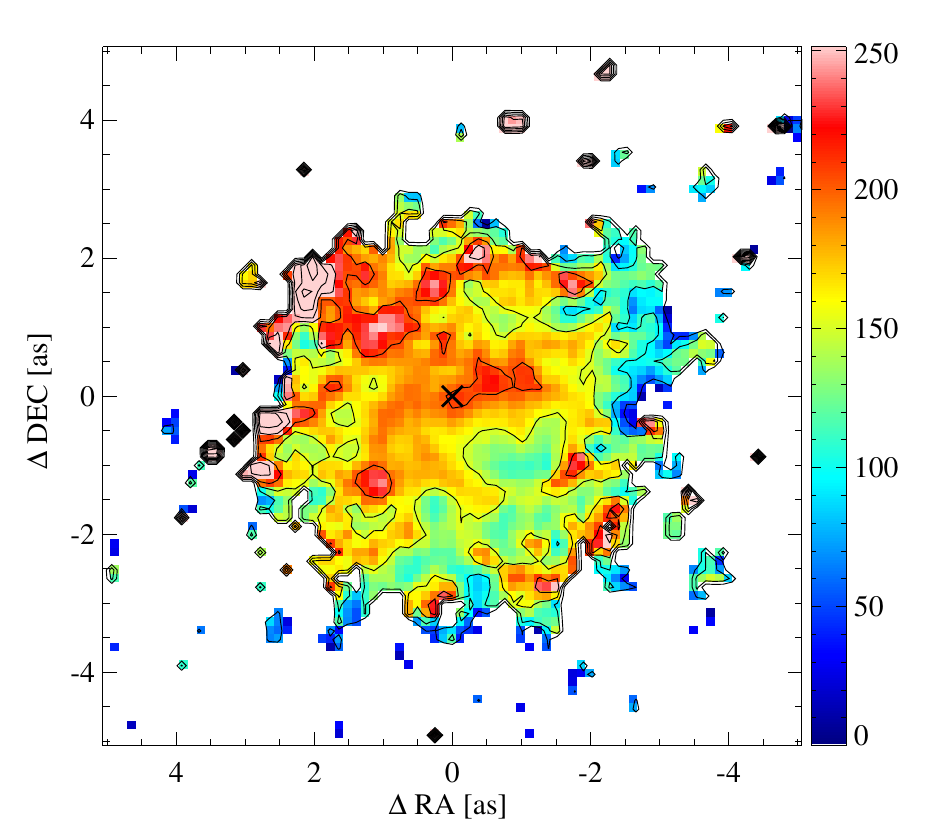}\label{fig:h21957fwhm}}
\subfigure[EW H$_2$(1-0)S(3)]{\includegraphics[width=0.33\textwidth]{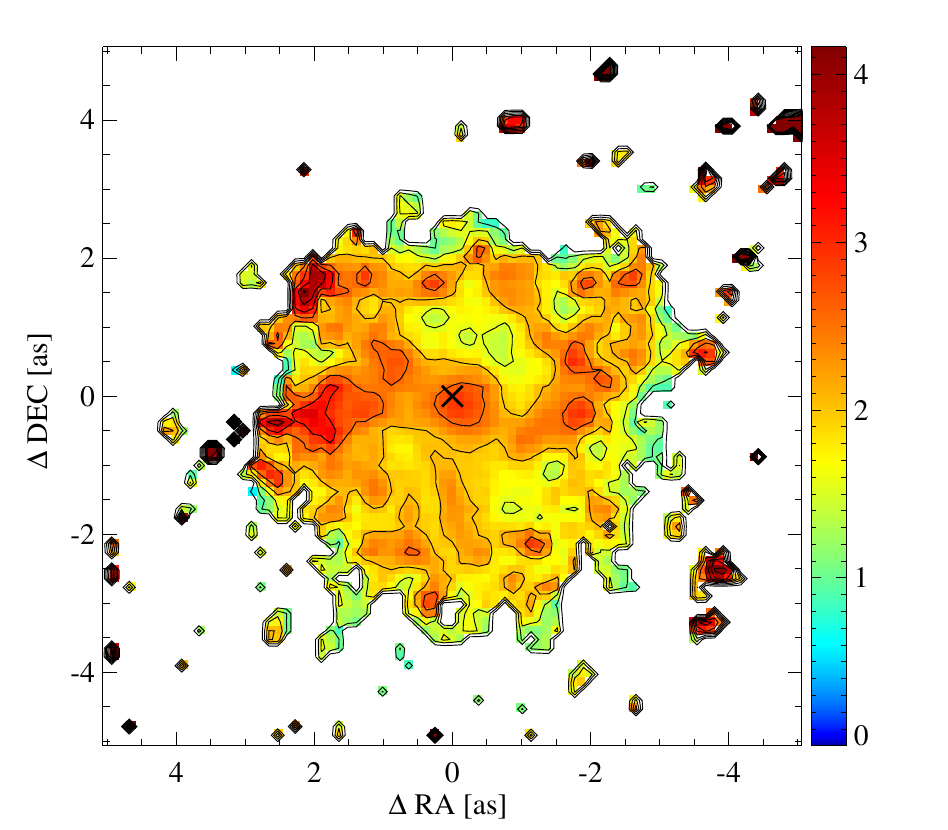}\label{fig:h21957ew}}
\subfigure[Flux H$_2$(1-0)S(2)]{\includegraphics[width=0.33\textwidth]{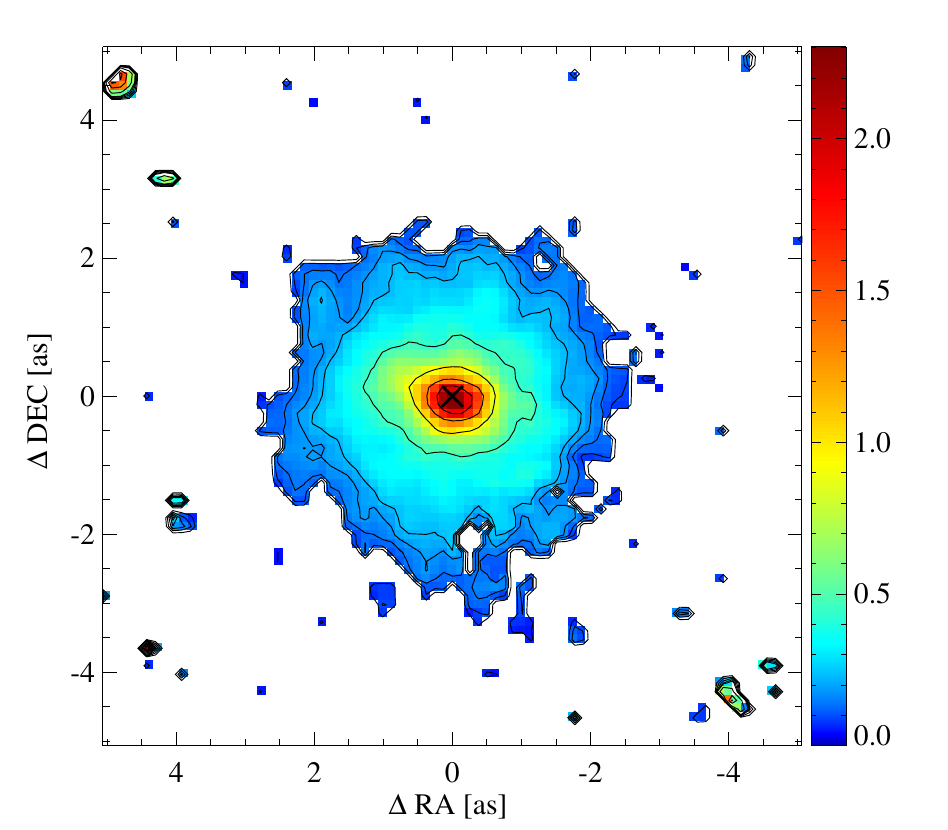}\label{fig:h203flux}}
\subfigure[FWHM H$_2$(1-0)S(2)]{\includegraphics[width=0.33\textwidth]{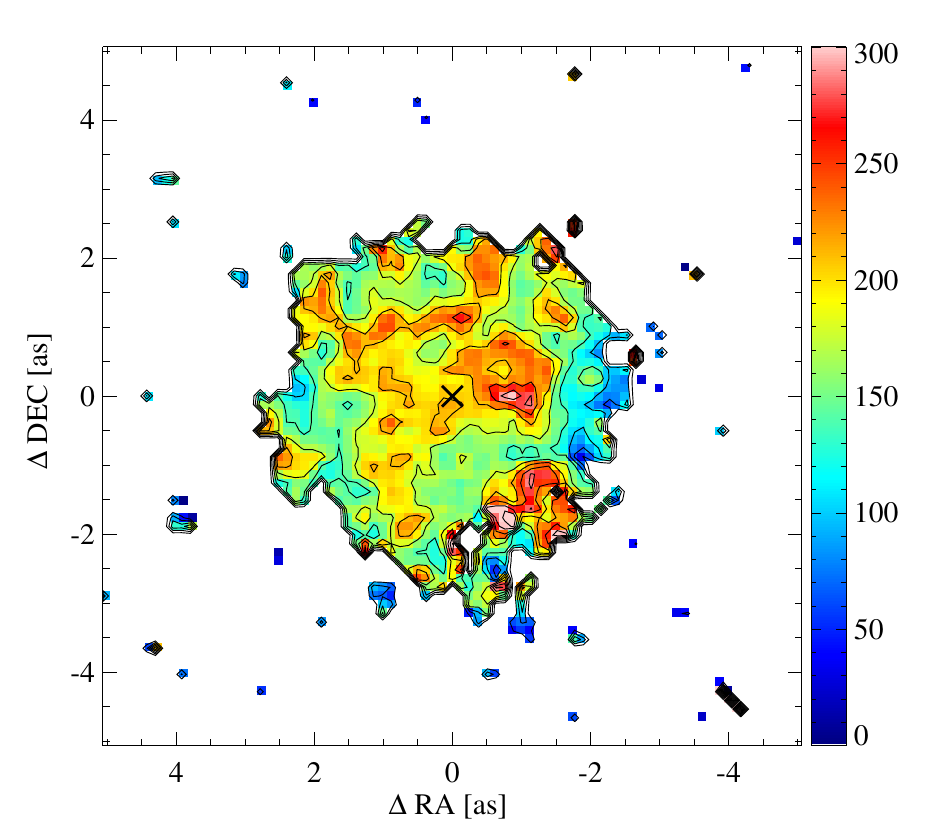}\label{fig:h203fwhm}}
\subfigure[EW H$_2$(1-0)S(2)]{\includegraphics[width=0.33\textwidth]{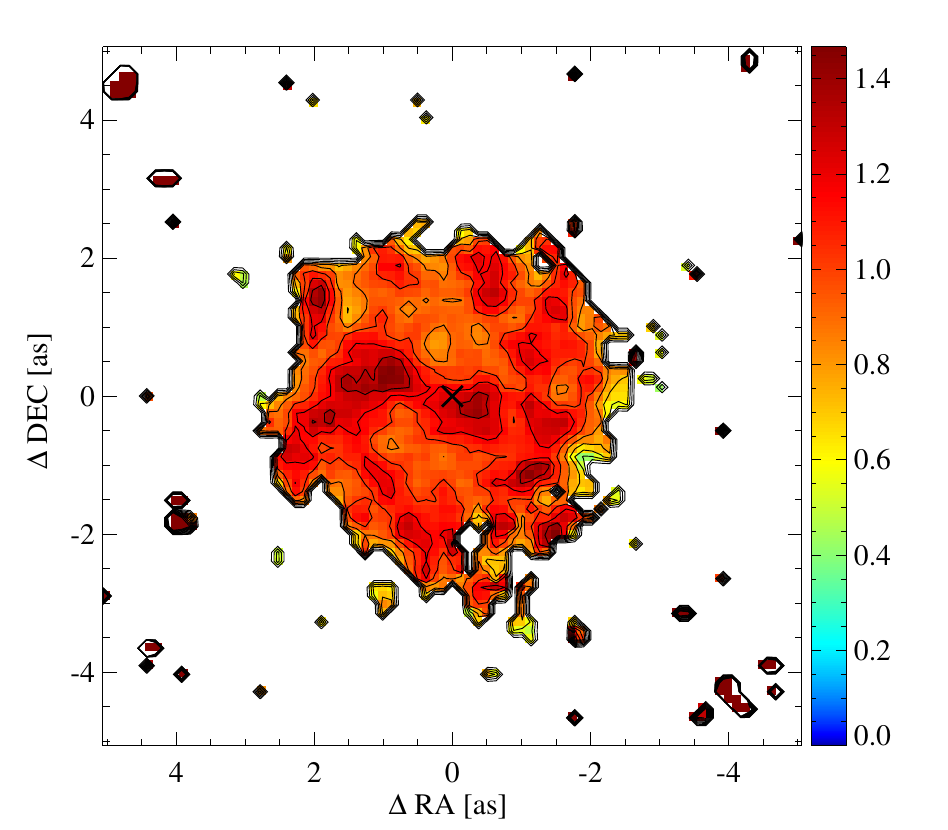}\label{fig:h203ew}}
\subfigure[Flux H$_2$(1-0)S(0)]{\includegraphics[width=0.33\textwidth]{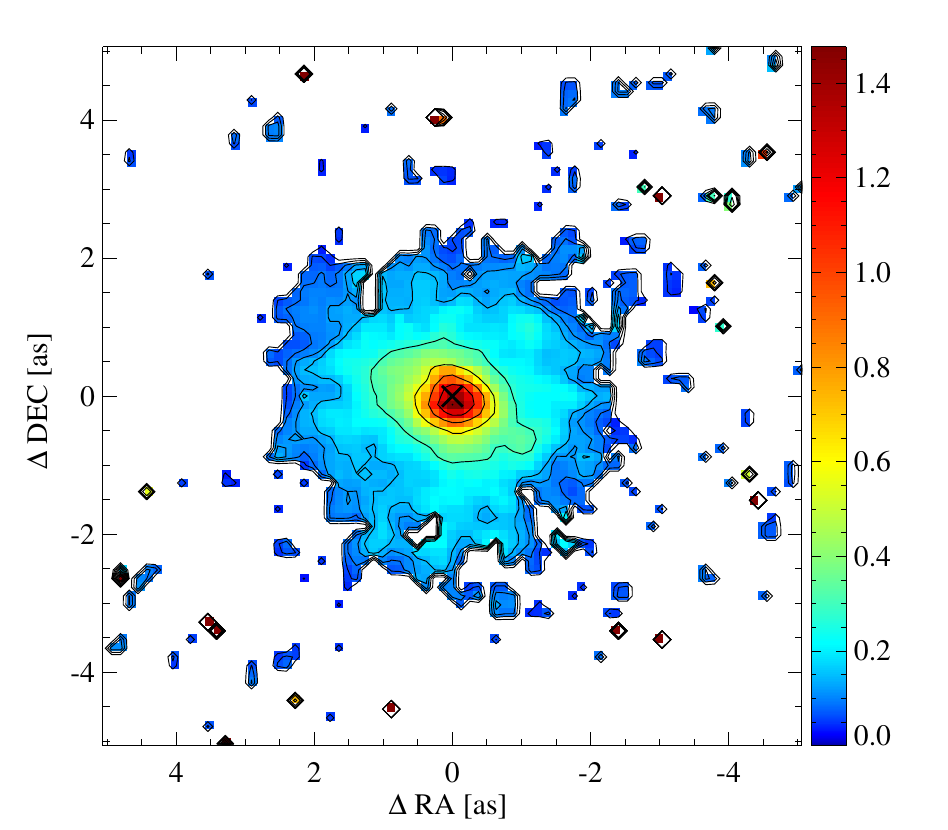}\label{fig:h222flux}}
\subfigure[FWHM H$_2$(1-0)S(0)]{\includegraphics[width=0.33\textwidth]{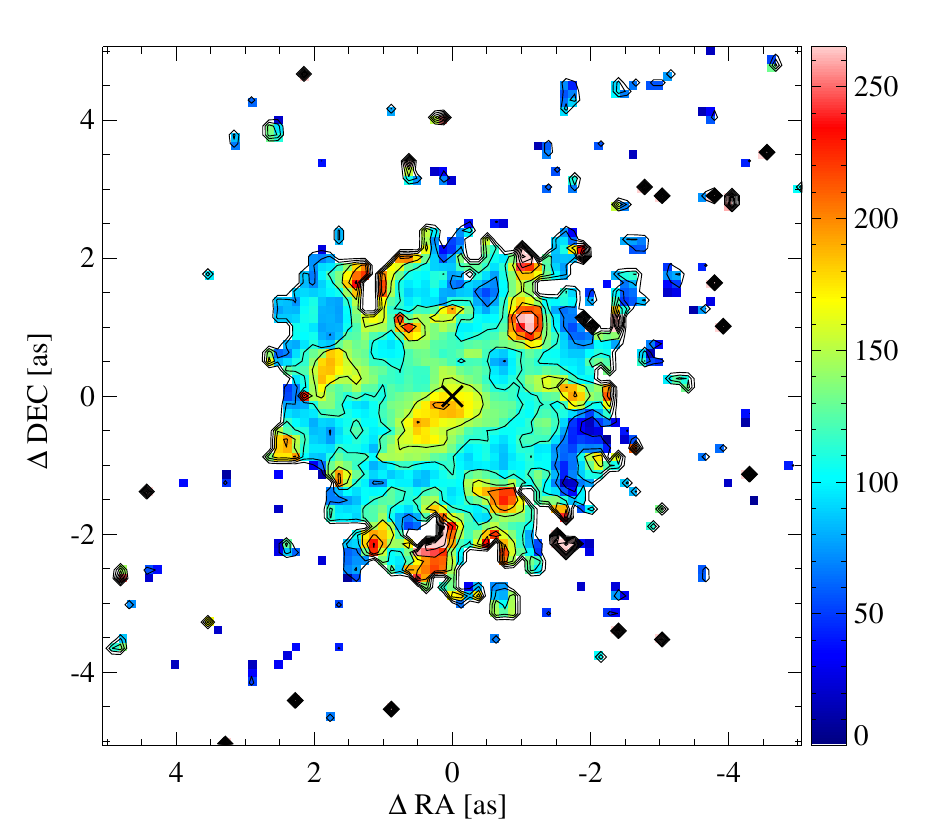}\label{fig:h222fwhm}}
\subfigure[EW H$_2$(1-0)S(0)]{\includegraphics[width=0.33\textwidth]{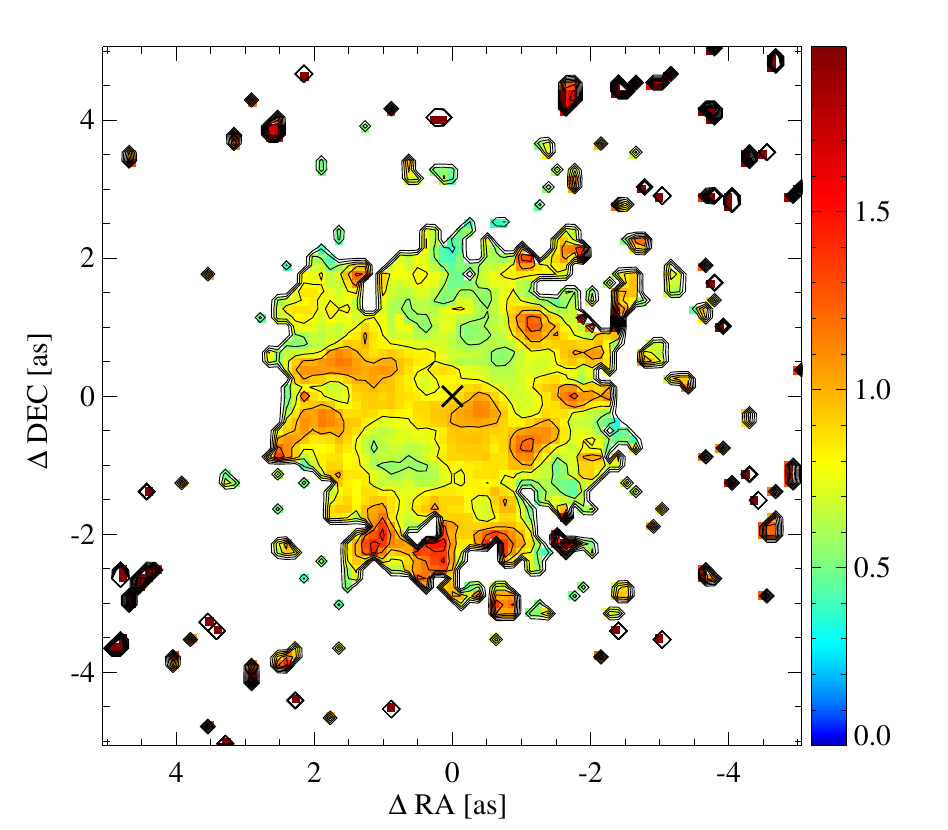}\label{fig:h222ew}}
\subfigure[Flux H$_2$(2-1)S(1)]{\includegraphics[width=0.33\textwidth]{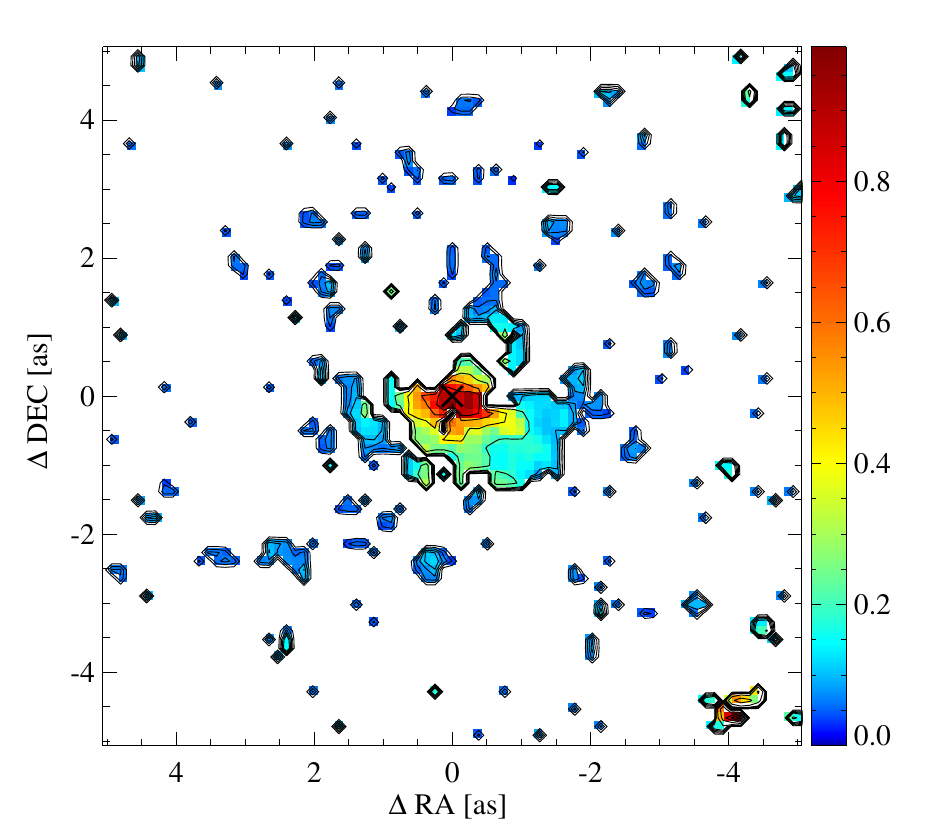}\label{fig:h2248flux}}
\subfigure[FWHM H$_2$(2-1)S(1)]{\includegraphics[width=0.33\textwidth]{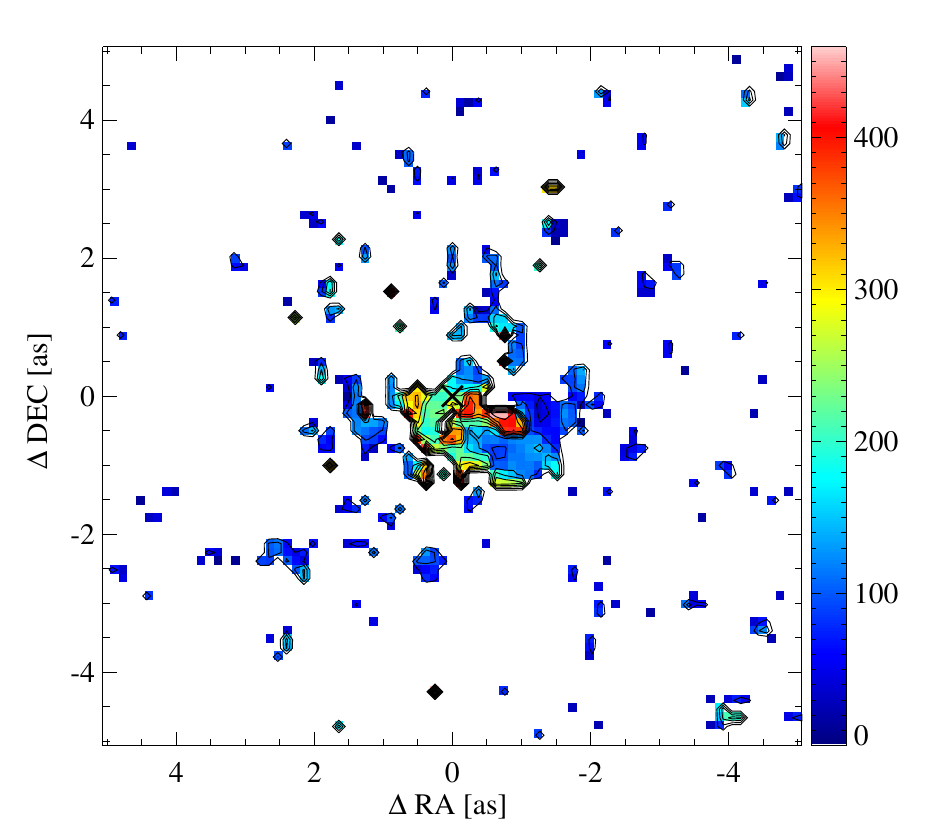}\label{fig:h2248fwhm}}
\subfigure[EW H$_2$(2-1)S(1)]{\includegraphics[width=0.33\textwidth]{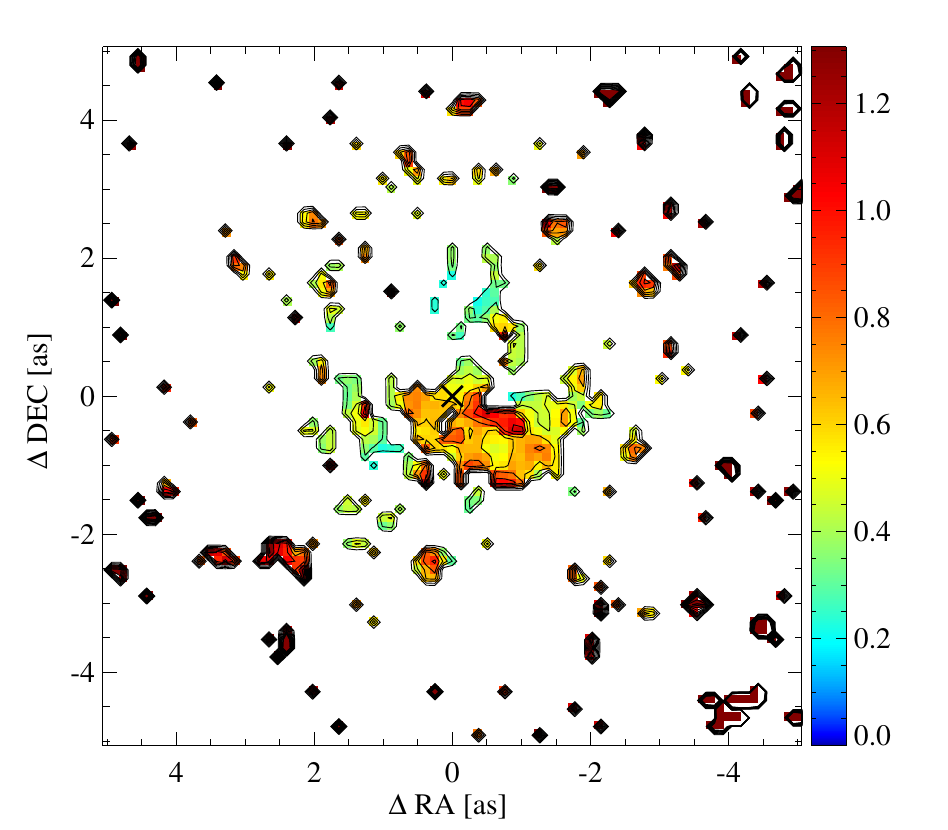}\label{fig:h2248ew}}
\caption{Molecular hydrogen emission lines. From left to right: flux [10$^{-20}$~W~m$^{-2}$], FWHM (corrected for instrumental broadening) [\kms] and EW [$\AA$] maps of, from top to bottom: H$_2$(1-0)S(3), H$_2$(1-0)S(2), H$_2$(1-0)S(0), and H$_2$(2-1)S(1).
}
\label{fig:lines2}
\end{figure*}

Figure \ref{fig:lines2} shows the flux, FWHM, and EW of all other detected H$_2$ lines detected up to $2.3$~\mic. The flux distribution is very similar in all H$_2$ lines and peaks on the nucleus. The FWHM is increased in the center along the minor axis and H$_2$(1-0)S(3) shows an increased width in the northeast and a decreased width in the northwest, similar to H$_2$(1-0)S(1). The nuclear spiral is detected in all EW maps which show an increased EW at the position of the spiral density wave.

\begin{figure*}[htbp]
\centering
\subfigure[{[\ion{Fe}{ii}]}]{\includegraphics[width=0.33\textwidth]{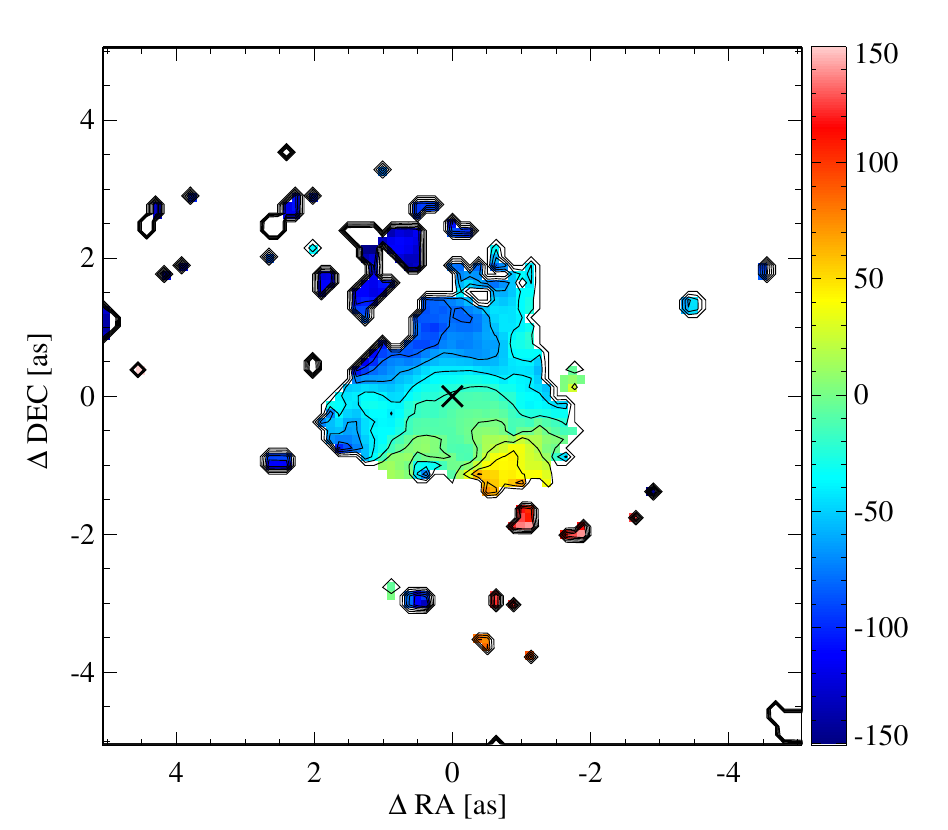}\label{fig:feiilosv}}
\subfigure[\ion{He}{i}]{\includegraphics[width=0.33\textwidth]{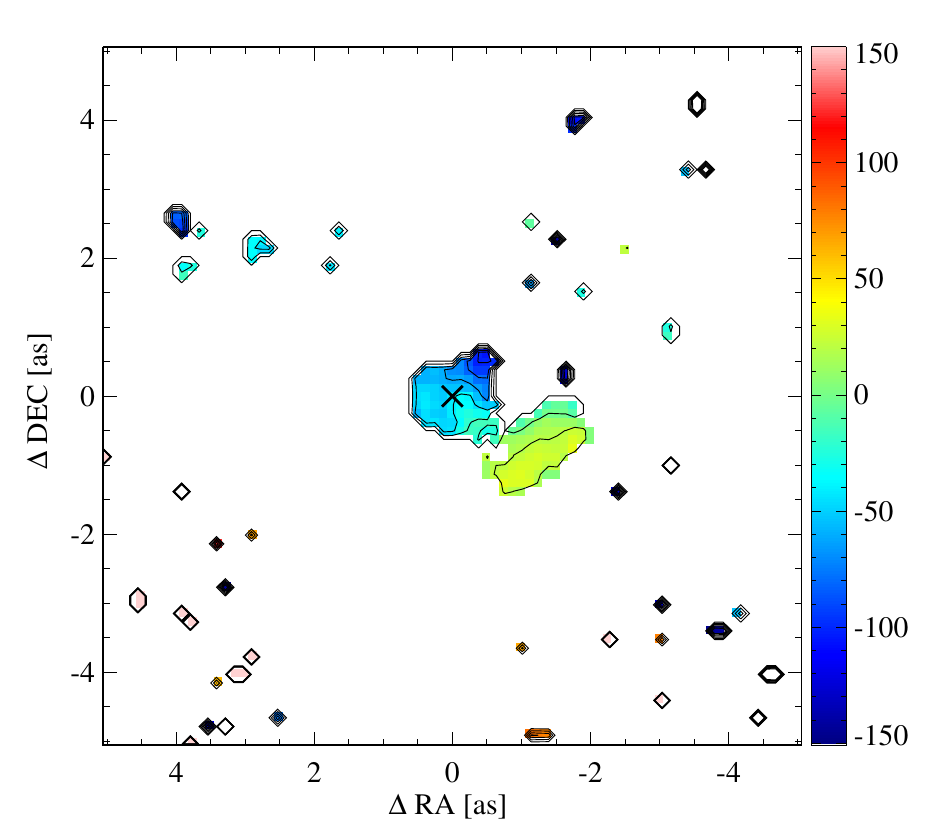}\label{fig:heilosv}}
\subfigure[\brg]{\includegraphics[width=0.33\textwidth]{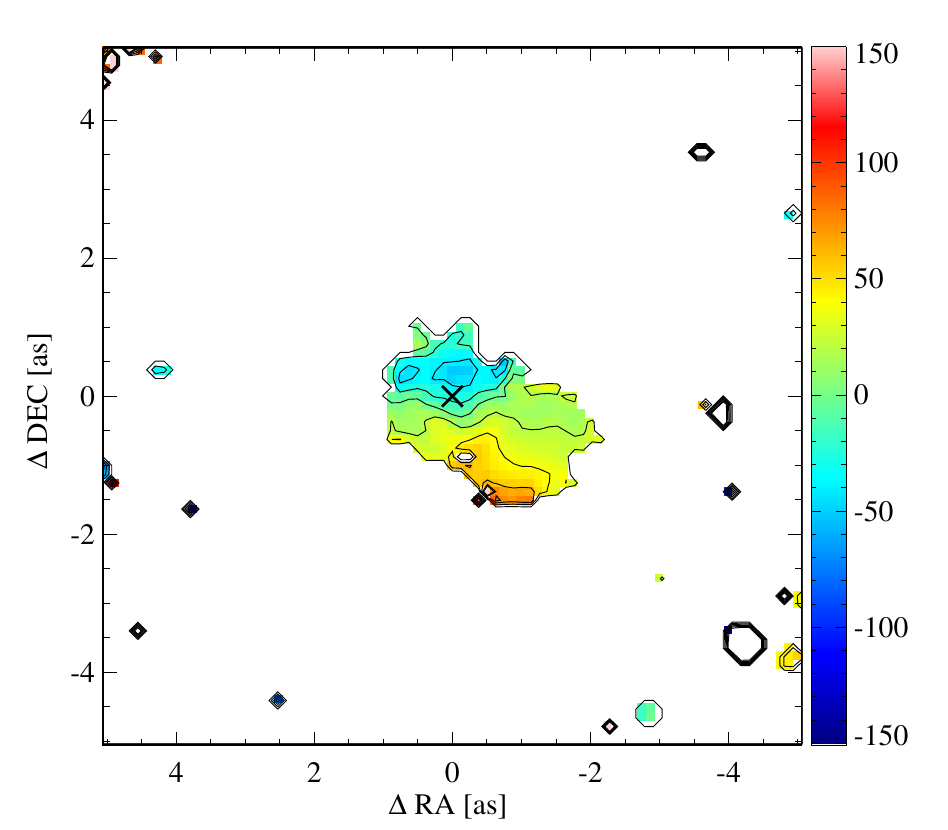}\label{fig:brglosv}}
\subfigure[H$_2$(1-0)S(3)]{\includegraphics[width=0.33\textwidth]{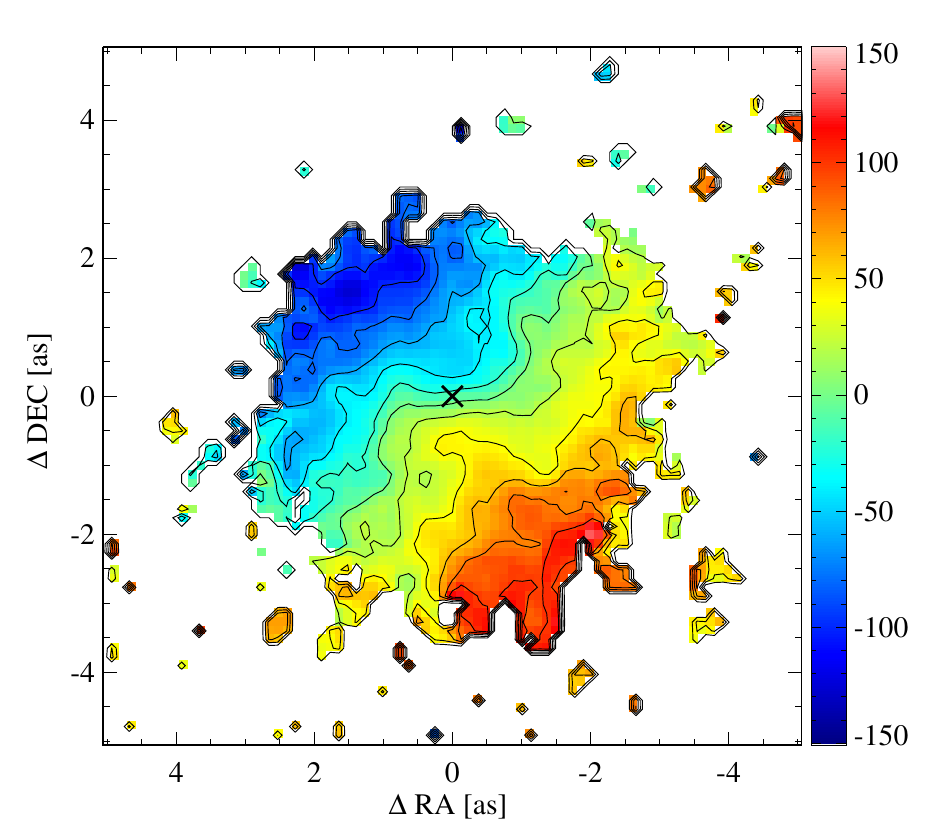}\label{fig:h21957losv}}
\subfigure[H$_2$(1-0)S(2)]{\includegraphics[width=0.33\textwidth]{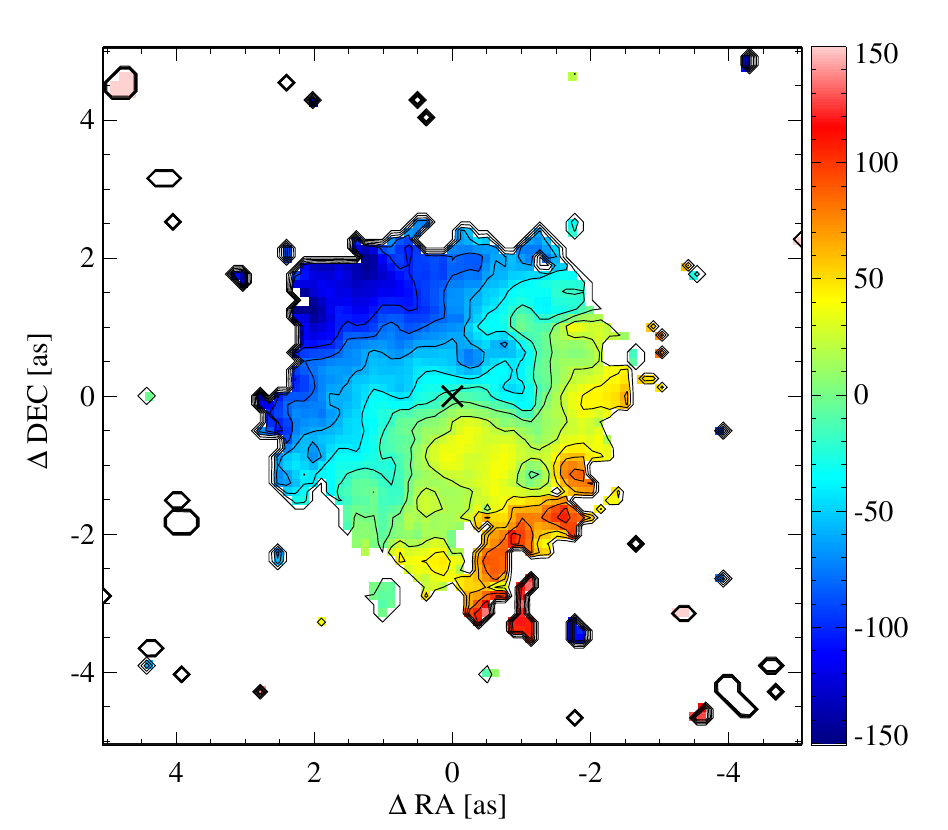}\label{fig:h203losv}}
\subfigure[H$_2$(1-0)S(0)]{\includegraphics[width=0.33\textwidth]{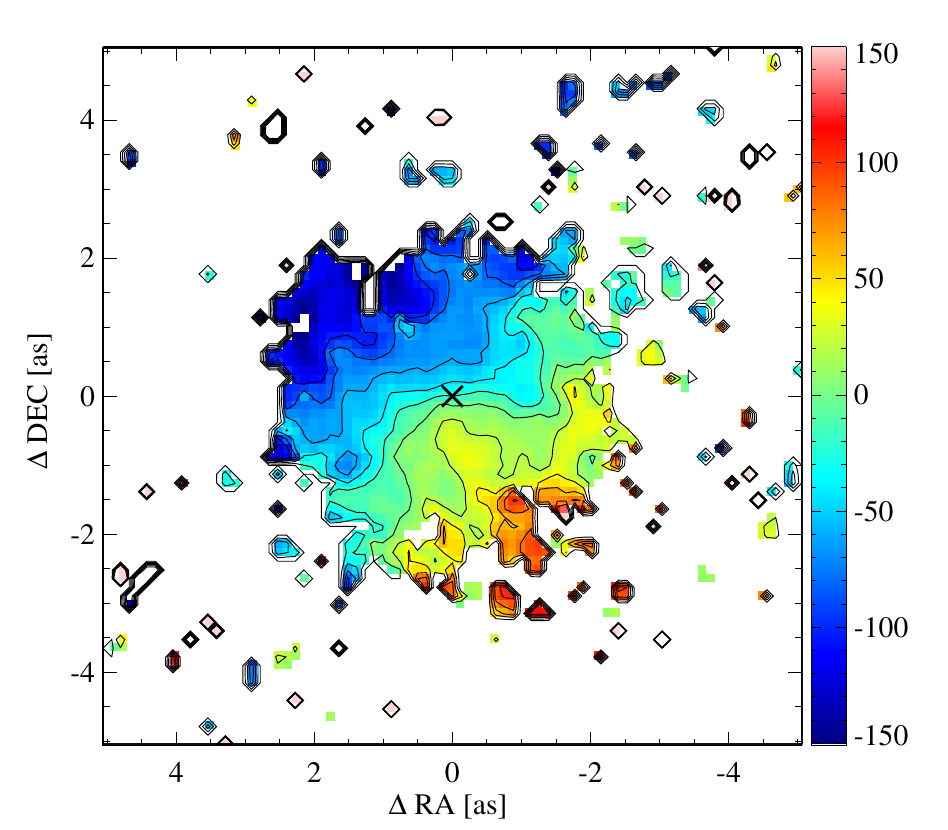}\label{fig:h222losv}}
\subfigure[H$_2$(2-1)S(1)]{\includegraphics[width=0.33\textwidth]{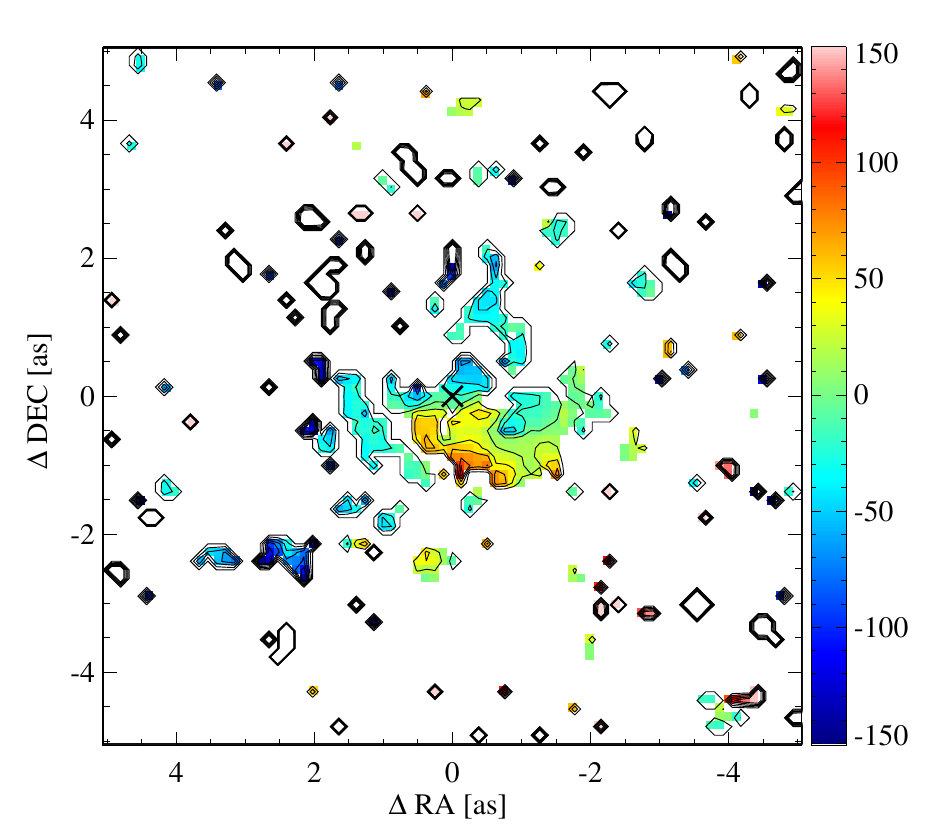}\label{fig:h2248losv}}
\caption{Presented are the LOSV [\kms] maps of, from left to right and top to bottom: \feii, \ion{He}{i},
narrow \brg, H$_2$(1-0)S(3), H$_2$(1-0)S(2), H$_2$(1-0)S(0), H$_2$(2-1)S(1).
}
\label{fig:LOSVlines}
\end{figure*}

Figure \ref{fig:LOSVlines} shows the LOSV of all H$_2$ lines from \ref{fig:lines2} and additionally the LOSV maps of the detected ionized gas. The LOSV in the molecular lines is very similar, as are flux and EW, indicating that they all originate from the same region, i.e. the nuclear molecular gas disk. For the ionized lines it is hard to tell because the region in which they are detected is not large enough to give information about the spatial velocity distribution. However, the LOSV fields do not look very different from the molecular LOSV fields (e.g., \brg). 

\end{appendix}

\end{document}

%% file: abstract.tex
\abstract{We present the results of near-infrared (NIR) H- and K-band European Southern Observatory SINFONI integral field spectroscopy (IFS) of the Seyfert galaxy \object{NGC 1566}. We investigate the central kpc of this nearby galaxy, concentrating on excitation conditions, morphology, and stellar content. NGC 1566 was selected from our NUGA (-south) sample and is a ringed, spiral galaxy with a stellar bar in northsouth direction (PA~$\sim5\degr$). The galaxy inhibits a very active Seyfert~1 nucleus but narrow line ratios from optical observations in the nuclear region are similar to Seyfert~2 galaxies. The recent strong activity phase, as inferred from strong variablity in X-ray to IR wavelengths, makes NGC 1566 an ideal candidate to look for feeding and feedback of a supermassive black hole.
We present emission and absorption line measurements in the central kpc of NGC 1566. Broad and narrow Br$\gamma$ lines were detected. The detection of a broad Br$\gamma$ component is a clear sign of a super-massive black hole in the center. Blackbody emission temperatures of $\sim1000$~K are indicative of a hot dust component, the torus, in the nuclear region. The molecular hydrogen lines, hydrogen recombination lines, and [\ion{Fe}{ii}] indicate that the excitation at the center is coming from an AGN. The central region is predominantly inhabited by molecular gas, dust, and an old K-M type giant stellar population. The molecular gas and stellar velocity maps both show a rotation pattern.
The molecular gas velocity field shows a perturbation toward the center that is typical for bars or spiral density waves. The molecular gas species of warm H$_2$(1-0)S(1) and cold $^{12}$CO(3-2) gas trace a nuclear gas disk of about $3\arcsec$ in radius with a nuclear spiral reaching toward the nucleus. From the equivalent width of H$_2$(1-0)S(1) a molecular ring with $r\lesssim3\arcsec$ can be inferred. This spiral seems to be an instrument that allows gas to fall toward the nucleus down to $<50$~pc scales. The excitation of molecular hydrogen in the nuclear gas disk is not clear but diagnostic diagrams show a distinction between the nuclear region and a $<9$~Myr old star forming region at the southwestern spiral arm. Possibly shocked gas is detected $\approx2\arcsec$ from the center, which is visible in dispersion maps of H$_2$(1-0)S(1) and $^{12}$CO(3-2) and in the 0.87~mm continuum.
}

%% file: introduction.tex
\section{Introduction}

The NUclei of GAlaxies (NUGA) project \citep{garcia-burillo_molecular_2003} started with the IRAM Plateau de Bure Interferometer (PdBI) and 30 m single-dish survey of nearby low-luminosity active galactic nuclei (LLAGN) in the northern hemisphere. The project is ideally suited to map the distribution and dynamics of (cool) molecular gas in the inner kpc of LLAGN and to study the possible mechanisms for gas fueling at a high angular resolution ($\approx$ 0\farcs5 -- 2\arcsec) and high sensitivity. The ongoing implementation of the Atacama Large Millimeter/submillimeter Array (ALMA) in the Atacama desert in Chile finally allows the NUGA project to expand to the southern hemisphere \citep{combes_ALMA_2013,combes_ALMA_2014} at an even higher angular resolution ($\sim 6-37$~mas, assuming the full array is used) and sensitivity. The Spectrograph for INtegral Field Observations in the Near Infrared (SINFONI) adds complementary information to the NUGA goal in the near-infrared \citep[NIR, ][]{smajic_ALMA-backed_2014}. By maping (hot) gas and the mass dominating stellar population we investigate star formation and the feeding and feedback of nearby LLAGN in the NIR.

\subsection{Feeding and feedback in AGN}


Unresolved, powerful, and highly ionizing emission detected in the centers of galaxies is thought to stem from accretion events onto a supermassive black hole (SMBH). These galaxies are said to have an active galactic nucleus (AGN). According to the unified model of AGN this active nucleus consists of a SMBH surrounded by an accretion disk at scales of up to a few lightdays which ionizes the surrounding gas on scales from several lightdays, in Seyfert galaxies, up to lightyears, in Quasi Stellar Objects, (e.g., broad line region - BLR) to several hundred parsecs (e.g., narrow line region - NLR) and up to even larger scales via jets. The torus, a dust and gas mantle, surrounds the AGN on parsec to tens of parsecs scales and is thought to be responsible for Seyfert~1 (torus almost face-on toward the observer) and Seyfert~2 (torus almost edge-on) classifications of AGN. The existence of the torus can be inferred from high column densities toward Seyfert~2 AGN and dust blackbody emission in Seyfert~1 AGN with temperatures up to the sublimation temperature of dust ($\approx1300$~K).

The host galaxy and its SMBH have been found to exhibit tight correlations. These correlations mostly apply to the stellar bulge \citep[e.g.,][]{magorrian_demography_1998,ferrarese_fundamental_2000,marconi_relation_2003}.
The correlations connect the mass of the central SMBH with the mass, luminosity, and kinematics of the bulge. Latest studies show that these correlations may depend on the galaxy classification \citep[e.g., barred galaxies,][]{graham_m_2013} and that several galaxies show an over-luminous bulge or an under-massive SMBH \citep{busch_low-luminosity_2014}. \citet{lasker_supermassive_2014} show that the total luminosity of the host galaxy seems a more robust tracer of the SMBH mass than the bulge luminosity.
The distribution of the gas in the host is essential to understand these correlations, since gas is the progenitor of stars which mainly contribute to mass, luminosity and kinematics of the bulge. Gravitational torques are one of the strongest form of force to act on the distribution of the gas on these scales.
Gravitational mechanisms that exert gravitational torques such as galaxy-galaxy interactions (e.g., galaxy merger) or non-axisymmetries within the galaxy potential (e.g., spiral density waves or stellar bars on large scales) can lead to loss of angular momentum in the gas. Hydrodynamical mechanisms such as shocks and viscosity torques introduced by turbulences in the interstellar medium (ISM) can remove angular momentum from gas, too. The original NUGA (north) project has already studied the gaseous distribution in more than ten nearby galaxies ($\approx4-40$~Mpc) with results that show a variety of morphologies in nuclear regions, including bars and spirals \citep{garcia-burillo_molecular_2005,boone_molecular_2007,hunt_molecular_2008,lindt-krieg_molecular_2008,garcia-burillo_molecular_2009}, rings \citep{combes_molecular_2004,casasola_molecular_2008,combes_molecular_2009} and lopsided disks \citep{garcia-burillo_molecular_2003,krips_molecular_2005,casasola_molecular_2010}.

On large scales, hundreds to thousands of parsecs, gravitational torques are the strongest mechanism to successfully transport gas close to the nucleus, whereas viscosity torques can take over on smaller scales \citep[$<200$~pc, e.g.,][]{combes_molecular_2004,van_der_laan_molecular_2011}. Therefore, large-scale stellar bars are an important feature to transport gas toward the inner Lindblad resonance (ILR) \citep[e.g.,][]{sheth_secular_2005} where the formation of nuclear spirals and rings is induced.

The NUGA sample studies the cold gas distribution.
We can use the cold gas distribution as a complement to our SINFONI NIR observations and compare it to the sites of hot molecular and ionized gas \citep{combes_ALMA_2013,combes_ALMA_2014,smajic_ALMA-backed_2014}.
This enables us to identify ongoing star formation sites and regions ionized by shocks (i.e. super novae (SN) or outflows). Furthermore, we are able to compare the distribution of cold (e.g., CO, HCN) and hot (e.g., H$_2$) molecular gas that will give us a more clear insight on feeding and feedback of the AGN through its ambient gas reservoir. \citet[][and references therein]{riffel_feeding_2013} assume that molecular gas (e.g., H$_2$) traces the ambient disk structure and in some cases the streaming motions, i.e. the feeding of the SMBH, whereas ionized gas often traces outflowing material that is above the galaxy disk plane, i.e. the feedback from the SMBH. However, molecular emission lines in the NIR (e.g., H$_2$(1-0)S(1) emission) are known shock tracer that are found in regions shocked by jets or outflows \citep[e.g.,][and references therein]{riffel_feeding_2014,davies_fueling_2014}. Lately, \citet[][]{garcia-burillo_molecular_2014} find that cold molecular gas (e.g., $^{12}$CO(3-2) emission) can be found in outflows.

Here, we analyze the interactions of nuclear star formation sites and the AGN with regard to fueling and feedback, of both participants. Using the differently excited H$_2$ lines (e.g., H$_2$~$\lambda$2.12~$\mu$m, 1.957~$\mu$m, 2.247~$\mu$m) and the hydrogen recombination line \brg\ in K-band and the forbidden transition [\ion{Fe}{ii}]~$\lambda1.644$~\mic\ in H-band we are able to constrain the excitation type (e.g., thermal, non-thermal) of the warm gas and the excitation temperature \citep{mouri_molecular_1994,rodriguez-ardila_molecular_2004,zuther_mrk_2007}. The cold gas information can then be compared to our results.

We will use the stellar absorption features (e.g., \ion{Si}{i}, CO(6--3), \ion{Mg}{i}, \ion{Na}{i}, CO(2--0)) to get insight into the star formation history of the nuclear region \citep[e.g.,][]{davies_close_2007}. Star formation, recent or ongoing, on scales of $0.1-1$~kpc around the nucleus is an important process which is frequently found in all types of AGN in contrast to quiescent galaxies \citep[e.g.,][]{cidfernandes_star_2004,davies_star-forming_2006,busch_low-luminosity_2015}. The debate if outflows from the AGN quench or initiate star formation is still going on, but it is most probable that outflows can show both effects. 

\subsection{NGC 1566}

\begin{figure}[htbp]
\centering
\includegraphics[width=0.47\textwidth]{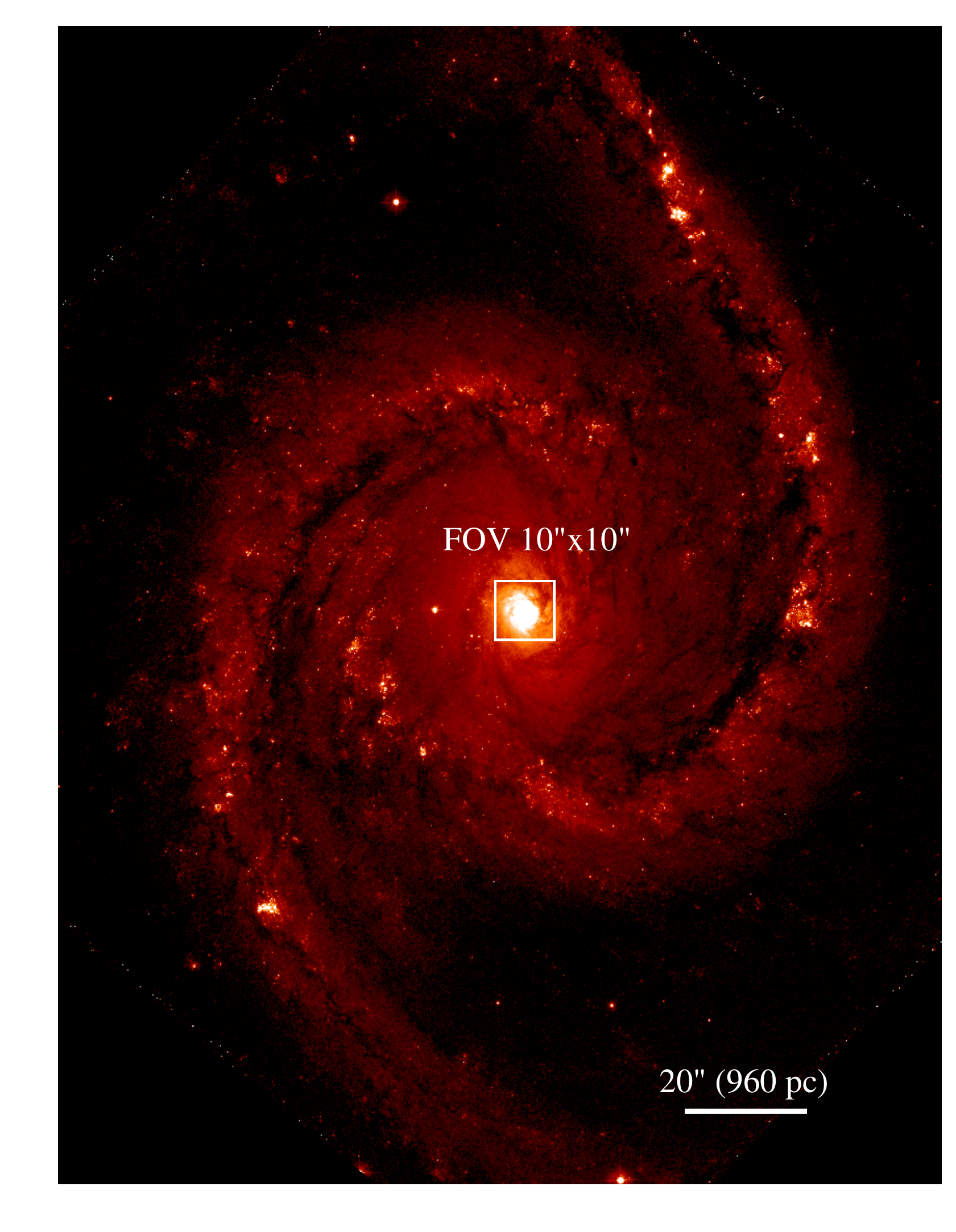}
\caption{
An HST image of NGC 1566 taken with the WFC3 UVIS2 F438W filter at a pivot wavelength of $4326$~$\AA$. North is up and east is left. The white rectangle marks our $10\arcsec\times10\arcsec$ field of view (FOV).
}
\label{fig:ngc1566}
\end{figure}

NGC 1566 is a barred, ringed, spiral SAB(s)bc galaxy in the Dorado group \citep{mulchaey_fueling_1997,reunanen_near-infrared_2002} harboring a low luminosity AGN (LLAGN) at a redshift of $z\approx0.005017$ \citep{koribalski_1000_2004}.  The bar is oriented in the north-south direction (PA$\sim5\degr$) and the ring has a diameter of 1.7 kpc (see Fig.~\ref{fig:ngc1566}). \citet{alloin_recent_1985} detect broad Balmer lines with widths of FWHM$_{\mbox{H}\beta}$ = 2400$\pm$300 km s$^{-1}$. They also measure an increase in flux of the broad H$\beta$ line by a factor of $4-5$ within 24 days. They find the narrow line ratio of H$\alpha$/H$\beta$ to be $\approx$ 3.1, which is typical for Seyfert~2 galaxies, however, the broad line flux variation and the broad to narrow line ratio H$\beta$(b)/H$\beta$(n) $\approx$ 10 shows that NGC 1566 is in a high activity state typical for Seyfert~1.2 galaxies. The light crossing time of the NLR is about $10^2 - 10^3$ years, hence, the NLR line ratios suggest that in the past few 100 years NGC 1566 has shown more the characteristics of a Seyfert~2 galaxy than that of a Seyfert~1. \citet{kriss_faint_1991} measure a redshift of the broad emission lines of 200 -- 1000 km s$^{-1}$. This might result from gravitational redshift \citep{netzer_profiles_1977} but can also be induced by an outflow of optically thick clouds at the far side or an infall of clouds at the near side of the BLR where the far side is obscured from our view. 

NGC 1566 shows variability from the X-ray to IR wavelengths. \citet{baribaud_variability_1992} conclude that X-ray flickering should occur on time scales of 5$\times$10$^{3}$ s. They also note that the NIR flux variation (heated dust reacting to UV flux changes) happens within the central 450 pc from the center at time scales similar to the broad H$\alpha$ variability time scale. The extinction toward the nuclear region of NGC 1566 from NLR and BLR line ratios seems to be negligible. \citet{baribaud_variability_1992} also calculate a dust mass M$_{\mbox{\tiny dust}}\approx$ 7$\times$10$^{-4}$~M$_{\odot}$ and a dust evaporation radius of about 47 light-days. \citet{reunanen_near-infrared_2002} used slit spectroscopy on the nucleus of NGC 1566. They detect [\ion{Fe}{ii}] in the H-band and the H$_2$ $\lambda$2.122 $\mu$m line in the K-band. They note that the lines and the continuum appear weaker along the NLR cone than perpendicular to it. They do not detect narrow \brg\ emission only a broad Br$\gamma$ component with an FWHM$_{\tiny Br\gamma}=2100$~km~s$^{-1}$.

NGC 1566 is an interesting candidate to investigate the feeding and feedback of the nuclear region and the AGN since the galaxy seems to be waking up from a low activity state in the past few hundred years (low ionization level in the NLR) and shows signs of high activity with strong variability and high ionization on scales smaller than the NLR. 
\citet{combes_ALMA_2014} find from ALMA $^{12}$CO(3-2) and HST data that gravitational torques are a very likely cause to drive gas infall in NGC 1566. They derive negative torques from 300~pc down to 50~pc. From this point dynamical friction can drive the gas to the nucleus.

We will investigate the central $10\arcsec\times10\arcsec$ to find if the hot gas distribution and kinematics are similar to the cold gas observed with ALMA. Are there any signs of a feeding of the SMBH (e.g., strong ionization of the nuclear region, streaming gas motion)? How strongly is it accreting? Is any strong feedback (e.g., jets) visible in the hot gas? Is gas infall accompanied by star formation or is star formation rather hindered by the torque budget found by \citet{combes_ALMA_2014}?\\

This paper is structured as follows: in Sect. 2 we present the observation and the data reduction. Section 3 states the results of our study and how these were derived. Section 4 discusses the results and compares them with literature. In Sect. 5 a summary of the results is given and the conclusions we take from our study are phrased.

Throughout the paper we adopt a systemic velocity of 1504~\kms\ and a distance of 10~Mpc for NGC 1566, following \citet{alloin_recent_1985,combes_ALMA_2014}.


%% file: obs_red.tex
\section{Observation and Data Reduction}
\label{sec:obsred}
\begin{figure}[htbp]
\centering
\includegraphics[width=0.35\textwidth,angle=90]{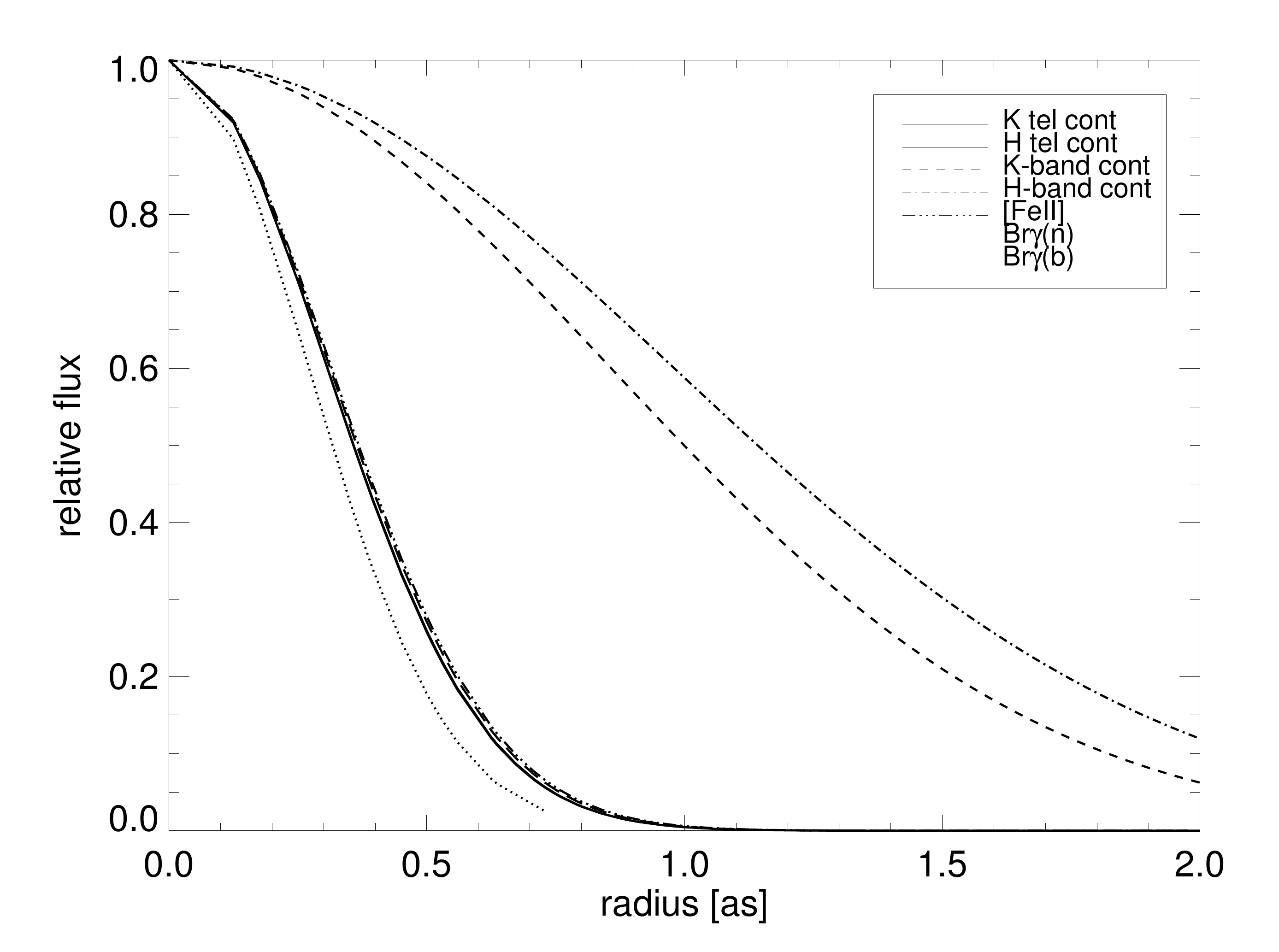}
\caption{The radial profiles of the telluric star in K- and H-band, the K- and H-band galaxy continuum, \feii, narrow \brg\ and broad \brg.}
\label{fig:PSF}
\end{figure}

In this paper we present the results of our ESO SINFONI \citep{eisenhauer_sinfoni_2003,bonnet_first_2004-1} observation of NGC 1566 with the Unit Telescope 4 of the Very Large Telescope in Chile. The 0\farcs25 plate scale with an $8\arcsec\times8\arcsec$ FOV without adaptive optics assistance was used. The average seeing was $\approx0\farcs5$. To increase the FOV to $12\arcsec\times12\arcsec$ and minimize the overlap of dead pixels in critical areas a dithering sequence of the FOV by $\pm2\arcsec$ was introduced. However, the outer $2\arcsec$ have too low quality data, hence for the analysis a FOV of 10$\arcsec\times$10$\arcsec$ was used. The dithering was done at 9 positions where the central $4\arcsec\times4\arcsec$ were observed with the full integration time. The gratings used are the H-band grating at a spectral resolution of R~$\approx 3000$ and the K-band grating at a resolution of R~$\approx 4000$. Both bands were observed at a digital integration time of 150 seconds and an TST... nodding sequence (T: target, S: sky), to increase on-source time. The overall integration time on the target source in H-band is 2550 seconds and in K-band 3000 seconds with an additional 1200 seconds in H-band and 1500 seconds in K-band on sky. 
The G2V star HIP 33144 was observed in H-band and in K-band within the respective science target observation. The observing strategy was to observe it twice (star in opposite corners of the FOV) with an integration time of 2 seconds. 
The ESO SINFONI pipeline was used for data reduction except for atmospheric OH line correction and final cube creation which were done manually. Detector specific corrections were performed manually, as well. For more details on the reduction see \citet{smajic_ALMA-backed_2014}.
The standard star was used to correct for telluric absorption in the atmosphere and to perform a flux calibration of the target. A high S/N solar spectrum was used to correct for the black body and intrinsic spectral features of an G2V star \citep{maiolino_correction_1996}. The solar spectrum was convolved with a Gaussian to adapt its resolution to the resolution of the standard star spectrum. The solar spectrum edges were interpolated by a black body with a temperature of T~$= 5800$~K. The telluric standard star spectrum was extracted by taking the total of all pixels within the radius of $3 \times$FWHM$_{\mbox{\tiny PSF}}$ of the point spread function \citep[PSF; ][]{howell_handbook_2000}, centered on the peak of a two-dimensional Gaussian fit. The flux calibration of the target source was performed during the telluric correction procedure. We referenced the standard star counts at $\lambda$1.662~$\mu$m and $\lambda$2.159~$\mu$m in H- and K-band respectively to the flux given by the 2MASS All-sky Point Source Catalogue\footnote{This publication makes use of data products from the Two Micron All Sky Survey, which is a joint project of the University of Massachusetts and the Infrared Processing and Analysis Center/California Institute of Technology, funded by the National Aeronautics and Space Administration and the National Science Foundation.}.
To determine the spatial resolution of our observation we looked at the radial profiles of the telluric star, the continuum emission in H- and K-band, and at the radial profiles of the emission lines \feii, narrow \brg, and broad \brg\ (see Fig.~\ref{fig:PSF}). 
We use the broad component of the hydrogen recombination line \brg\ to determine an accurate value of the FWHM of the PSF for the K-band because we can measure this value from the science data itself. Figure \ref{fig:PSF} shows that the narrow component of \brg\ and \feii\ as well as the telluric stars in H- and K-band show about the same spatial extent. From this we infer that the spatial resolution in H- and K-band is similar. We measure a FWHM of the broad \brg\ component of $\sim0\farcs59$ which corresponds to $\sim$~29~pc. Spatially the PSF shows an elongation in the east-west direction slightly rotated by 10$\degr$ (Fig.~\ref{fig:lines1}).

The correction of the OH lines, the cube alignment, the final cube creation, the linemap and spectra extraction was conducted using our own IDL routines.

We use calibrated 350 GHz ALMA data and compare these with our NIR results.
The final product data is publicly available from the ALMA archive under project ID 2011.0.00208.S (PI: Combes). Details on observational setup, calibration, imaging and quality can be found in \citet{combes_ALMA_2014} and in the project reports in the ALMA archive.
The line cube comprises 50 channels of 10~km/s width and a beam of $0\farcs64\times0\farcs43$ at an PA of $123\degr$. The rms achieved is $0.05$~mJy~beam$^{-1}$ in the continuum and $\sim1.3$~mJy~beam$^{-1}$ in the line cubes.
Imaging and parts of the analysis have been conducted with the CASA software \citep[v3.3][]{mcmullin_casa_2007}.

%% file: res_and_disc.tex
\section{Results and discussion}

We resolve the central 485~pc of NGC 1566 at a seeing limited spatial resolution of 29~pc using the integral field spectrograph SINFONI at the VLT. Narrow and broad components of the hydrogen recombination line Br$\gamma$~$\lambda$2.166~$\mu$m and several rovibrational molecular hydrogen lines (e.g. H$_2$(1-0)S(1)~$\lambda2.12$~$\mu$m) are identified. Additionally, several stellar absorption features in K-band (e.g. CO(2-0)~$\lambda$2.29~$\mu$m, NaD~$\lambda$2.207~$\mu$m, CaT~$\lambda$2.266~$\mu$m) are detected. The H-band shows a variety of stellar absorption features (e.g. \ion{Si}{i}~$\lambda$ 1.59~$\mu$m, CO(6-3)~$\lambda$1.62~$\mu$m) and it harbors the [\ion{Fe}{ii}]~$\lambda$1.644~$\mu$m line, which is important for NIR line diagnostics. The flux and the FWHM of all detected emission lines are summarized in table \ref{tab:region} for the regions: center $r=1\arcsec$, center $r=5\arcsec$, center $r={\rm PSF}$, star formation region (SFr), star formation region at PSF sized aperture (SF PSF), and center at PSF sized aperture (cPSF), see Fig. \ref{fig:lines1} for more detail. The emission line FWHMa presented in this paper
are corrected for instrumental broadening.
The cold molecular gas tracer $^{12}$CO(3-2) and 0.87~mm emission observed with ALMA are used to compare hot NIR gas and dust with cold gas and dust in the sub-mm.

\subsection{The distribution of gas}
We describe and compare the distribution of the observed gas, i.e., ionized \feii\ and \brg\ gas and molecular H$_2$(1-0)S(1) and $^{12}$CO(3-2) gas.

\subsubsection{Ionized gas}
\label{sec:ionizedgas}

Ionized gas is detected in the \feii~$\lambda$1.644~$\mu$m, \brg, and \hei~$\lambda$2.06~$\mu$m emission lines in the NIR.

\begin{figure*}[htbp]
\centering
\subfigure[Flux $\mbox{[}$\ion{Fe}{ii}$\mbox{]}$]{\includegraphics[width=0.33\textwidth]{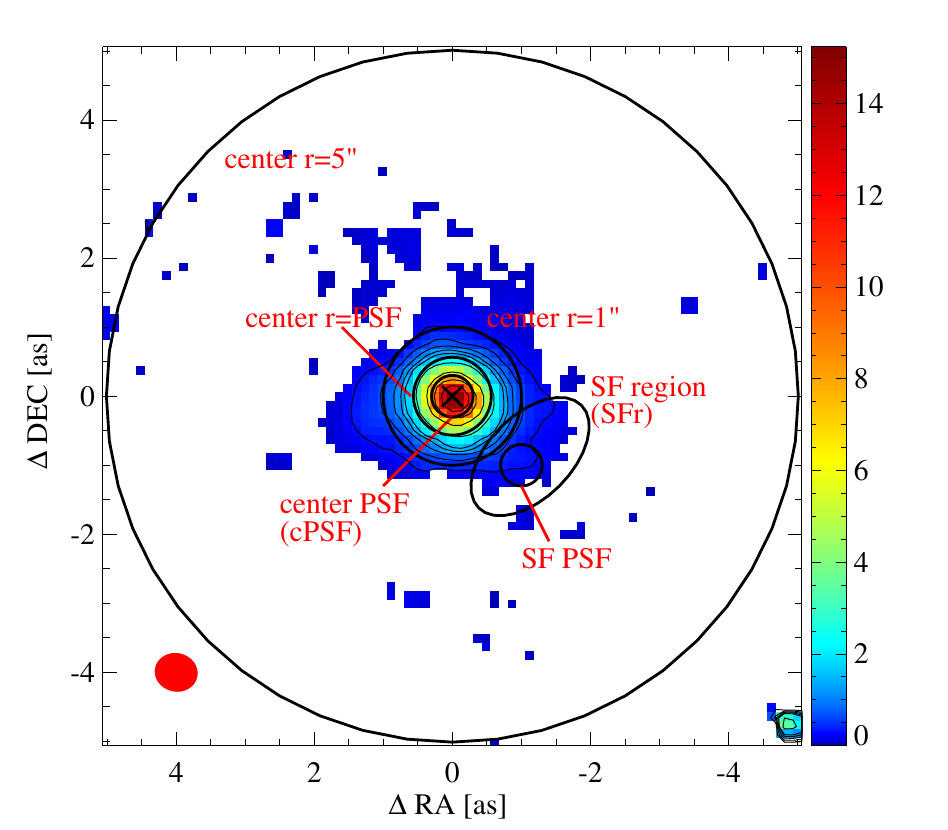}\label{fig:feiiflux}}
\subfigure[FWHM $\mbox{[}$\ion{Fe}{ii}$\mbox{]}$]{\includegraphics[width=0.33\textwidth]{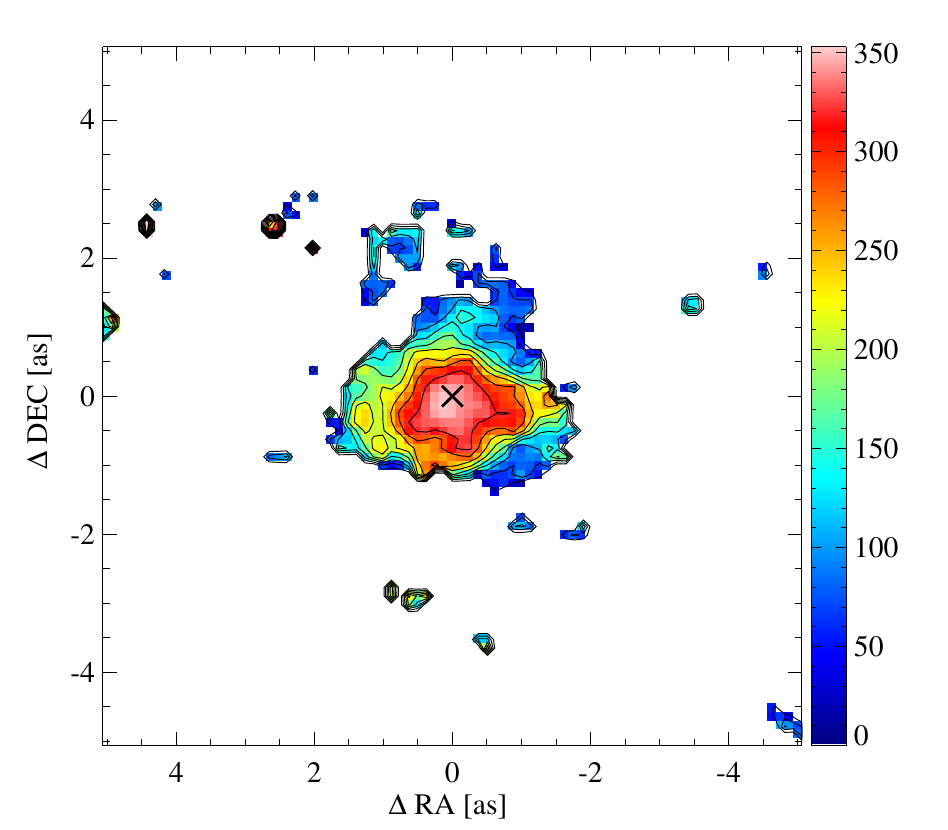}\label{fig:feiifwhm}}
\subfigure[EW $\mbox{[}$\ion{Fe}{ii}$\mbox{]}$]{\includegraphics[width=0.33\textwidth]{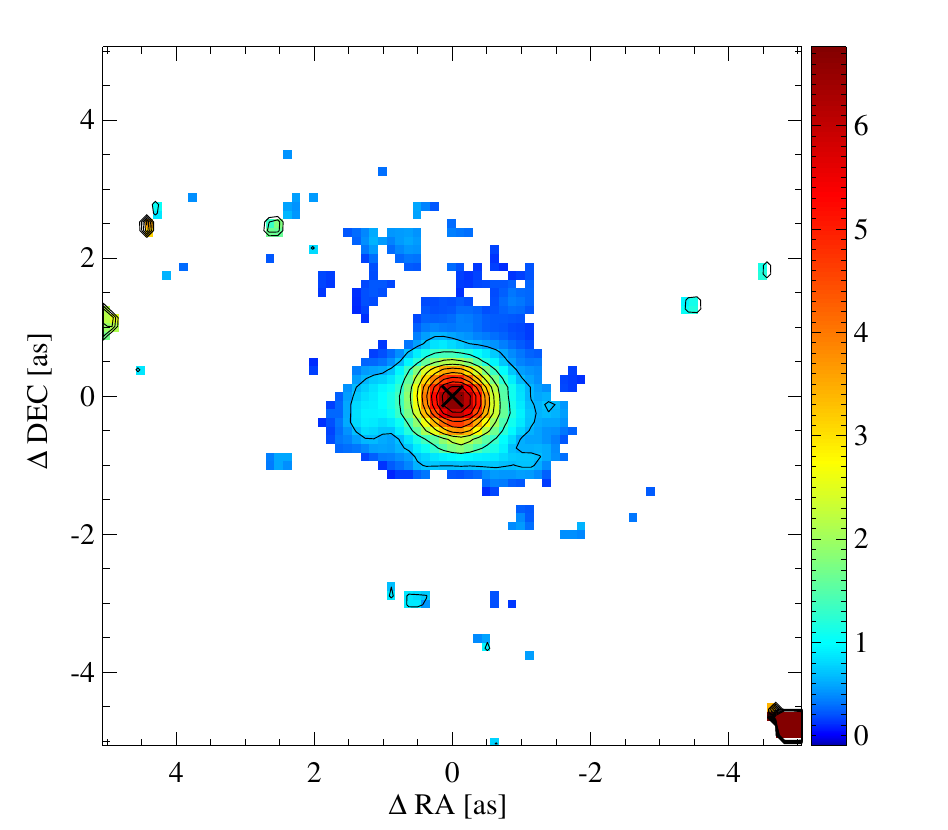}\label{fig:feiiew}}
\subfigure[Flux \ion{He}{i}]{\includegraphics[width=0.33\textwidth]{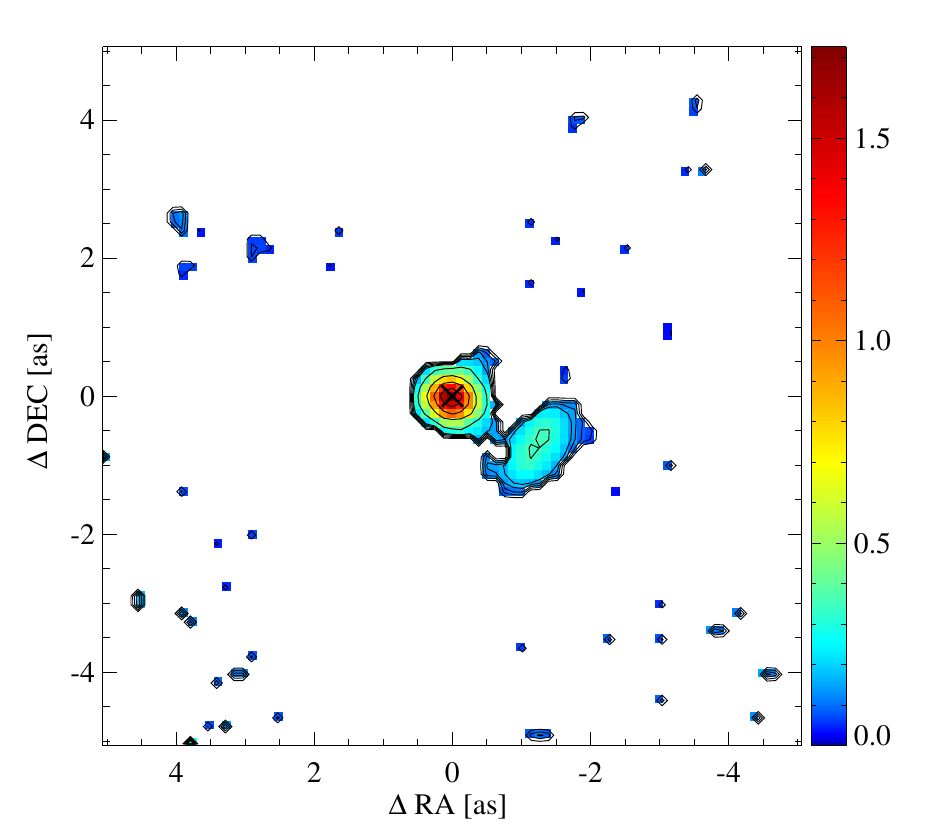}\label{fig:heiflux}}
\subfigure[FWHM \ion{He}{i}]{\includegraphics[width=0.33\textwidth]{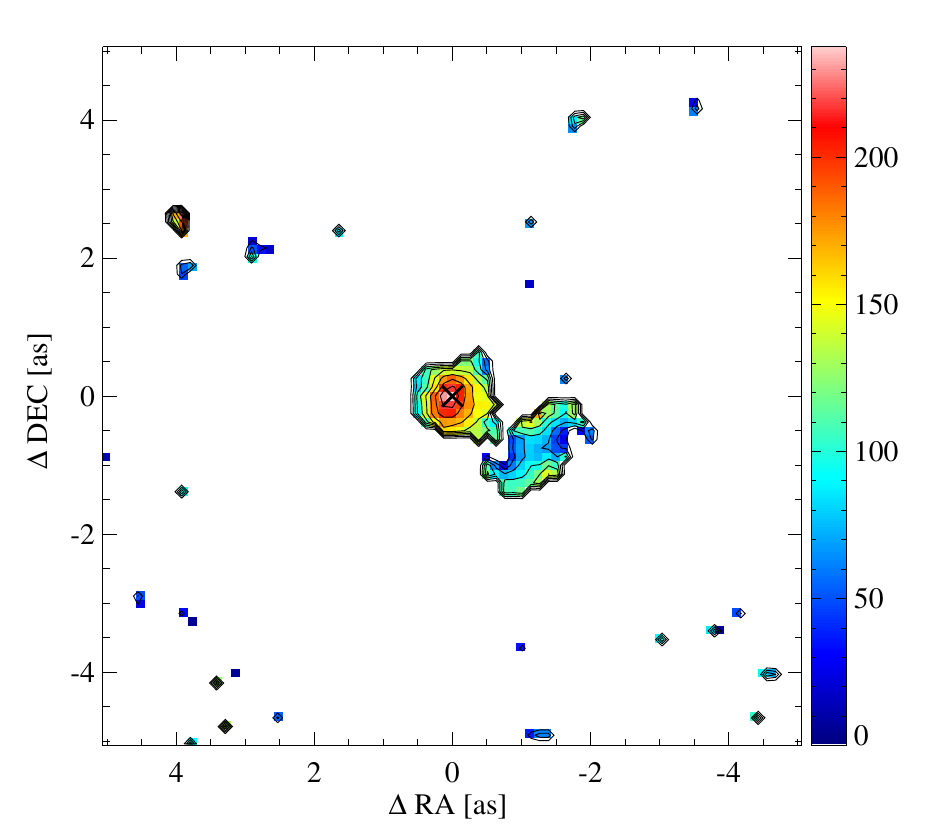}\label{fig:heifwhm}}
\subfigure[EW \ion{He}{i}]{\includegraphics[width=0.33\textwidth]{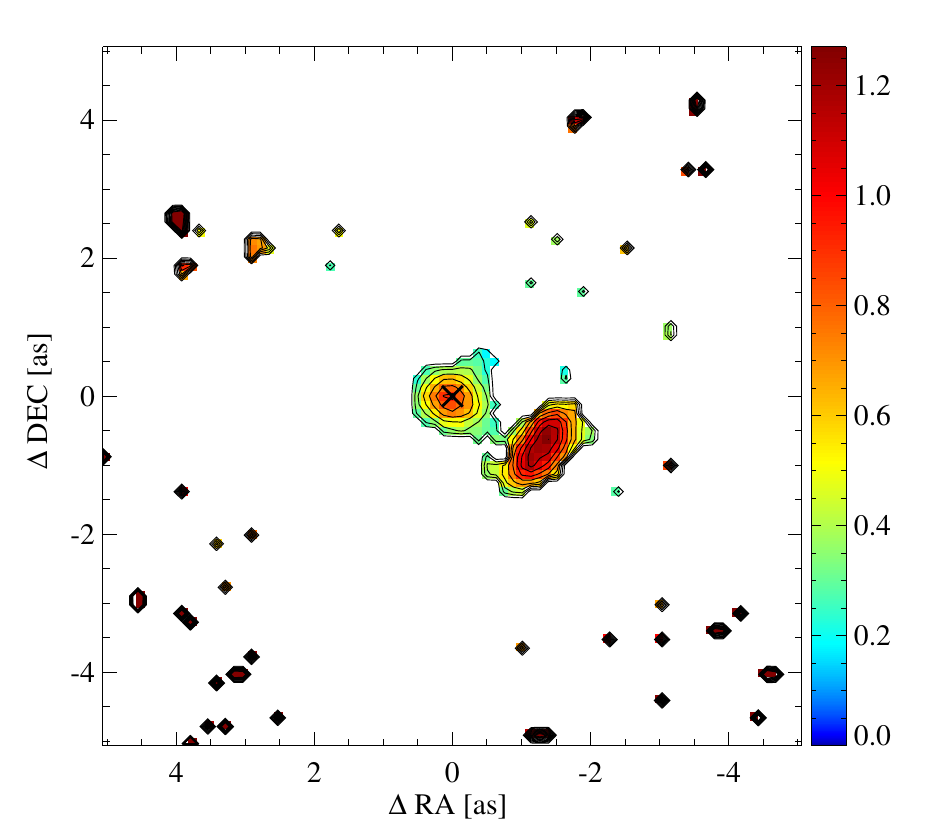}\label{fig:heiew}}
\subfigure[Flux H$_2$(1-0)S(1)]{\includegraphics[width=0.33\textwidth]{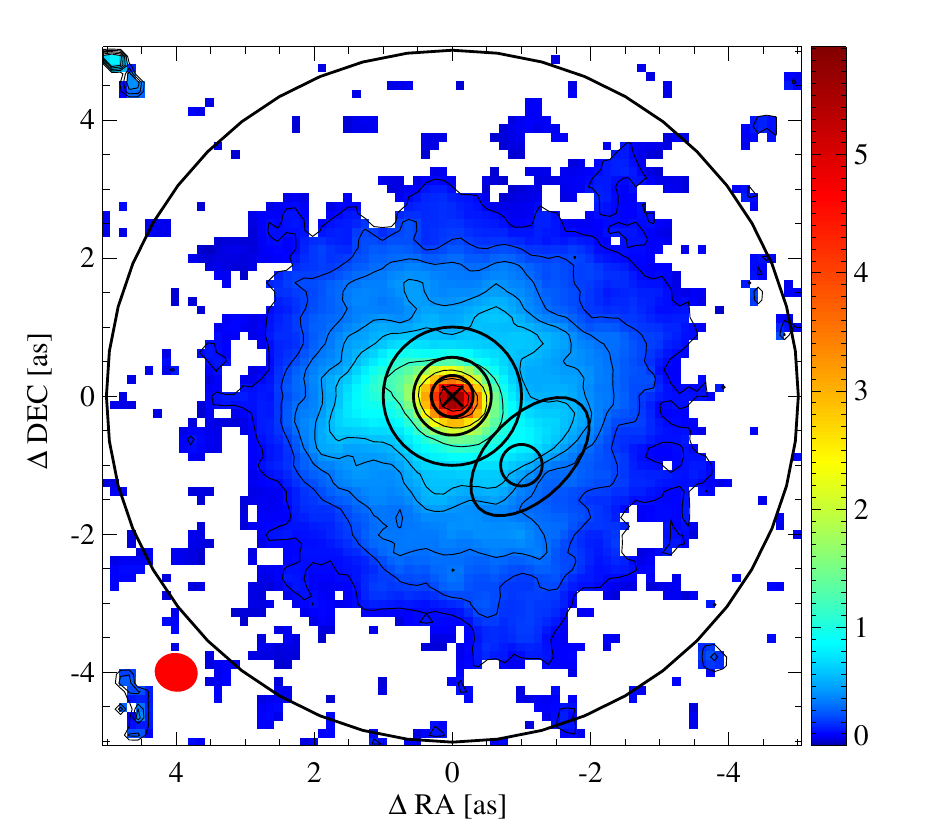}\label{fig:h212flux}}
\subfigure[FWHM H$_2$(1-0)S(1)]{\includegraphics[width=0.33\textwidth]{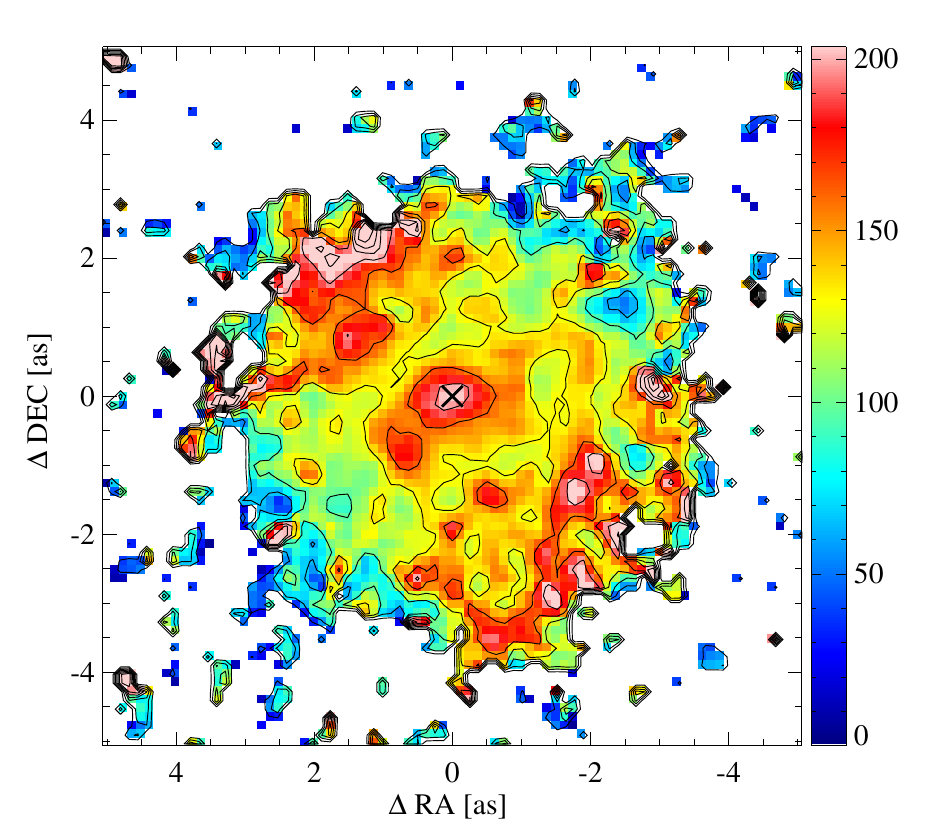}\label{fig:h212fwhm}}
\subfigure[EW H$_2$(1-0)S(1)]{\includegraphics[width=0.33\textwidth]{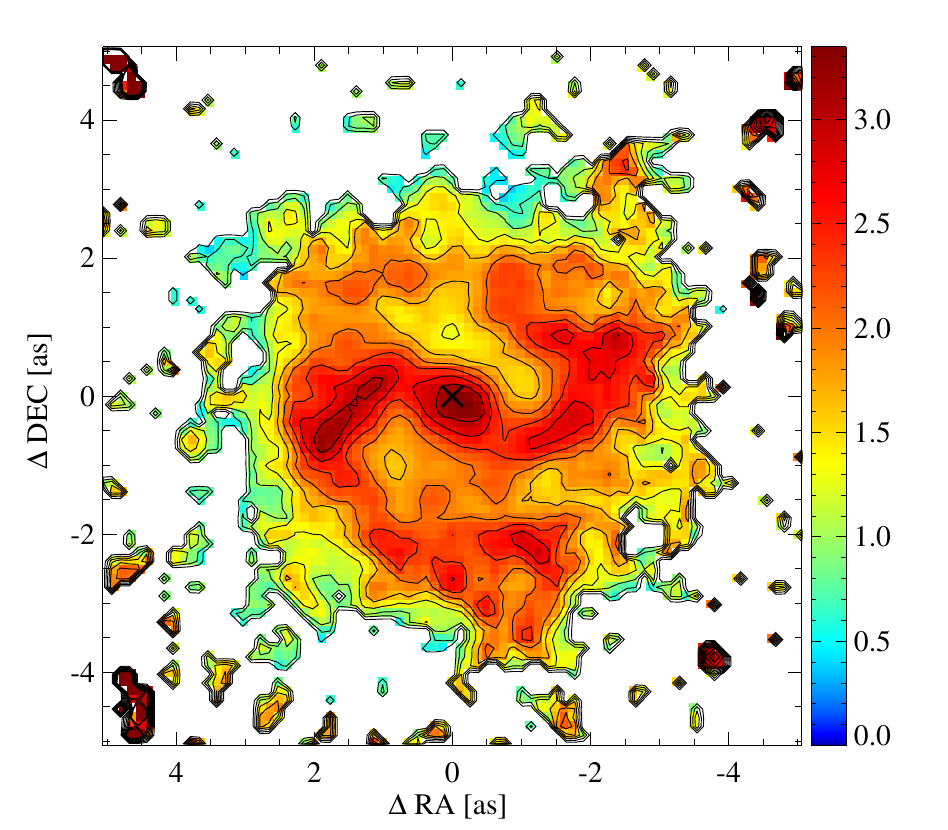}\label{fig:h212ew}}
\subfigure[Flux Br$\gamma$]{\includegraphics[width=0.33\textwidth]{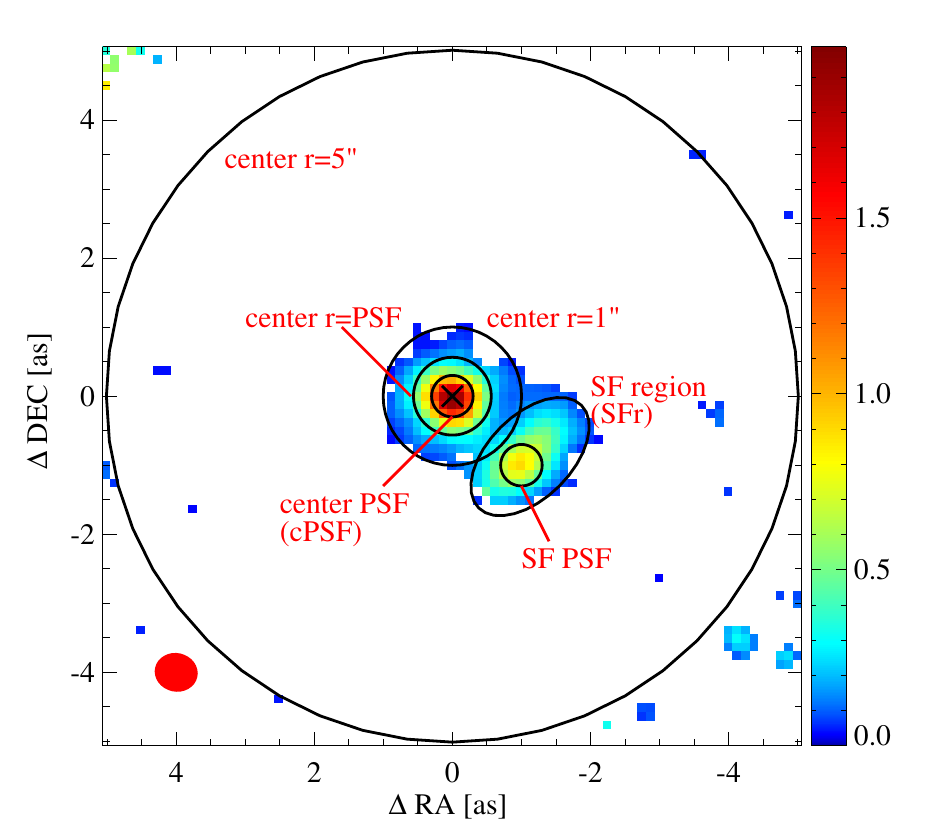}\label{fig:brgflux}}
\subfigure[FWHM Br$\gamma$]{\includegraphics[width=0.33\textwidth]{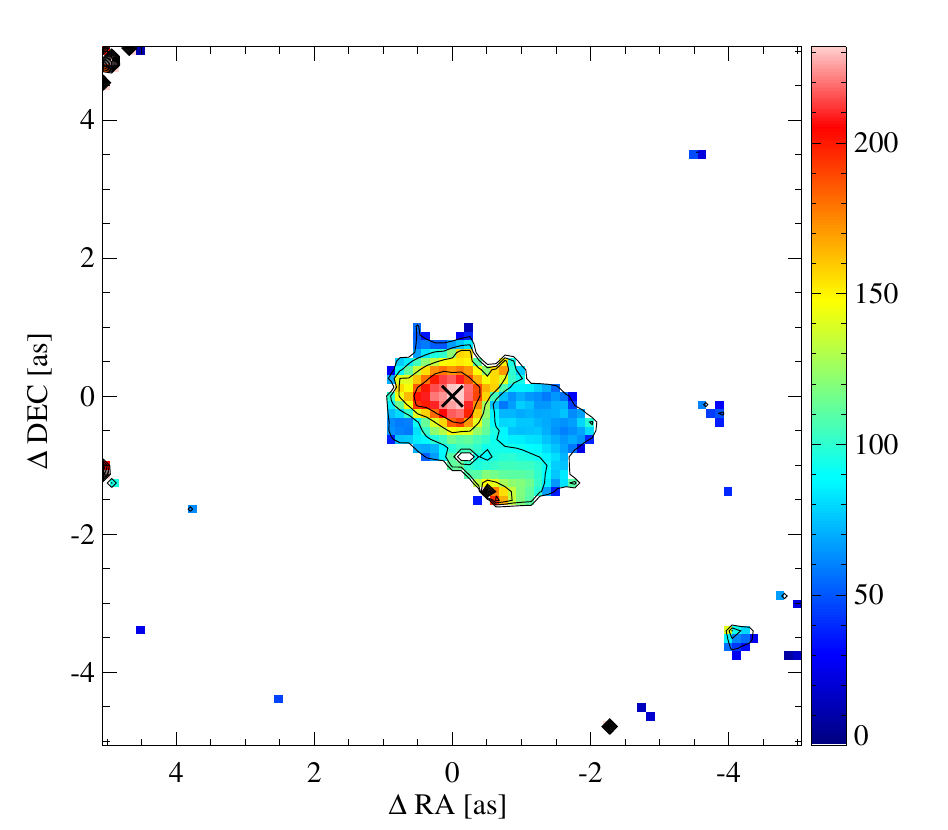}\label{fig:brgfwhm}}
\subfigure[EW Br$\gamma$]{\includegraphics[width=0.33\textwidth]{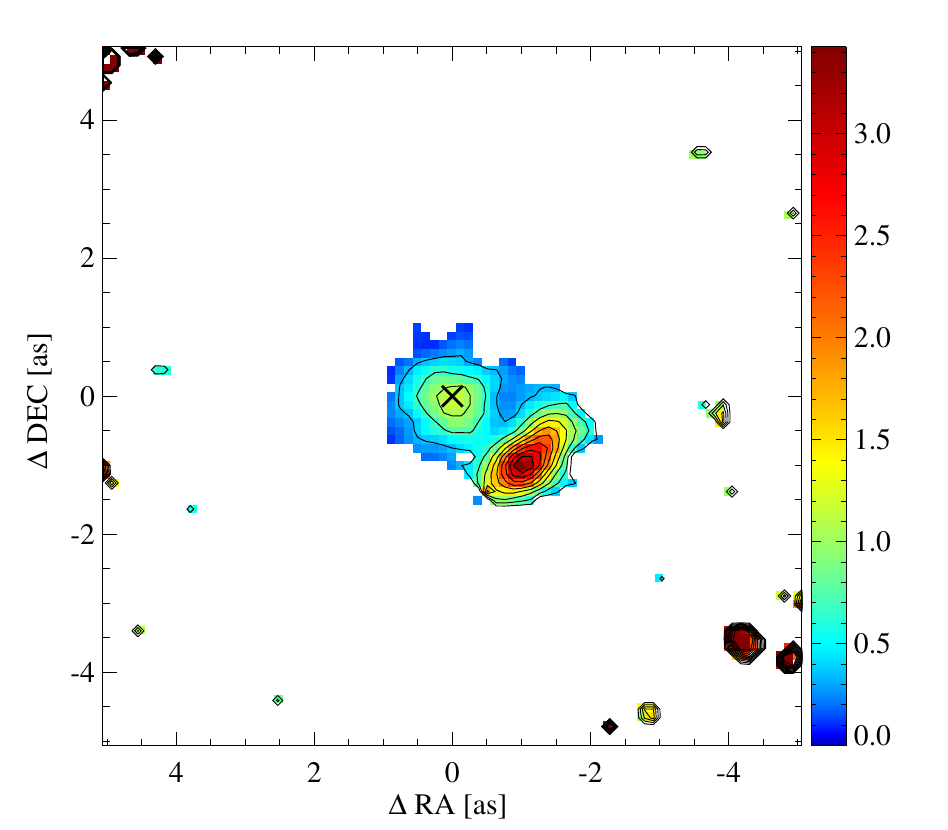}\label{fig:brgew}}
\caption{From left to right: flux [10$^{-20}$~W~m$^{-2}$], FWHM (corrected for instrumental broadening) [\kms] and EW [$\AA$] maps of, from top to bottom: \feii, \ion{He}{i}, H$_2$(1-0)S(1), and narrow \brg\ emission lines. The regions discussed in this paper are marked in the flux maps \subref{fig:feiiflux}, \subref{fig:h212flux}, and \subref{fig:brgflux}.}
\label{fig:lines1}
\end{figure*}

The forbidden transition [\ion{Fe}{ii}]~$\lambda$1.644~$\mu$m is slightly blended by the CO(7-4)~$\lambda$1.641~$\mu$m absorption feature in H-band but is very strong. With an FWHM of 350 km s$^{-1}$ it is the broadest narrow emission line at the center (Figs. \ref{fig:feiiflux} \& \ref{fig:feiifwhm}). 
The line map of the [\ion{Fe}{ii}] emission shows a triangular shape pointing in north, east and southwest directions. The FWHM and equivalent width (EW) maps both show elongated features in the east-west direction. Toward the north the FWHM decreases quickly to about 100~km~s$^{-1}$. To the south-west it stays for about one arcsecond above 300~km~s$^{-1}$ before it drops to $\sim150$~km~s$^{-1}$. The eastern part extends to about 1\farcs5 with an FWHM of about 250 km s$^{-1}$. The [\ion{Fe}{ii}] emission is point-like on the nucleus with a stronger eastern wing and a weaker south-western wing best seen in the EW map (Fig.~\ref{fig:feiiew}).
The EW is $6.8$~$\AA$ on the nucleus and only 0.6~$\AA$ in the south-west. A small plateau is detected to the east with an EW of $\sim0.9$~$\AA$.

The \brg\ line shows, spatially and spectrally, two components. We detect a narrow and a broad \brg\ emission at the center. The broad component shows an FWHM of about 2000 km~s$^{-1}$ and is redshifted by 340~\kms.
From the spatial distribution of the BLR emission a reliable value for the PSF of $\sim$~29~pc is determined (Sect. \ref{sec:obsred}). The broad component is a clear indicator for a SMBH. The narrow component at the center has an FWHM of about 230 km s$^{-1}$ and suggests a region size of ~13.5 pc, after deconvolution with the PSF width.

We also detect a spatially resolved strong narrow \brg\ emission about $1\arcsec$ southwest from the nucleus (Fig.~\ref{fig:brgflux}). The off-nuclear \brg\ emission shows a rather elliptical shape in the south-east to north-west direction. The FWHM in this region is $\sim100$\kms\ with a maximum of 150\kms\ in the south-east and a minimum of 70\kms\ in the north-west. The EW of the narrow Br$\gamma$ line at the nucleus is about 1.0~$\AA$. At the off-nuclear region it is about 2.6~$\AA$ in the brightest spot and falls to about 2~$\AA$ along the ellipse.

In addition, the \ion{He}{i}~$\lambda$2.06 $\mu$m emission line is detected. The emission is strongest at the center but is also very prominent in the southwestern region where the strong narrow Br$\gamma$ emission is detected. The FWHM peaks on the center with 250\kms\ whereas the south-western region shows a broadness $\sim120$~km~s$^{-1}$ similar to that of the Br$\gamma$. The flux distribution and with it the EW of both lines is not as similar. The southwestern emission region shows a peak in Br$\gamma$ in its lower part, whereas \ion{He}{i} is uniformly distributed over the whole ellipse. The EW of \hei\ is up to 0.8~$\AA$ at the nucleus and up to 1.3~$\AA$ at the southwestern emission region.

\subsubsection{Molecular gas}

\begin{figure*}[htbp]
\centering
\subfigure[Flux $^{12}$CO(3-2)]{\includegraphics[width=0.33\textwidth]{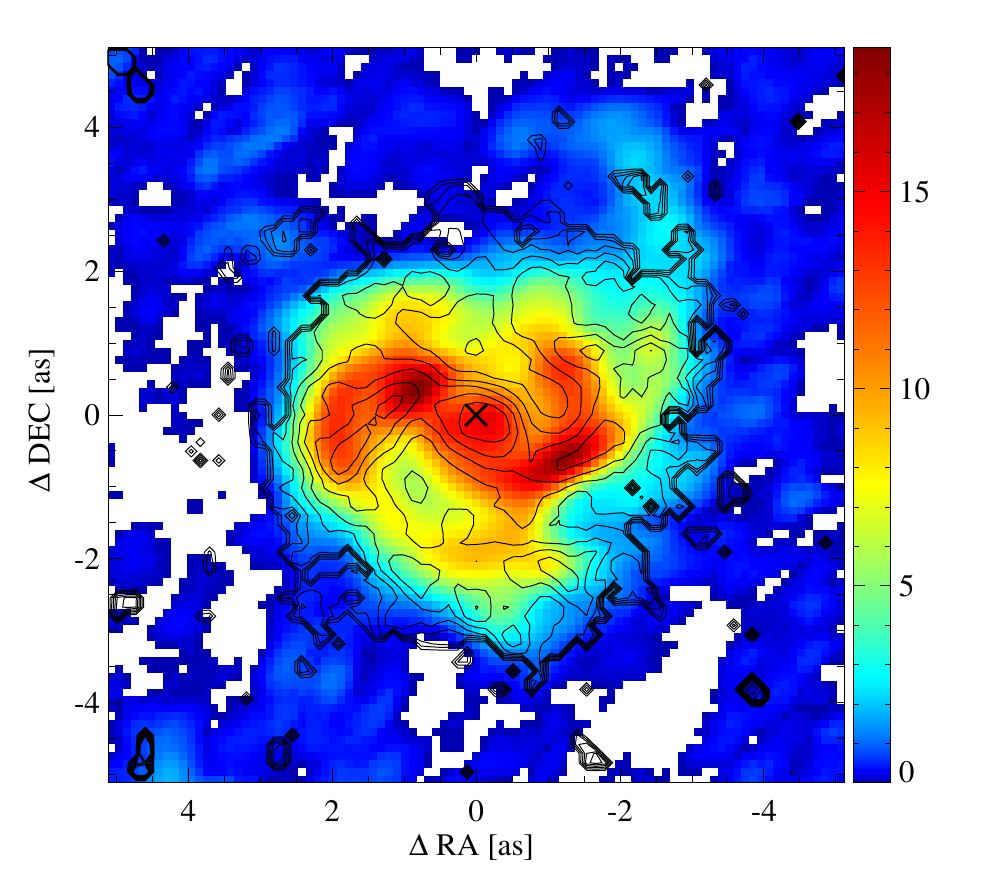}\label{fig:ALMACO32}}\qquad\qquad
\subfigure[LOSV $^{12}$CO(3-2)]{\includegraphics[width=0.33\textwidth]{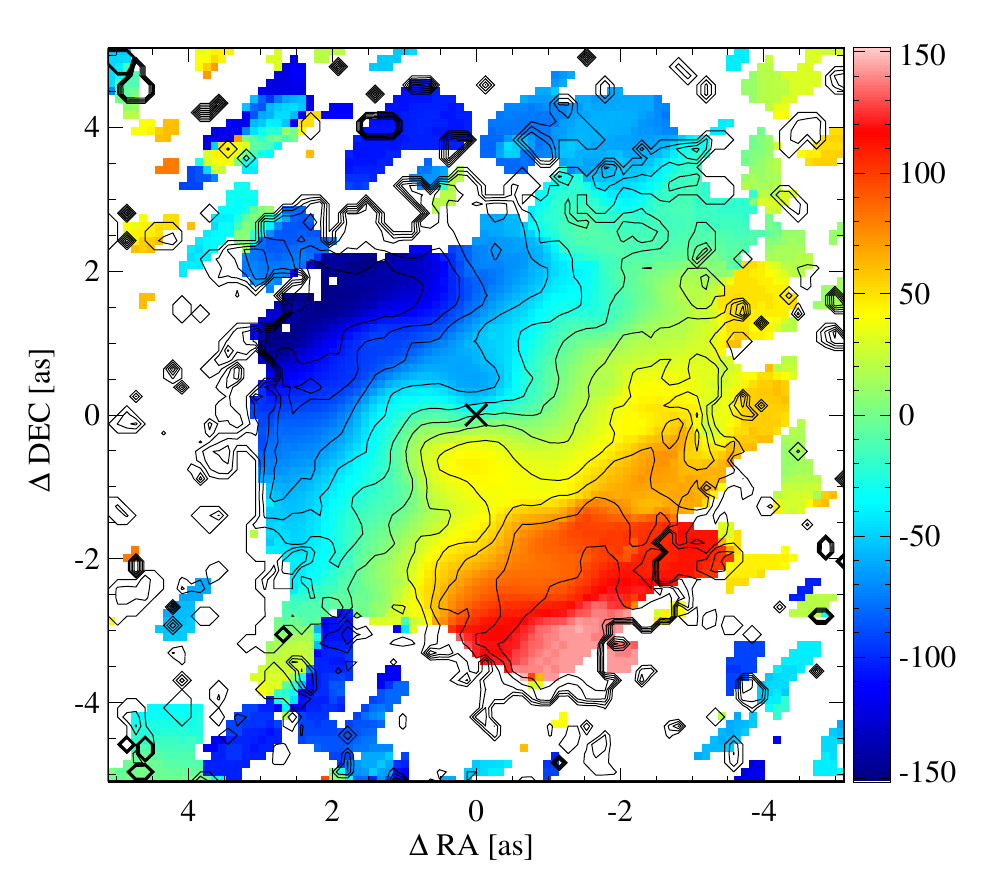}\label{fig:ALMACO32mom1}}\\
\subfigure[Dispersion $^{12}$CO(3-2)]{\includegraphics[width=0.33\textwidth]{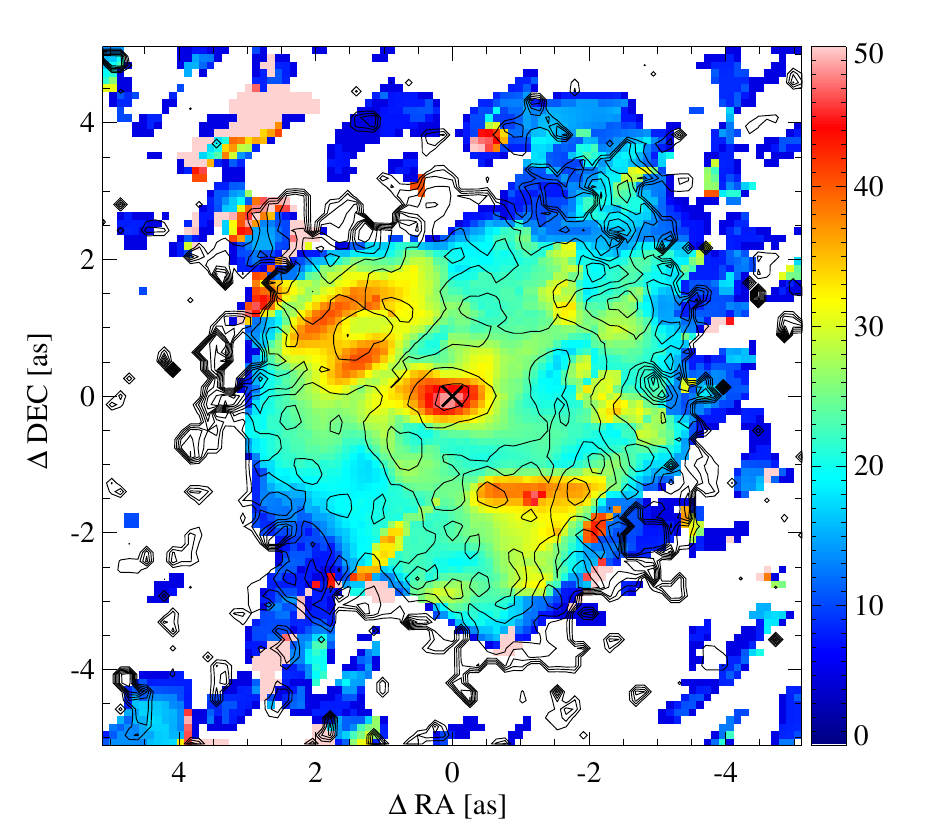}\label{fig:ALMACO32mom2}}\qquad\qquad
\subfigure[0.87~mm continuum]{\includegraphics[width=0.33\textwidth]{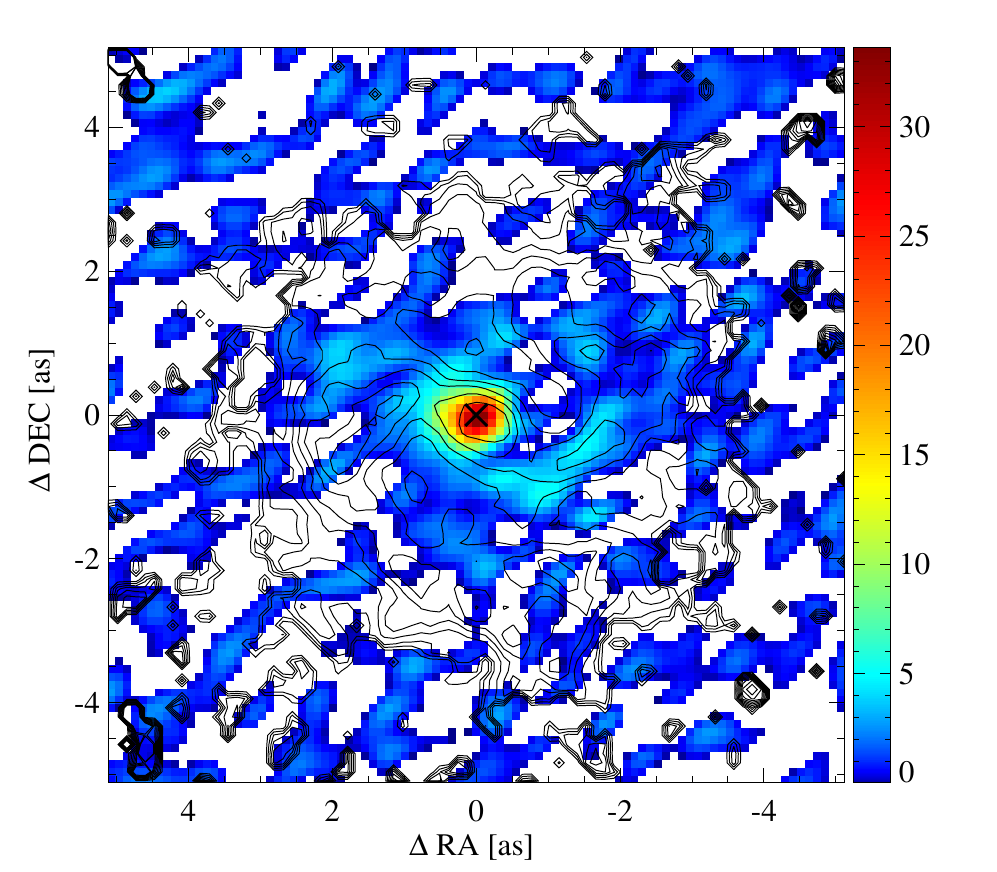}\label{fig:ALMAcont}}
\caption{Panel \subref{fig:ALMACO32} \& \subref{fig:ALMAcont} show the $^{12}$CO(3-2) and continuum at 0.87~mm flux maps in [Jy~beam$^{-1}$~\kms] and [Jy~beam$^{-1}$], respectively, overlayed with H$_2$(1-0)S(1) EW contours. Panels \subref{fig:ALMACO32mom1} \& \subref{fig:ALMACO32mom2} show the $^{12}$CO(3-2) first and second moment maps in [\kms] overlayed with H$_2$(1-0)S(1) LOSV and FWHM contours, respectively.
}
\label{fig:ALMA}
\end{figure*}

The most prominent molecular hydrogen lines that we detect in K-band are the H$_2$(1-0)S(1) $\lambda$2.12 $\mu$m and H$_2$(1-0)S(3) $\lambda$1.96 $\mu$m lines. Other detected molecular hydrogen lines are the H$_2$(1-0)S(2) $\lambda$2.03 $\mu$m, H$_2$(1-0)S(0) $\lambda$2.22 $\mu$m, H$_2$ (2--1)S(1) $\lambda$2.248 $\mu$m, H$_2$(1-0)Q(1) $\lambda$2.41 $\mu$m H$_2$(1-0)Q(3) $\lambda$2.42 $\mu$m lines.

The molecular emission line H$_2$(1-0)S(1) shows strong emission on the nucleus and reveals a nuclear spiral structure within an $r=3\arcsec$ nuclear disk (Figs. \ref{fig:h212flux}, \ref{fig:h212ew}).
The EW map reveals the full extent of the nuclear structure. The arms are clearly visible, with the eastern arm having a higher EW than the western, which is visible in the flux map as well. In EW the center looks like a gaseous bar. The spiral does not seem to become narrow here, however, this might be a resolution effect. The strong off-nuclear Br$\gamma$ emission is situated along the southern edge of the western arm, where the EW is lower.
Both arms are oriented at a PA of $\sim90\degr$ and point counter-clockwise. The eastern arm and the nuclear region both show the maximum in EW of about 3.4~$\AA$, outside these regions the EW drops to $\sim1.75$~$\AA$. Both arms show concentrated cigar-shaped emission in the parts connected to the nucleus and then turn over by about $\sim90\degr$ into more diffuse emission. The arms do not look like a geometrical spiral close to
the center. Inferred from the EW map a bar seems to connect the spiral arms toward the nucleus.
In the outer part the two-arm spiral becomes flocculate and forms a ring-like structure at a 2$\arcsec$ radius. This ring might correspond to an inner ILR due to a secondary nuclear bar (Sect.~\ref{sec:nirconti}) or is rather created by gas falling in from the $200-300$~pc scales toward the lower angular momentum transport region at $\le150$~pc scales \citep{combes_ALMA_2014}.

The line of sight velocity (LOSV) shows a rotation at a PA of $\sim45\degr$ that reaches velocities of about $\pm$150~km~s$^{-1}$ at a radius of about 2\farcs5 (for more details see Sect.~\ref{sec:gaskin}). In addition, there is a strong gradient in the central region with the line of nodes at a PA $\sim0\degr$. This change in the PA is indicative of a bar or spiral density wave.
The FWHM reaches a velocity of 200\kms\ at the center and in regions $\sim2\farcs5$ to the northeast and to the southwest of the nucleus. Along the spiral arms the FWHM is $\sim130$\kms. Along the minor axis of the galactic rotation the FWHM does not fall to the width of the spiral arms but stays at about 160\kms. In the northwest and southeast the FWHM drops down to 70\kms\ (Fig.~\ref{fig:h212fwhm}).


H$_2$(1-0)S(3) and all other detected molecular hydrogen emission lines look very similar to H$_2$(1-0)S(1) in shape and value (e.g., similar velocities, similar flux distribution, see Figs. \ref{fig:lines2} and \ref{fig:LOSVlines}).
Therefore, we use the H$_2$(1-0)S(1) emission line as the general description of all molecular hydrogen lines.


In the mm-regime we use the $^{12}$CO(3-2) line to compare the cold molecular gas distribution to our hot molecular gas distribution derived from ro-vibrational H$_2$ line emission described above. In general, the $^{12}$CO(3-2) emission is very similar to the H$_2$(1-0)S(1) emission. In both lines an $r=3\arcsec$ disk with a nuclear spiral is detected. The nuclear spiral looks almost identical when comparing the $^{12}$CO(3-2) emission and the H$_2$(1-0)S(1) EW maps (Fig.~\ref{fig:h212ew} and \ref{fig:ALMACO32}). The difference lies in the location of the emission maxima. The emission line $^{12}$CO(3-2) peaks at connection points of the spiral arms to the center.

The LOSV-field is identical in shape, i.e., maxima and disturbances, and value, i.e., both gases show max/min velocities of $\pm150$~\kms, see Figs.~\ref{fig:ALMACO32mom1}~\&~\ref{fig:h212losv}. The dispersion of the $^{12}$CO(3-2) gas is similar in distribution but differs in value, i.e., the NIR H$_2$ dispersion is lower by $\sim30$~\kms.

\input{region5}

\subsection{Gas masses}

From the detected molecular hydrogen lines the warm H$_2$ gas mass can be determined using the luminosity of the H$_2$(1-0)S(1), $L_{\mbox{\tiny H$_2$(1-0)S(1)}}$, and the equation
\begin{equation}
\mbox{M}_{\mbox{\tiny H$_2$}}=4.243\times10^{-30}\left(\frac{L_{\mbox{\tiny H$_2$(1-0)S(1)}}}{\mbox{W}}\right)\;\mbox{M}_\odot
\label{eqn:}
\end{equation}
following \citet{turner_mass--light_1977,scoville_velocity_1982,wolniewicz_quadrupole_1998,riffel_mapping_2008}. The warm H$_2$ gas mass in a 5$\arcsec$ radius aperture, which is all of the warm H$_2$ in our FOV, is derived to 57~M$_{\odot}$. To estimate the cold gas mass we use the conversion factor derived by \citet{mazzalay_molecular_2013}
\begin{equation}
\frac{M_{\mbox{\tiny H}_2(\mbox{\tiny cold})}}{M_{\mbox{\tiny H}_2(\mbox{\tiny warm})}} = (0.3-1.6)\times 10^6.
\label{eqn:}
\end{equation}
We find a cold H$_2$ gas mass in the central $10\arcsec\times10\arcsec$ of $(1.7-9.1)\times10^{7}$~M$_{\odot}$. \citet{combes_ALMA_2014} detect $7\times10^7M_{\odot}$ in their $r=18\arcsec$ FOV from CO(3-2) observations with ALMA. The values are in good agreement since the bulk of the molecular mass in NGC 1566 is located in the inner 6$\arcsec$ \citep[Fig.~\ref{fig:lines1},~\ref{fig:ALMA} and ][ and their Fig. 3]{combes_ALMA_2014}.

The cold gas masses of the central $r=3\arcsec$ disk and the cPSF region are estimated from H$_2$(1-0)S(1) and $^{12}$CO(3-2) emission (see Fig.~\ref{fig:h212flux} and \ref{fig:ALMACO32}). We measure a $^{12}$CO(3-2) flux of $\sim570$~Jy~km~s$^{-1}$ for the central $r=3\arcsec$ gas disk and $\sim12$~Jy~km~s$^{-1}$ for the cPSF region. Furthermore, we estimate masses of $\sim6.6\times10^7$~M$_\odot$ and $\sim1.4\times10^6$~M$_\odot$ respectively, using the Milky Way conversion values from \citet{bolatto_co--h2_2013}. Using H$_2$(1-0)S(1) luminosities we estimate cold gas masses for the central $r=3\arcsec$ gas disk and cPSF region of $(1.4-7.5)\times10^7$~\msun\ and $(1.5-7.8)\times10^6$~\msun\ respectively.



\subsection{Emission line regions}
\label{sec:emlinediag}
The detection of several narrow emission lines gives us the opportunity to analyze the emission at the center of NGC~1566.
We analyze the ratios of the narrow ionized and molecular emission lines with the goal of finding the nature of their excitation. We investigate apertures centered on the nucleus and on the ionization region situated $\sim1\farcs5$ southwest of the nucleus.

\subsubsection{Emission line ratios}
The narrow \brg\ emission line as well as [\ion{Fe}{ii}] and H$_2$(1-0)S(1) transitions can be used in a diagnostic diagram to disentangle photoionization by young, bright stars and shock ionization (e.g., supernovae). Young and bright stars can be found in systems with recent and strong star formation like starburst galaxies. LINER galaxies exhibit high [\ion{Fe}{ii}] and H$_2$(1-0)S(1) fluxes. These species are good shock tracer as they are often found in regions of supernovae or outflows/jets.

\begin{figure}[h!]
\centering
\includegraphics[height=0.5\textwidth, angle=90]{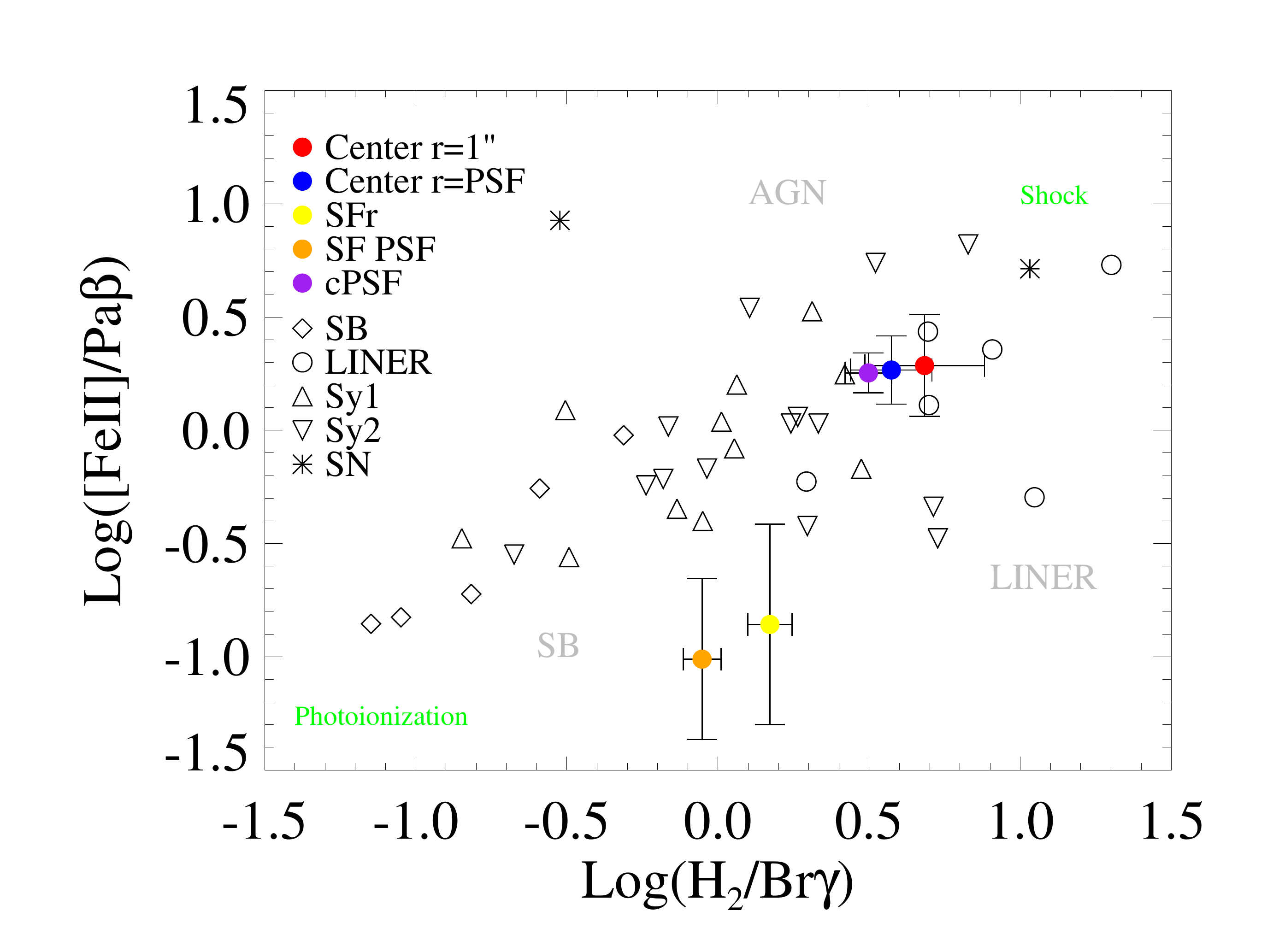}
\caption{Diagnostic diagram of $\log{[\ion{Fe}{ii}]/\mbox{Pa}\beta}$ and $\log{\mbox{H}_2/\mbox{Br}\gamma}$ for the central $r=$~PSF, $r=1\arcsec$ and the star formation region (SFr). The conversion factor of $0.744$ was used for [\ion{Fe}{ii}]$\lambda1.644$~\mic\ over [\ion{Fe}{ii}]$\lambda1.257$~\mic\ \citep{nussbaumer_transition_1988}. For the conversion of \brg\ to Pa$\beta$ the case B ratio of $0.17$ was used. Open symbols correspond to literature values from \citet[][LINER]{larkin_near-infrared_1998}, \citet[][SB]{dale_near-infrared_2004}, \citet[][Sy1, Sy2]{rodriguez-ardila_molecular_2004,rodriguez-ardila_molecular_2005}.}
\label{fig:ddiag1}
\end{figure}

We find that the nuclear regions cPSF, $r=$~PSF, and $r=1\arcsec$ lie on the linear transition relation from SB over AGN to LINER (see Fig.~\ref{fig:ddiag1}). The $r=$~PSF region is well situated in the AGN regime indicating mixed ionization mechanisms, typical for AGN.

The region SFr lies off the linear correlation seen in Fig.~\ref{fig:ddiag1}. It is situated in the AGN regime, but under the relation where the Seyfert galaxies reside.
The ratio $\log{[\ion{Fe}{ii}]/\mbox{Pa}\beta}$ puts SFr into the \ion{H}{ii} galaxy region, however, the ratio $\log{\mbox{H}_2/\mbox{Br}\gamma}$ shifts it to the AGN regime. This behavior can be explained by an H$_2$ overabundance. As mentioned above, the high H$_2$ fluxes shift the central $r=1\arcsec$ aperture toward the LINER like excitation regime. This is the case for SFr as well. The high H$_2$ flux shifts the SFr region from the photo ionization or star formation regime toward the mixed ionization or AGN regime. The H$_2$ over \brg\ ratio is five to ten times higher than in typical \ion{H}{ii} galaxies.



The rich molecular gas disk in the central $r=3\arcsec$ shows a variety of H$_2$ transitions in the NIR (see Tab. \ref{tab:region}).
There are three main excitation mechanisms for molecular hydrogen in the NIR \citep{mouri_molecular_1994,rodriguez-ardila_molecular_2005} which can be discriminated with the detected H$_2$ species:\\
\begin{enumerate}[i)]

	\item {\it UV fluorescence} (non-thermal) can occur in warm high-density gas where highly energetic UV photons from the Lyman-Werner band ($912-1108$~$\AA$) are re-emitted by the H$_2$ molecules. To distinguish the UV pumping (non-thermal) from collisional excitation (thermal) higher level transitions need to be detected since the lower levels are populated by collisions.
	\item {\it X-ray heating} (thermal) is responsible for H$_2$ excitation in regions with temperatures of $<1000$~K. At higher temperatures collisional excitation populates the lower levels.
	\item {\it Shocks} (thermal) can collisionaly populate the electronic ground levels of H$_2$ molecules. The rovibrational transitions are populated following a Boltzmann distribution where kinetic temperatures can be higher than $2000$~K \citep{draine_theory_1993}.

\end{enumerate}


\begin{figure*}[]
\centering
\subfigure[]{\includegraphics[height=0.49\textwidth, angle=90]{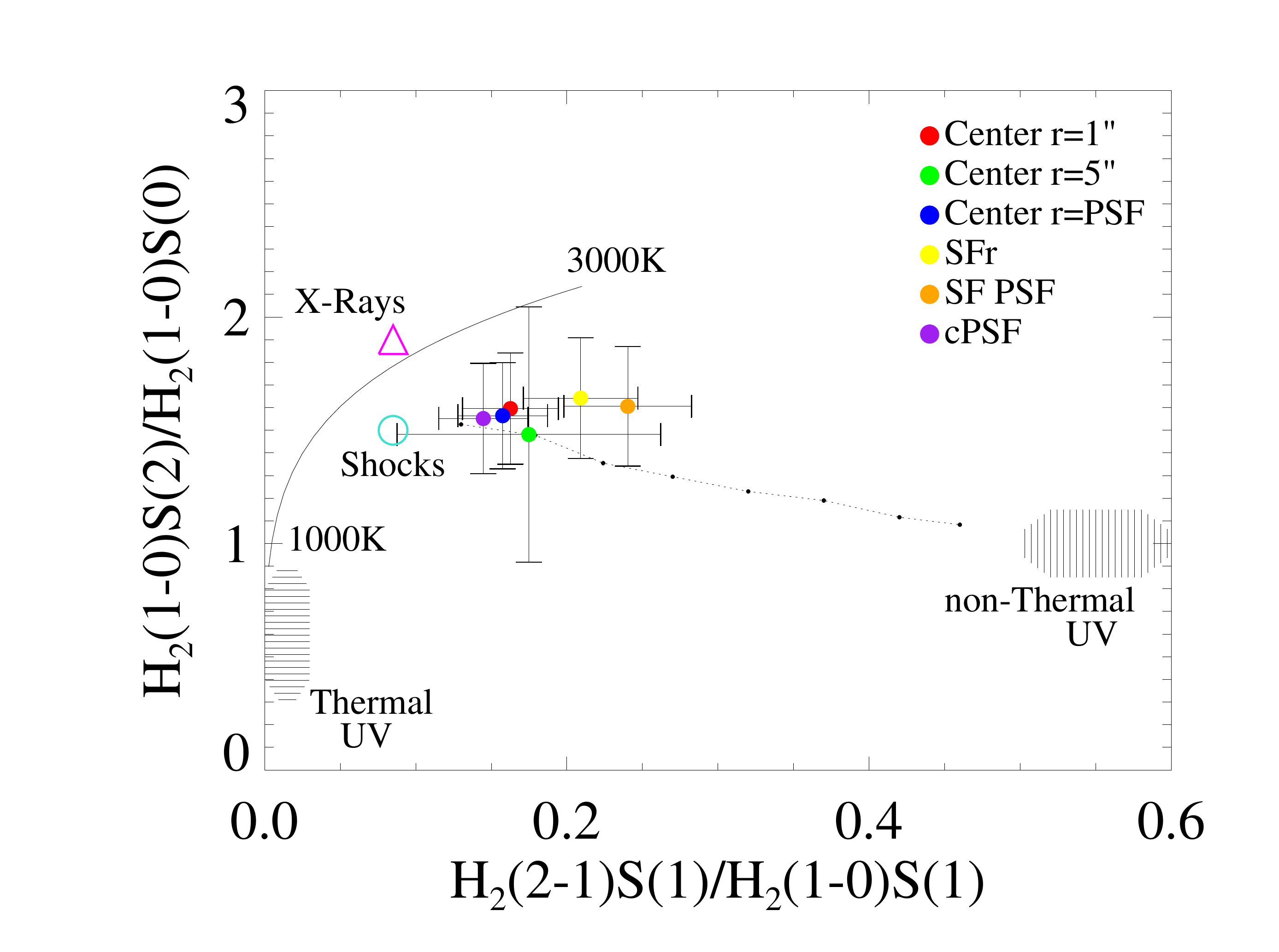}\label{fig:h2diag1}}
\subfigure[]{\includegraphics[height=0.49\textwidth, angle=90]{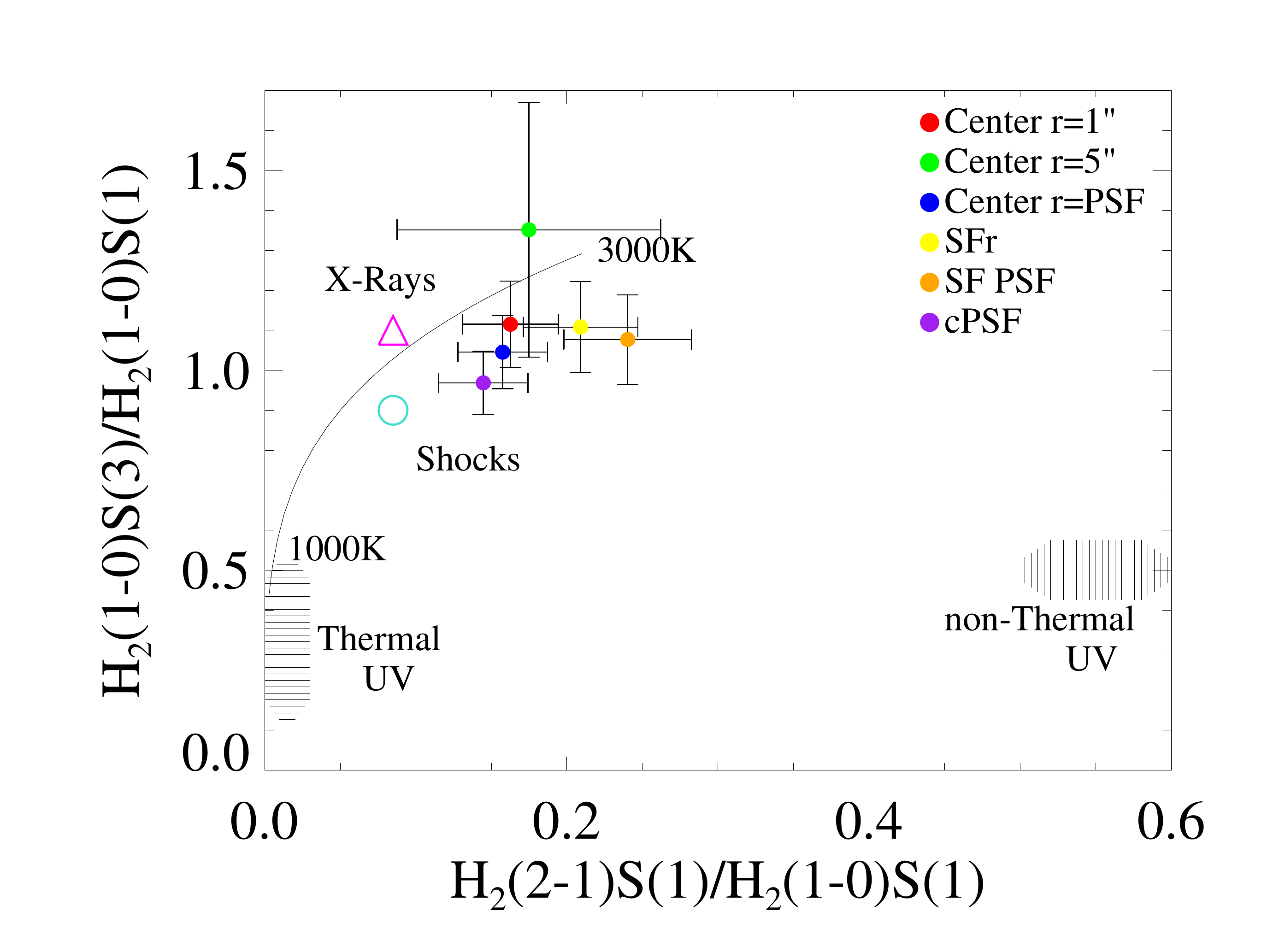}\label{fig:h2diag2}}
\caption{Molecular hydrogen line ratio diagrams. The ratios of H$_2$(2-1)S(1)/H$_2$(1-0)S(1) versus H$_2$(1-0)S(2)/H$_2$(1-0)S(0) are shown in \subref{fig:h2diag1}. The ratios of H$_2$(2-1)S(1)/H$_2$(1-0)S(1) versus H$_2$(1-0)S(3)/H$_2$(1-0)S(1) are shown in \subref{fig:h2diag2}. The curves represent the thermal emission at $1000-3000$~K. Vertical stripes represent the region where models by \citet{black_fluorescent_1987} predict non-thermal UV excitation. Horizontal stripes are thermal UV excitation models by \citet{sternberg_infrared_1989}. The open, magenta triangle represents thermal X-ray models by \citet{draine_h2_1990} and the open, turquoise circle represents a shock model from \citet{brand_constancy_1989}. The filled dark bullets connected with a dotted line in Fig. \subref{fig:h2diag1} are the predicted line ratios from a mixture of thermal and low-density fluorescence models of \citet{black_fluorescent_1987}. The first bullet from the left represents $10\%$ non-thermal and $90\%$ thermal UV-fluorescence, the second $20\%$ non-thermal and $80\%$ thermal UV-fluorescence and so on.}
\label{fig:h2diag}
\end{figure*}

Both diagnostic diagrams (Fig.~\ref{fig:h2diag}) show similar behavior for the investigated regions. The H$_2$(2-1)S(1)/H$_2$(1-0)S(1) ratio ranges from $\sim0.15$ to $\sim0.25$ for all regions. The central regions, i.e. center $r=1\arcsec$, center $r=$~PSF, and cPSF, show the lowest ratio in H$_2$(2-1)S(1)/H$_2$(1-0)S(1) declining with aperture, i.e., lower non-thermal UV component. The star forming region, SFr, and the PSF sized aperture taken here (SF PSF) exhibit the highest value and the ratio is increasing with smaller apertures, i.e., higher non-thermal UV component.
All regions are situated beneath the thermal Boltzmann distribution indicating rather a shock than an X-ray origin of the thermally excited molecular emission. 
The line ratio measured for the full FOV ($r=5\arcsec$) is not very reliable. This ratio exhibits the highest error bars, probably due to contamination of the low flux continuum in the outer parts of the FOV by OH line correction residuals.

\subsubsection{Level population of the H$_2$ gas}


The ro-vibrational levels will be populated according to the Boltzmann equation when we assume thermal excitation. Then the excitation temperature $T_{exc}$ can be derived from the inverse of the gradient of the line fitted to the thermalized levels in the graph shown in Fig.~\ref{fig:h2lvlpop}. These will be on a linear relation if the excitation is thermal. The estimate of the population density can be inferred from the observed column density \citep{lester_properties_1988}:

\begin{equation}
N_{col}=\frac{f}{A_{ul}}\frac{\lambda}{hc}\times\frac{4\pi}{\Omega},
\end{equation}
with flux $f$ in W~m$^{-2}$, $A_{ul}$ the transition probability \citep{wolniewicz_quadrupole_1998}, $\lambda$ the rest frame line wavelength, $h$ the Planck constant, $c$ the speed of light, and $\Omega$ the aperture size in radian. In thermal equilibrium the ratio of two levels can be written as

\begin{equation}
\frac{N^\prime}{N^{\prime\prime}}\frac{g^{\prime\prime}_J}{g^{\prime}_J}=\exp\left({\frac{-\Delta E}{k_B T}}\right),
\end{equation}
with column densities $N^\prime$ and $N^{\prime\prime}$, statistical weights $g^\prime_J$ and $g^{\prime\prime}_J$, Boltzmann constant $k_B$, and the temperature of the thermal equilibrium $T$.

The level population diagram in Fig.~\ref{fig:h2lvlpop} compares our measured level population of the H$_2$ emission lines to UV-excitation models derived by \citet{davies_molecular_2003}.

Model~1 is a low density model ($n_H=10^3$~cm$^{-3}$) with cool $T=100$~K gas and a relatively weak FUV field. In model~2 density and UV field are increased by one order of magnitude and a thermal profile is adopted for the temperature with $T_{\rm max}=1000$~K. Model~3 is the same as model~2 but with a maximum temperature of $T_{\rm max}=2000$~K. Model~4 is the same as model~2 but the FUV field is increased by a factor of 100. And model~5 is the high density model with an $n_H=10^6$~cm$^{-3}$ and temperature and FUV field as in model~4.
For more details on the models see \citet{davies_molecular_2003}.

\subsubsection{Line emission at the nuclear region}


The nuclear line emssion is well situated in the AGN regime in the diagnostic diagram in Fig. \ref{fig:ddiag1}. Interestingly, the $r=1\arcsec$ region with its slightly bigger aperture moves further toward the LINER regime.
This trend is caused by aperture effects only.
The \brg\ flux at the center stems from a deconvolved region of 13.5~pc, whereas the H$_2$ emission stems as well from the extended $r=3\arcsec$ molecular gas disk.
The \brg\ emission at $r=1\arcsec$ is the flux in the wings of the PSF whereas the H$_2$ emission is present in the central gas disk of up to $r<3\arcsec$ distance from the center.
Hence, larger apertures will shift the H$_2$ over \brg\ ratio toward higher H$_2$ fluxes and in this case toward the LINER domain.

\begin{figure}[htbp]
\centering
\includegraphics[height=0.5\textwidth, angle=90]{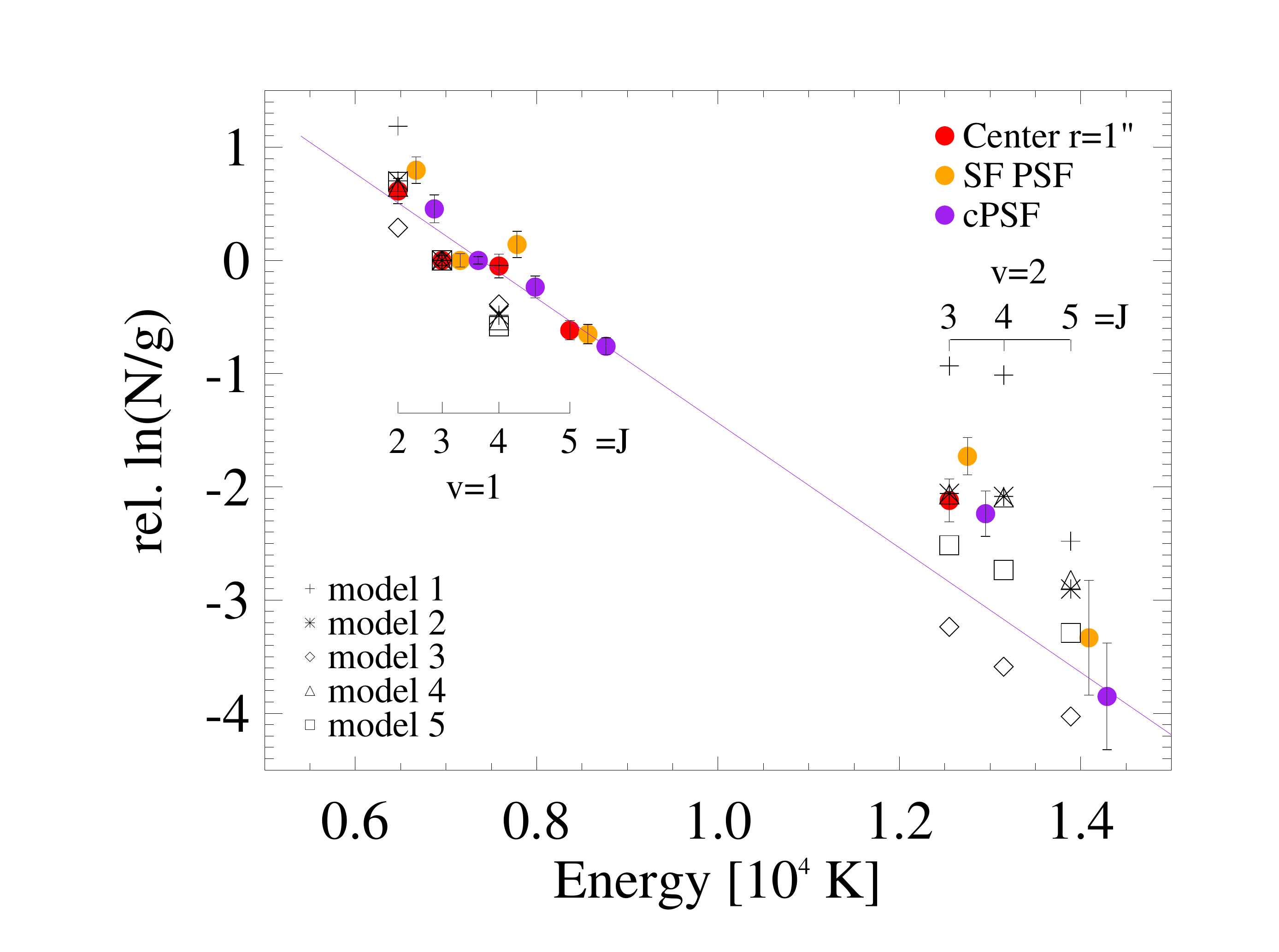}
\caption{Molecular hydrogen level population diagram relative to the H$_2$(1-0)S(1) transition. The column density N is given in $[$m$^{-2}]$. The center $r=1\arcsec$ region marks the energy of the level, the other regions are shifted for a better presentation in the plot. The models are for dense UV-excited gas taken from \citet{davies_molecular_2003}. The linear fit is to the $v=1$ levels of region cPSF which yields a kinetic excitation temperature of $T_{exc}\sim1800$~K.}
\label{fig:h2lvlpop}
\end{figure}

The diagrams in Fig.~\ref{fig:h2diag} infer that the central regions move with smaller aperture toward the shock model at $\sim2000$~K indicating a stronger thermal ionization close to the nucleus. However, dense gas ionized by UV-fluorescence can show similar emission. The $v=1$ transitions are thermalized by collisions and with higher density of the gas the $v=2$ transitions are thermalized as well and hence underpredicted with respect to lower density gas excited by UV-fluorescence \citep[e.g.,][]{sternberg_infrared_1989,sternberg_chemistry_1995,sternberg_ratio_1999,davies_molecular_2003,davies_molecular_2005}.
Therefore we compare the level population of H$_2$ to models from \cite{davies_molecular_2003}. The $v=1$ transitions in the central regions, e.g. cPSF, seem to be thermalized with an excitation temperature of $T_{exc}\sim1800$~K (Fig.~\ref{fig:h2lvlpop}). However, none of the apertures taken from the center exhibit values of purely thermalized gas. This is shown by the $v=2$,~$J=3$ level which lies off the excitation temperature line fitted to the $v=1$ levels. This should not be the case for a thermal ionization process. Hence, other ionization processes, e.g. UV-fluorescence, have to be taken into account. The $v=2$,~$J=5$ level seems suppressed with regard to the $v=2$,~$J=3$ level and fits the thermal equilibrium fit. This effect is seen in the X-ray models of \citet{draine_h2_1990} which predict a decrement in the $v=2$~$J=5$ level.

We are not able to disentangle ionization and excitation of the gas by the AGN or by stars.


\subsubsection{Star formation}
\label{sec:sf}

The star formation history of the central $10\arcsec\times10\arcsec$ of NGC~1566 does not seem to have been involved in recent star formation due to the lack of \ion{H}{ii} regions. In fact, \brg\ emission is only detected on the nucleus and in one off nuclear region. 
The luminosity of \brg, $L_{\mbox{\tiny\brg}}$, is proportional to the Lyman continuum flux and can be used as a measure for the star formation rate (SFR) \citep{panuzzo_dust_2003,valencia-s._is_2012} in the emitting region
\begin{equation}
\mbox{SFR}\,=\,\frac{L_{\mbox{\tiny Br}\gamma}}{1.585\times 10^{32}\mbox{W}}\;\mbox{M$_{\odot}$ yr}^{-1}.
\label{eqn:}
\end{equation}
The two regions of interest are at the nucleus with an 1$\arcsec$ radius and a luminosity of $L_{\mbox{\tiny\brg}}=8.45\times10^{29}$~W and the lenticular region 1$\farcs$5 southwest from the center with a luminosity of $L_{\mbox{\tiny\brg}}=4.15\times10^{29}$~W over an area of $\sim1.33$ arcsec$^2$. We derive a SFR of $\sim5.3\times10^{-3}$~M$_{\odot}$~yr$^{-1}$ at the center and SFR~$\sim2.6\times10^{-3}$~M$_{\odot}$~yr$^{-1}$ at the southwestern region. 
Additionally, we can estimate the supernova rate (SNR) in this region using the \feii\ emission. We follow \citet{bedregal_near-ir_2009} and use two different calibrations
\begin{equation}
\mbox{SNR}_{\mbox{\tiny Cal97}}=5.38\;\frac{L_{\mbox{\tiny[\ion{Fe}{ii}]}}}{10^{35}\mbox{W}}\;\mbox{yr}^{-1}
\label{eqn:}
\end{equation}
after \citet{calzetti_reddening_1997} and
\begin{equation}
\mbox{SNR}_{\mbox{\tiny AlH03}}=8.08\;\frac{L_{\mbox{\tiny[\ion{Fe}{ii}]}}}{10^{35}\mbox{W}}\;\mbox{yr}^{-1}
\label{eqn:}
\end{equation}
after \citet{alonso-herrero_[fe_2003}. The luminosity of \feii\ is measured to be $L_{\mbox{\tiny\feii}}=7.1\times10^{30}$~W at the center and $L_{\mbox{\tiny\feii}}=2.51\times10^{29}$~W at the southwest. The SNRs are $\sim3.82\times10^{-4}$~yr$^{-1}$, and $\sim5.74\times10^{-4}$~yr$^{-1}$, respectively, at the center and $\sim1.35\times10^{-5}$~yr$^{-1}$, and $\sim2.03\times10^{-5}$~yr$^{-1}$, respectively, in the southwest. The estimates at the central region are upper limits since star formation and the AGN are responsible for the excitation of \brg\ and \feii\ and their respective contributions can not be distinguished.

\begin{figure}[t]
\centering
\includegraphics[width=0.33\textwidth]{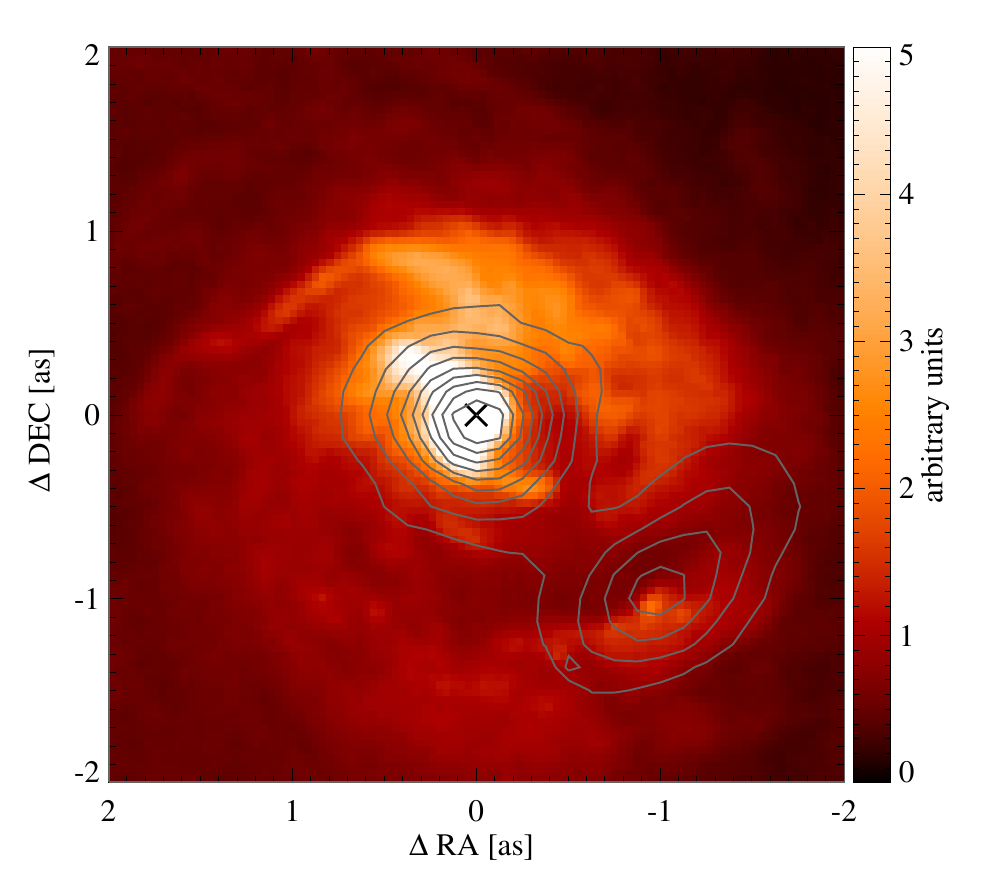}
\caption{\brg\ contours plotted over the HST image at 4326~$\AA$. Note the bright emission spots at the SFr emission region at (-1,-1).}
\label{fig:brgohst}
\end{figure}

The off-nuclear Br$\gamma$ emission in region SFr is a strong indicator for star formation. The EW of Br$\gamma$ is relatively high here.
The H$_2$(1-0)S(1) EW in that region goes down with respect to the same region in the eastern spiral arm indicating additional continuum emission, e.g., young star formation. The diagnostic diagram in Fig.~\ref{fig:ddiag1} places this region into the AGN regime, however, it is off the linear correlation.
This is an aperture effect.
PSF smearing due to the earlier mentioned \ion{H}{ii} deficiency at the center of NGC 1566 will introduce the shift in the diagnostic diagram of this region.
The $\log{\mbox{H}_2/\mbox{\brg}}$ line ratio has values of down to $-0.2$ at the position of the brightest spots in the HST images (Figs. \ref{fig:h212obrg}, \ref{fig:brgohst}). These at least three distinct bright emission regions in the HST image are probably the brightest or least attenuated star formation regions. However, the elongated shape of the \brg\ emission there indicates more star formation behind the dust and molecular gas of the nuclear spiral.

In the molecular line ratio diagrams the estimated ratios are close to the predicted ratios of \citet{black_fluorescent_1987} for a mixture of thermal and low-density fluorescence models (Fig.~\ref{fig:h2diag1}). The two off nuclear regions, SFr and SF~PSF, lie clearly off the thermal excitation curve with a tendency towards non-thermal UV excitation at smaller apertures.
The estimated ratio of non-thermal to thermal excitation for the region SFr would imply a contribution of $\sim30\%$.
This is a strong hint at young star formation taking place at regions SFr and SF~PSF.
Since strong \brg\ emission is detected in the SF~PSF region and the [\ion{Fe}{ii}] emission is low here, and therefore the SFR to SNR ratio is high with $>10^2$, it is probable that the star formation here is very young, e.g. $<9$~Myr.

The contribution of non-thermal excitation for the central regions is about $10\%-20\%$ (cPSF, $r=$~PSF and $r=1\arcsec$). This is lower than in regions SF~PSF and SFr but still a significant value that might hint at star formation at the nucleus.

The H$_2$ level population shows that the $v=1$ transitions scatter around the higher density models (model 2,3,4,5). The region SFr tends toward lower density gas, e.g. model 1, and indicates the characteristic ortho-to-para shift of fluorescent excitation \citep{sternberg_ratio_1999}. The $v=2$ transitions show a similar effect but the differences in the models are here more evident than for the $v=1$ transitions. 



\subsection{Continuum}
\label{sec:continuum}

We analyze the emission of the 0.87~mm continuum observed with ALMA and the NIR continuum observed with SINFONI.

\subsubsection{The millimeter continuum}

The mm-continuum at $0.87$~mm (Fig.~\ref{fig:ALMAcont}) peaks in the same region as the NIR continuum but the distribution is different compared to the NIR (see Sect.~\ref{sec:nirconti}). The $0.87$~mm emission is distributed similar to the molecular lines. Apart from the peak in the center, 0.87~mm emission is detected in the south-western spiral arm at the position of narrow \brg\ emission and in the north-east coinciding with the regions that show an increased width in the molecular lines (Fig.~\ref{fig:ALMACO32mom2}). \citet{combes_ALMA_2014} find that the $0.87$~mm continuum is dominated by dust emission in the full $18\arcsec\times18\arcsec$ FOV. To decide on the dominating mechanism in local emission regions is not possible due to the lack of high resolution radio data at other frequencies.

\subsubsection{The NIR continuum}
\label{sec:nirconti}

\begin{figure}[htbp]
\centering
\includegraphics[width=0.35\textwidth]{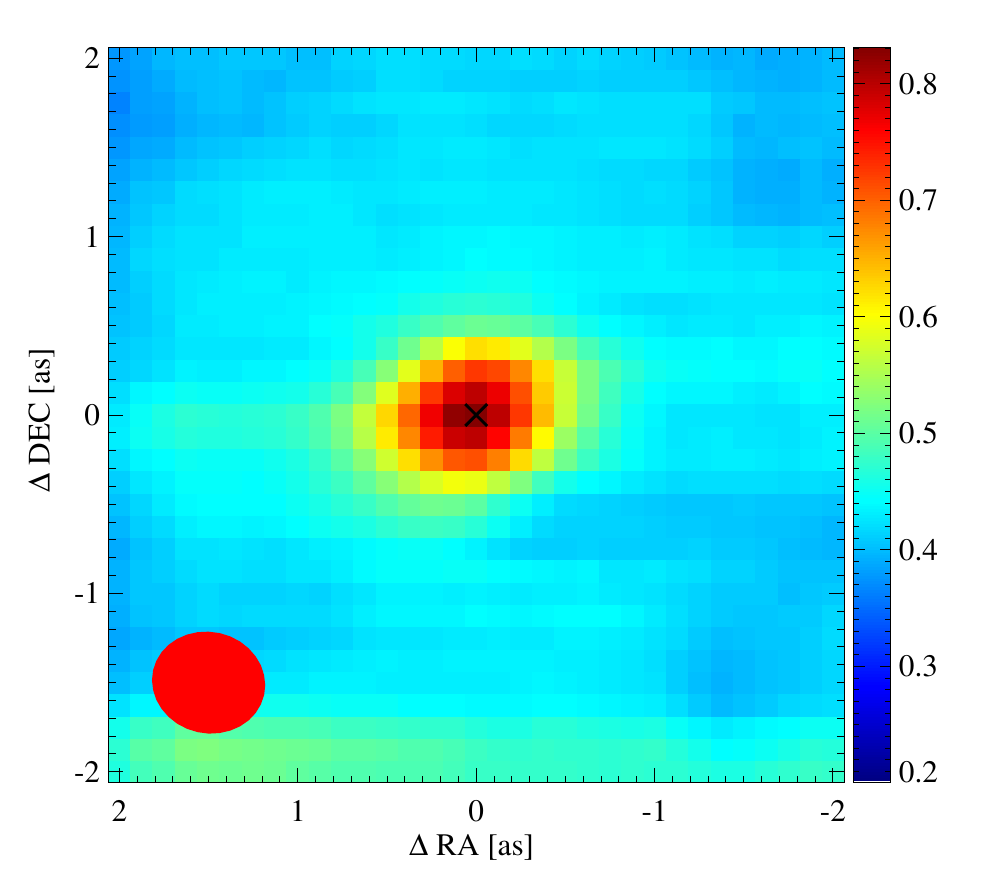}
\caption{The H-K color diagram in magnitudes of the central $4\arcsec\times4\arcsec$ of NGC 1566. The red ellipse at ($1\farcs5,-1\farcs5$) shows the beam size. For details see Sect. \ref{sec:continuum}.}
\label{fig:hkcont}
\end{figure}

In the NIR the H- and the K-band emission was observed. The continuum flux density is stronger in H- than in K-band. Towards the center the continuum becomes redder, as is expected in a Seyfert~1 galaxy, hence the slope becomes flatter but it is not inverted. The H-K map (Fig.~\ref{fig:hkcont}) shows a clear reddening toward the center with an H-K value of $>0.8$~mag at the very center.
This implies that we see warm to hot dust emission in the galaxy center \citep{fischer_nearby_2006,busch_low-luminosity_2014}. Since we see a broad \brg\ component we assume that we as well see the inner edge of the dust torus surrounding the AGN.

To analyze the NIR continuum further a decomposition was performed. The continuum components: stellar template, hot dust blackbody, power-law and an overlaying extinction component were used to determine the continuum composition. The stellar component was fitted using resolution adapted template stars from \citet{winge_gemini_2009}. Since these stars are only available in K-band from 2.2~\mic\ to 2.4~\mic\ the decomposition had to be performed in this wavelength range. For more details on the decomposition see \citet{smajic_unveiling_2012} and \citet{smajic_ALMA-backed_2014}.

\begin{figure*}[t]
\centering
\subfigure[Center $r=1\arcsec$]{\includegraphics[height=0.33\textwidth, angle=90]{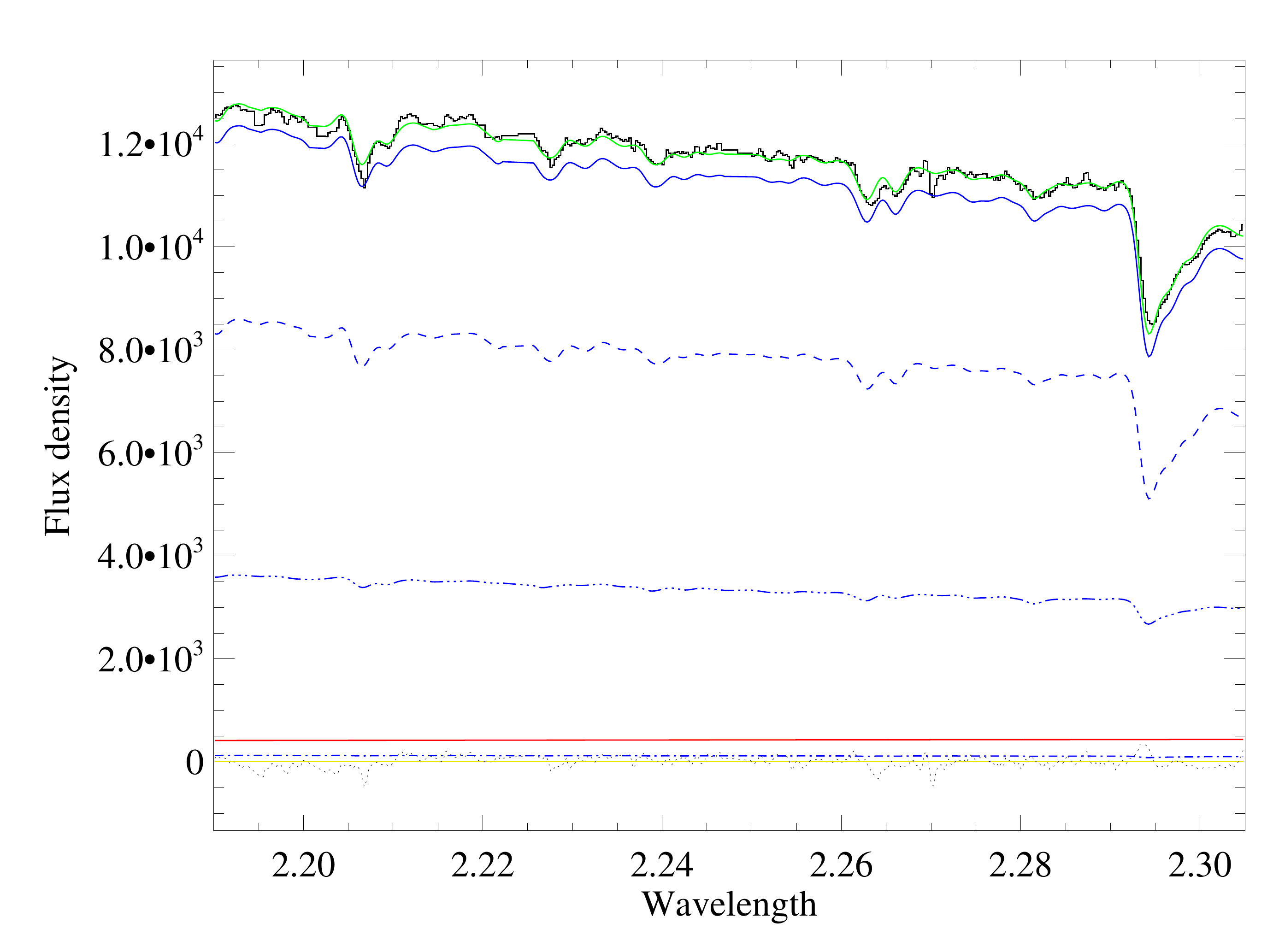}\label{fig:region1k}}
\subfigure[Center $r=5\arcsec$]{\includegraphics[height=0.33\textwidth, angle=90]{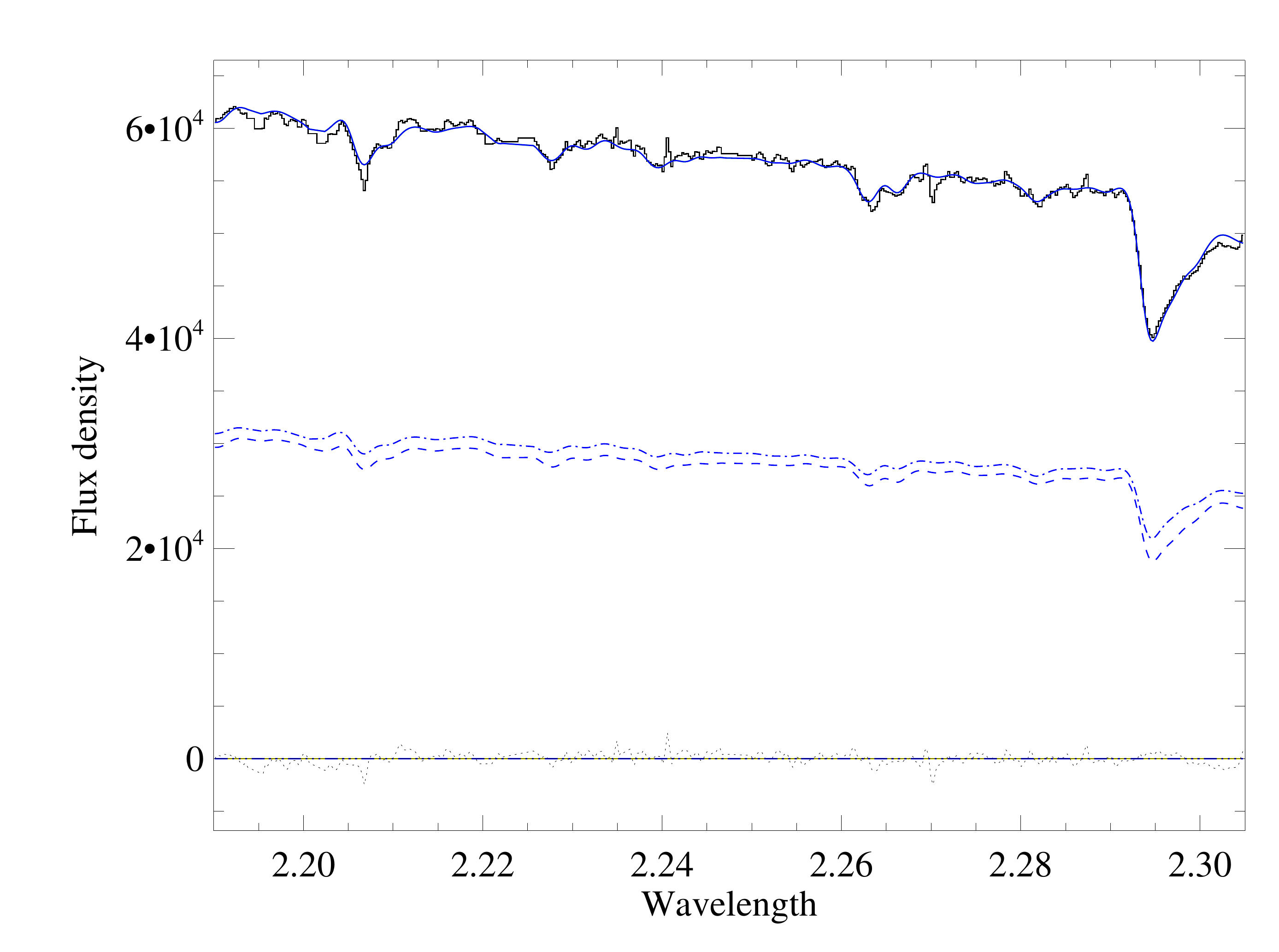}\label{fig:region2k}}
\subfigure[Center $r=$~PSF]{\includegraphics[height=0.33\textwidth, angle=90]{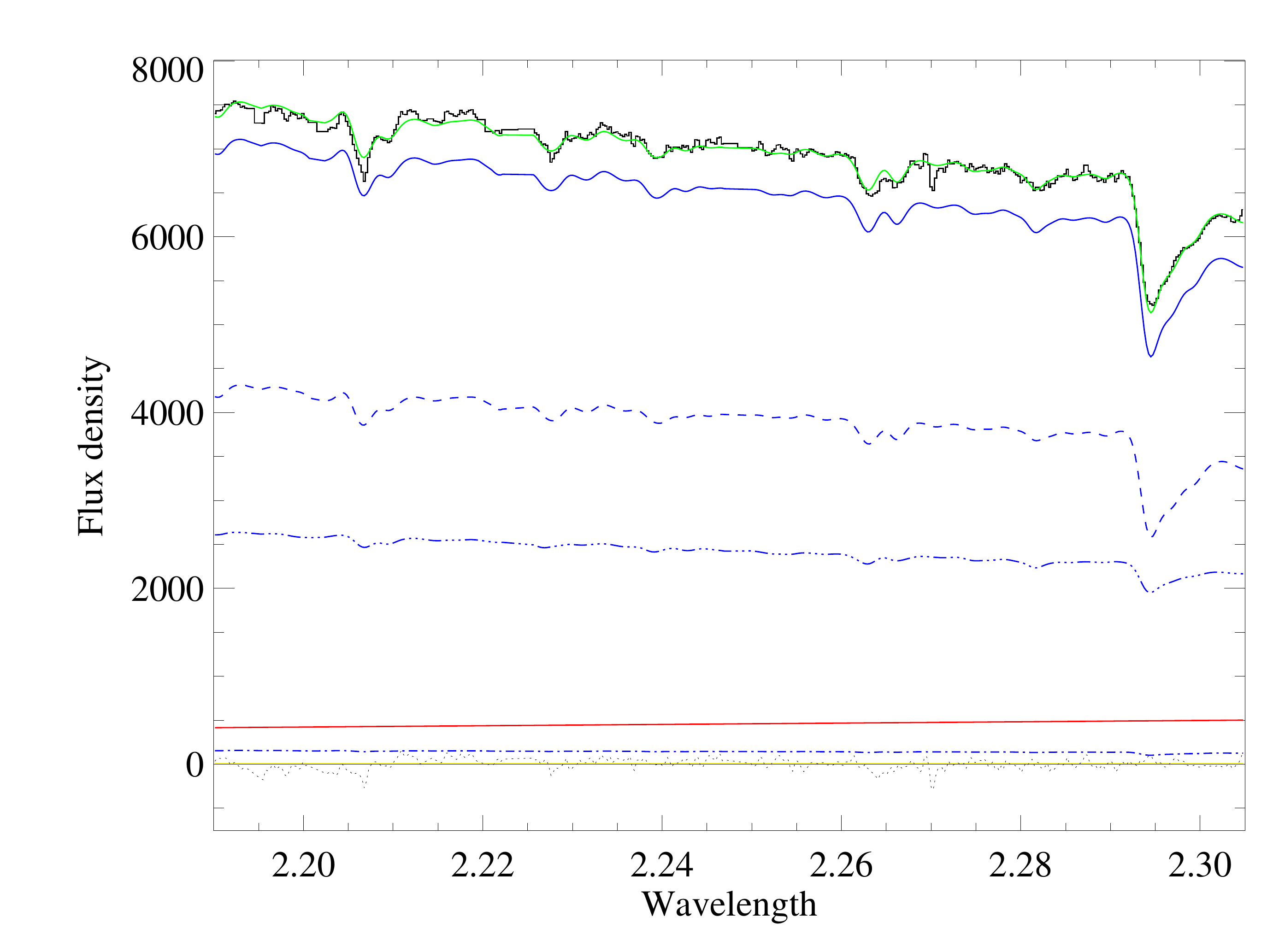}\label{fig:region3k}}
\subfigure[SFr]{\includegraphics[height=0.33\textwidth, angle=90]{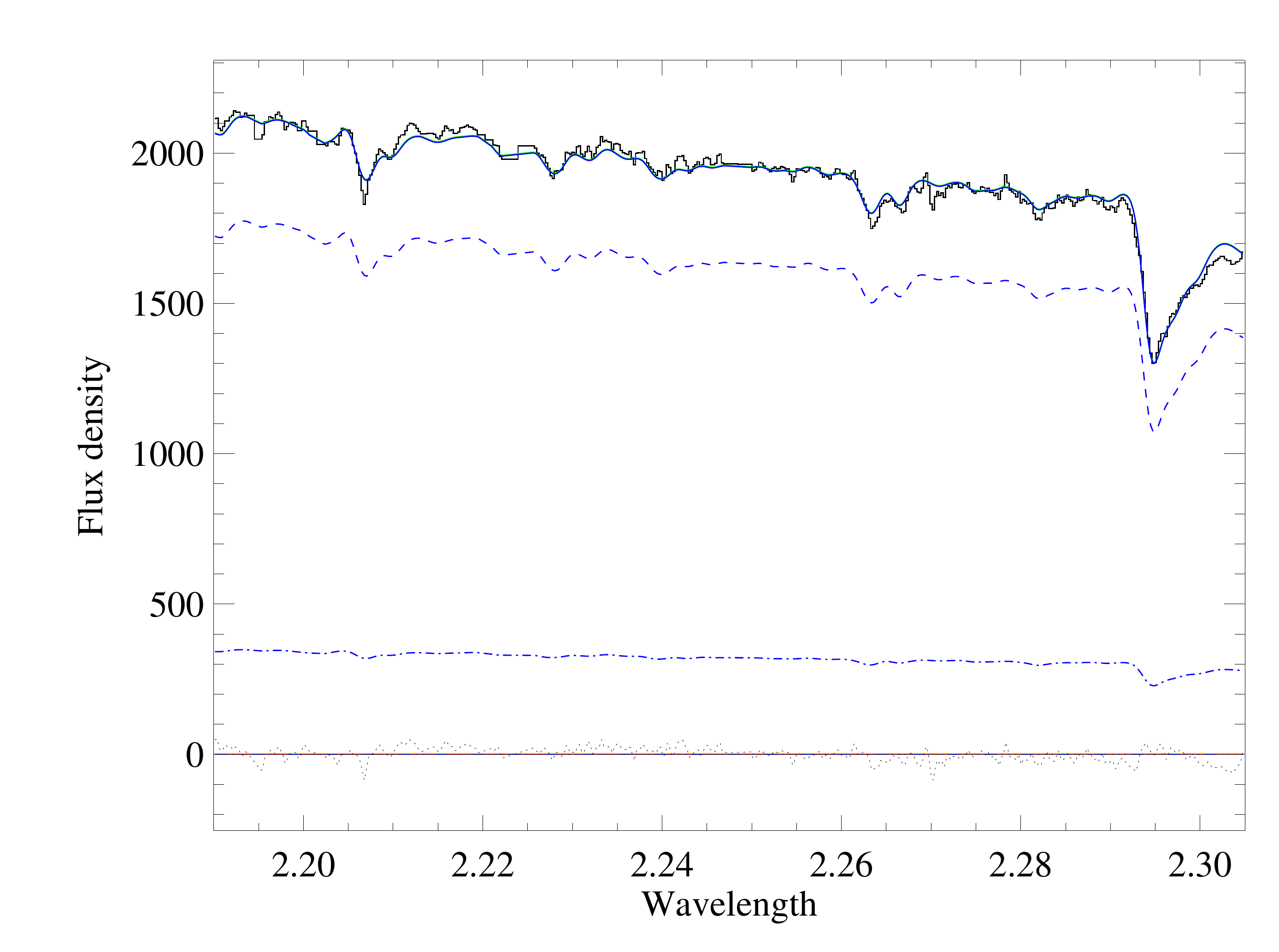}\label{fig:region4k}}
\subfigure[SF PSF]{\includegraphics[height=0.33\textwidth, angle=90]{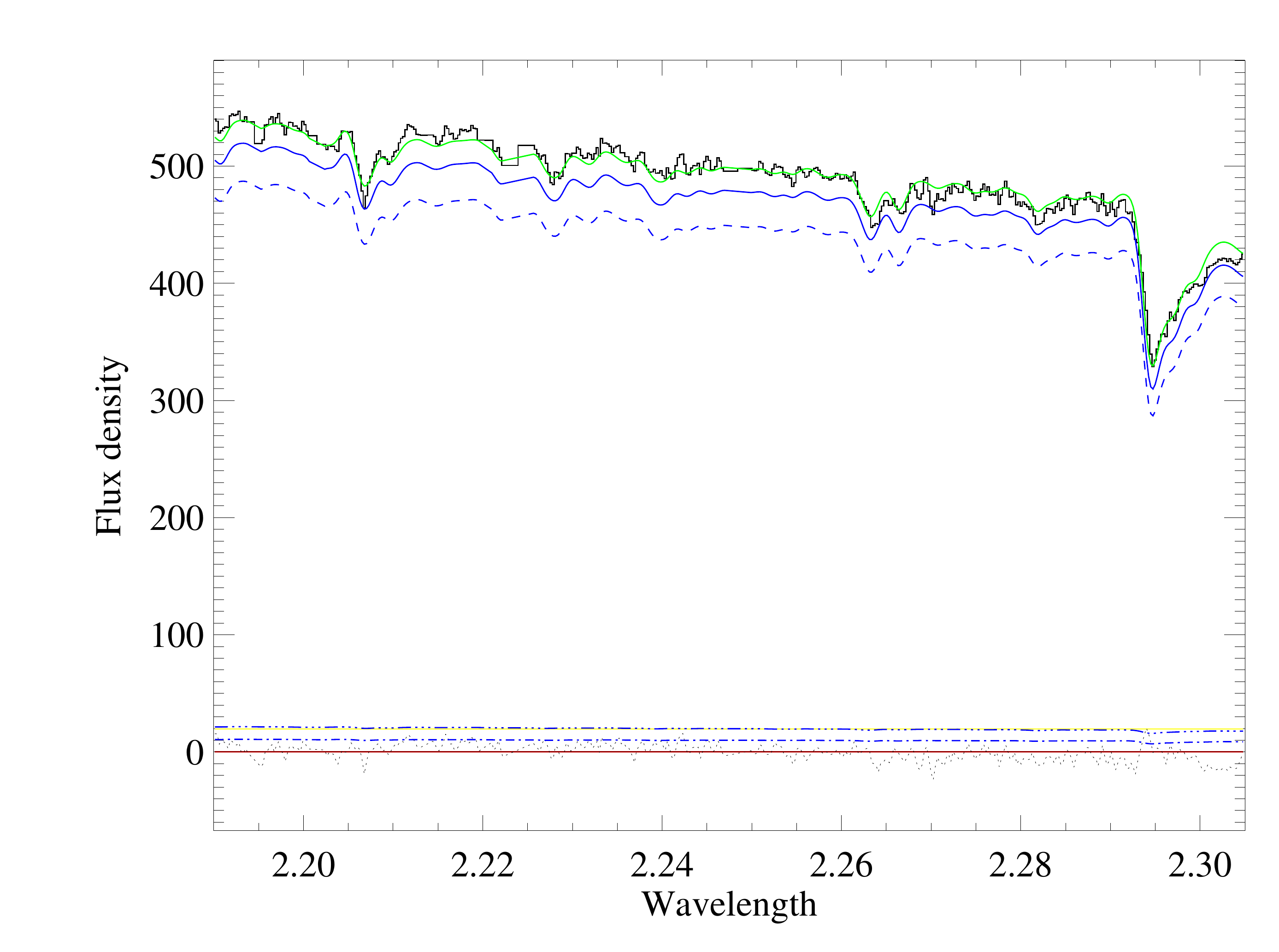}\label{fig:region5k}}
\subfigure[cPSF]{\includegraphics[height=0.33\textwidth, angle=90]{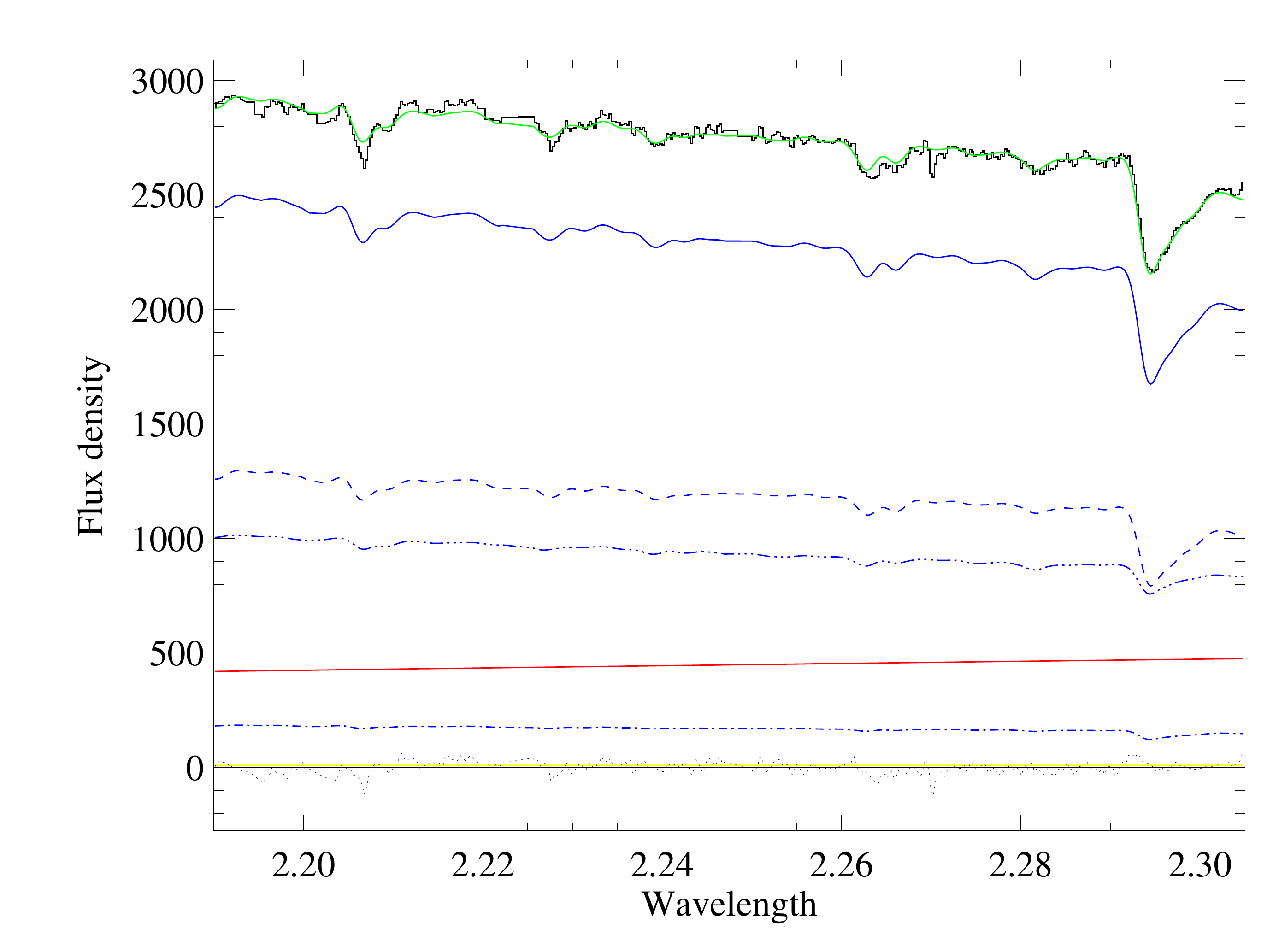}\label{fig:region6k}}
\caption{The continuum decomposition for six different regions. The axes are flux density [$10^{-18}$~W~m$^{-2}$~\mic$^{-1}$] versus wavelength [\mic]. The components are: {\it Red curve}: hot dust component; {\it Yellow curve}: power-law component; {\it Full blue curve}: total stellar component; {\it Lined blue curve}: M0III star; {\it Line-dot blue curve}: K3III star; {\it Triple-dot-lined blue curve}: G5III star. The combined components curve is in green and the original input spectrum is in black. The dotted black curve around zero is the residual of spectrum minus combined curve. Note that the emission lines at $\lambda$2.223~\mic\ and $\lambda$2.247~\mic\ were masked.}
\label{fig:regfit}
\end{figure*}

The decomposition provides two interesting results, see Fig.~\ref{fig:regfit}\subref{fig:region1k},\subref{fig:region3k},\subref{fig:region6k}. The blackbody, responsible for fitting hot dust, returns a hot $\sim1000$~K component with a significant contribution at the very center.
The stellar component produces best results when HD2490, an M0III star, representing the old stellar population, and HD1737, a G5III star are used. The contribution of G5III-like stellar emission becomes apparent at the center.

In the NIR, the continuum emission of NGC~1566 in the central $10\arcsec\times10\arcsec$ is mainly produced by stars. The continuum decomposition, however, reveals a hot dust component at the nucleus. The fits at the center, $r=1\arcsec$ (Fig.~\ref{fig:region1k}), $r=$~PSF (Fig.~\ref{fig:region3k}), and center PSF (cPSF) (Fig. \ref{fig:region6k}), need a hot dust component. The hot dust exhibits only $4\%$ of the flux density in the $r=1\arcsec$ region, whereas the $r=$~PSF region exhibits $7\%$ of the flux density and the cPSF region exhibits even 16\% flux density as hot dust emission at a temperature of $\sim1000$~K. The hot dust emission is not strong enough to create a steep red continuum in K-band. The difference in percentage here is caused by the aperture size of the central regions. All regions exhibit a hot dust flux density of $\sim500\times10^{-18}$~W~m$^{-1}$~\mic$^{-1}$ (Fig.~\ref{fig:regfit}). This hot dust emission at the nucleus is visible in the H-K map as well (Fig.~\ref{fig:hkcont}). The nucleus of NGC 1566 shows characteristics of Seyfert~1 nuclei on smaller scales
, i.e. broad hydrogen lines, and hot dust continuum emission. The effect of aperture on the features is essential. Apertures of more than 100~pc or 2$\arcsec$ would probably not be able to detect and measure the hot dust component. NGC 1566 would then show the characteristics of a quiescent galaxy without any narrow \ion{H}{ii} emission at the nuclear region, the broad \brg\ emission might be detected further out, depending on its strength \citep[e.g.,][]{reunanen_near-infrared_2002}.

The stellar continuum in our FOV is best fitted by an M0III star, a giant. The central $r\sim1\arcsec$ needs a G5III giant as well for the fit. The G-type star contributes one third of the flux in the central $r=1\arcsec$ region. Outside this region the G-type star is not needed. 
The differences between the two stars are the temperature, i.e. the spectral slope, and the depth of the absorption features, e.g. CO(2-0). The G5III stellar contribution indicates a need for bluer continuum emission or for shallower features in the continuum emission at the center of NGC 1566. The need for bluer continuum emission might stem from uncertainties in the hot dust contribution, however, the hot dust flux stays constant over several apertures. The main argument for a G-type star here is the EW of the CO absorption feature. The EW in CO(2-0) of the M0III star is too high, i.e. the CO absorption feature is too deep, although a hot dust contribution is taken into account. A higher hot dust contribution can be excluded since the spectral shape of the residual galaxy spectrum, after subtraction of the hot dust black body emission, becomes too blue for an K- or M-type star to fit. Star formation at the center of Seyfert galaxies and in the torus itself is not unusual \citep[e.g.][]{davies_close_2007} and can account for additional featureless continuum.
\begin{figure*}[htbp]
\centering
\subfigure[K-band continuum]{\includegraphics[width=0.33\textwidth]{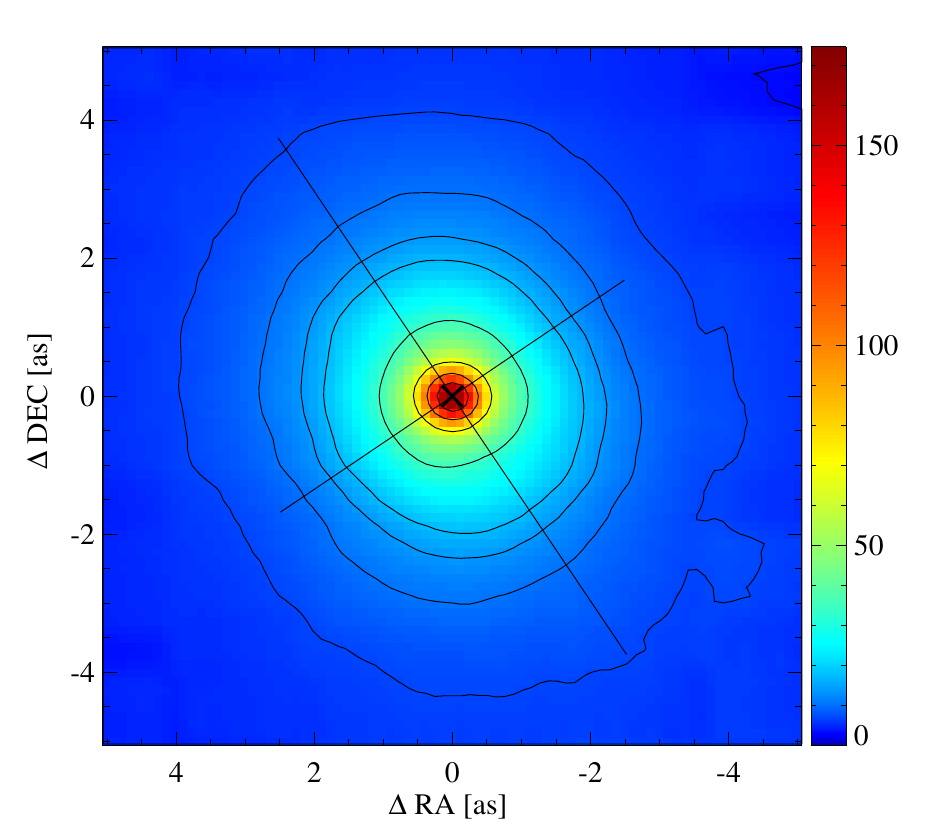}\label{fig:kcont}}
\subfigure[Stellar continuum]{\includegraphics[width=0.33\textwidth]{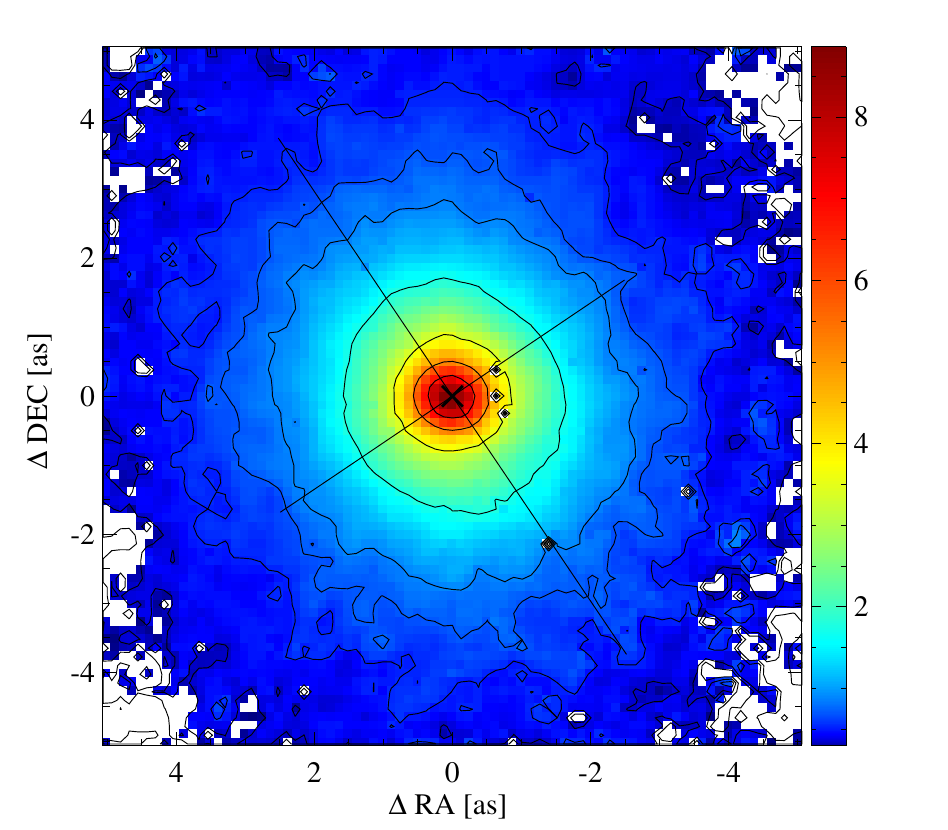}\label{fig:stellcont}}
\subfigure[Stellar dispersion]{\includegraphics[width=0.33\textwidth]{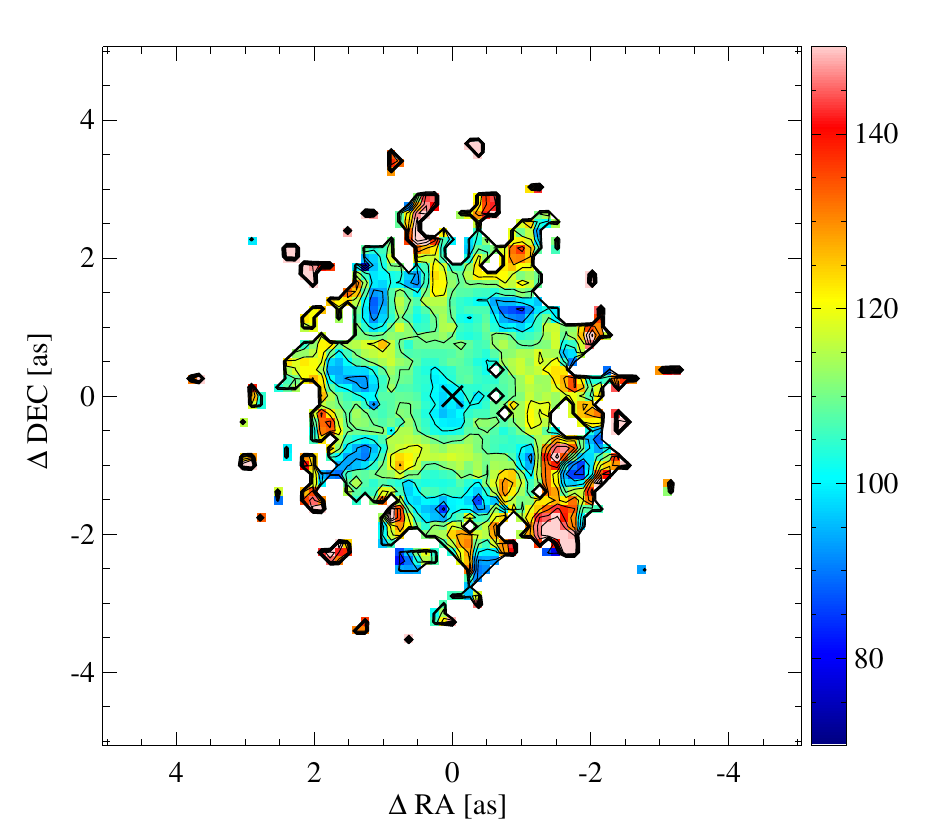}\label{fig:co20disp}}
\caption{From left to right: K-band continuum [$10^{-18}$~W~m$^{-2}$~$\mu$m$^{-1}$], the fitted stellar continuum [arbitrary units] and the SVD [\kms] obtained from the continuum decomposition. For more details, see Sects. \ref{sec:continuum}.
}
\label{fig:conti}
\end{figure*}

The spatial shape of the continuum emission is roundish with a slight elongation at a PA of $34\degr$. This is the same PA as the line-of-nodes of the stellar velocity field. Hence, the stars in the center do not follow the main stellar bar distribution, which is at a PA of $\sim0\degr$. The $34\degr$ that we measure are either created by a small nuclear bar within the $r=9\farcs7$ pseudo-ring \citep{comeron_ainur:_2010} or the angle stems from the bulge population and is affected by projection effects. However, to disentangle this a sophisticated decomposition of the galaxy structure using high resolution imaging in the NIR is required \citep[e.g.][]{busch_low-luminosity_2014}.

\begin{figure}[htbp]
\centering
\includegraphics[height=0.35\textwidth, angle=90]{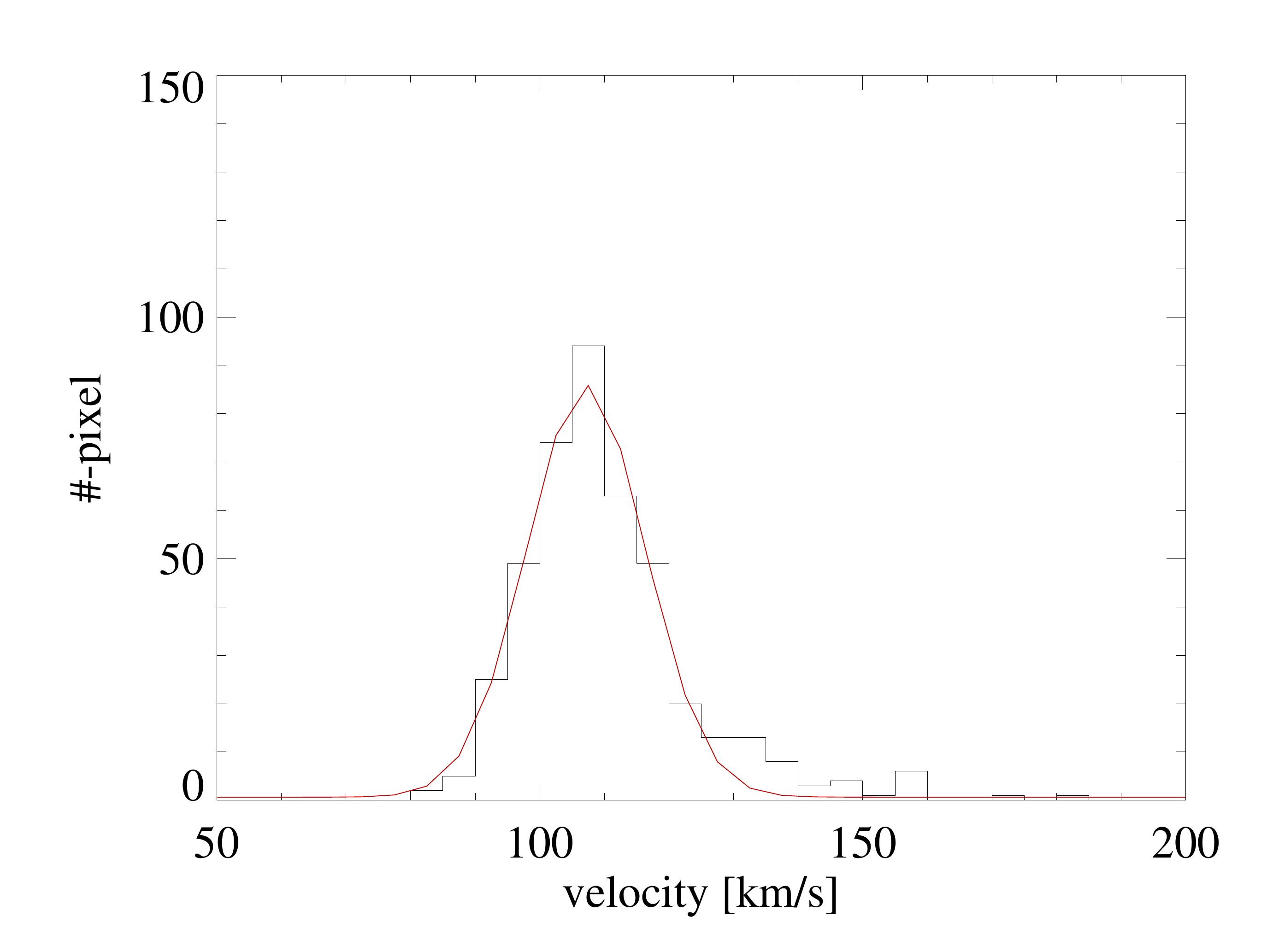}
\caption{
Stellar velocity dispersion histogram. The X-axis shows the fitted dispersion in $5$~\kms\ bins. The Y-axis shows the total number of spatial pixel that correspond to the dispersion bin. The red curve is a Gaussian fit to the distribution.}
\label{fig:disp}
\end{figure}

\begin{figure*}[htbp]
\centering
\subfigure[Stellar LOSV]{\includegraphics[width=0.33\textwidth]{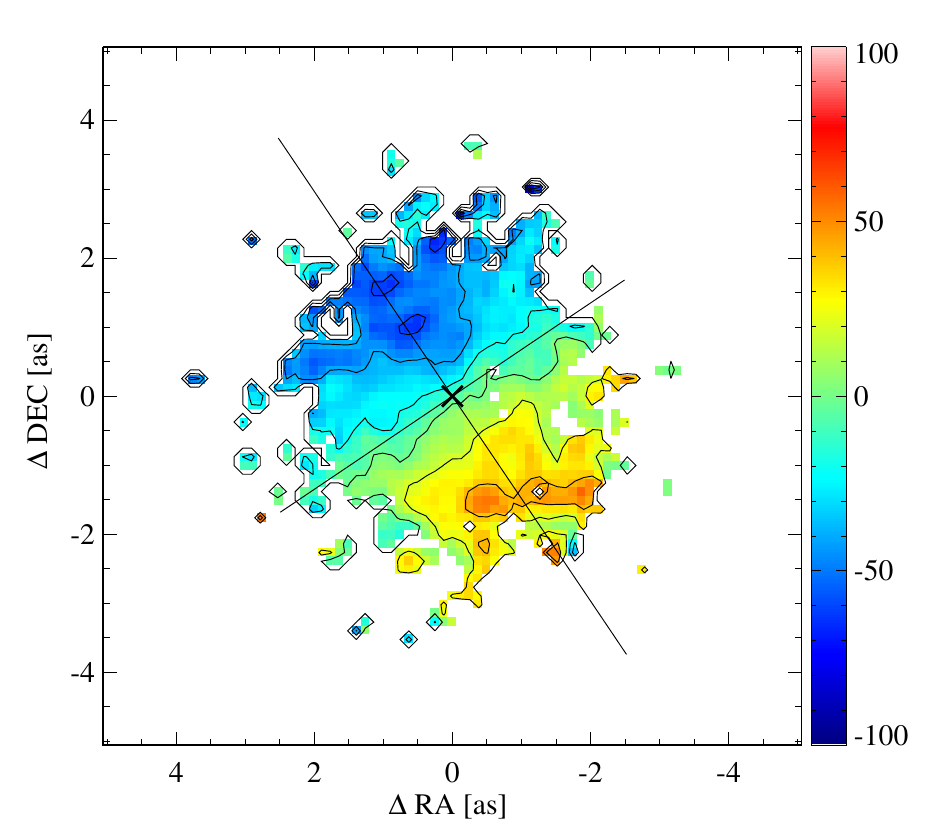}\label{fig:co20losv}}
\subfigure[Model LOSV]{\includegraphics[width=0.33\textwidth]{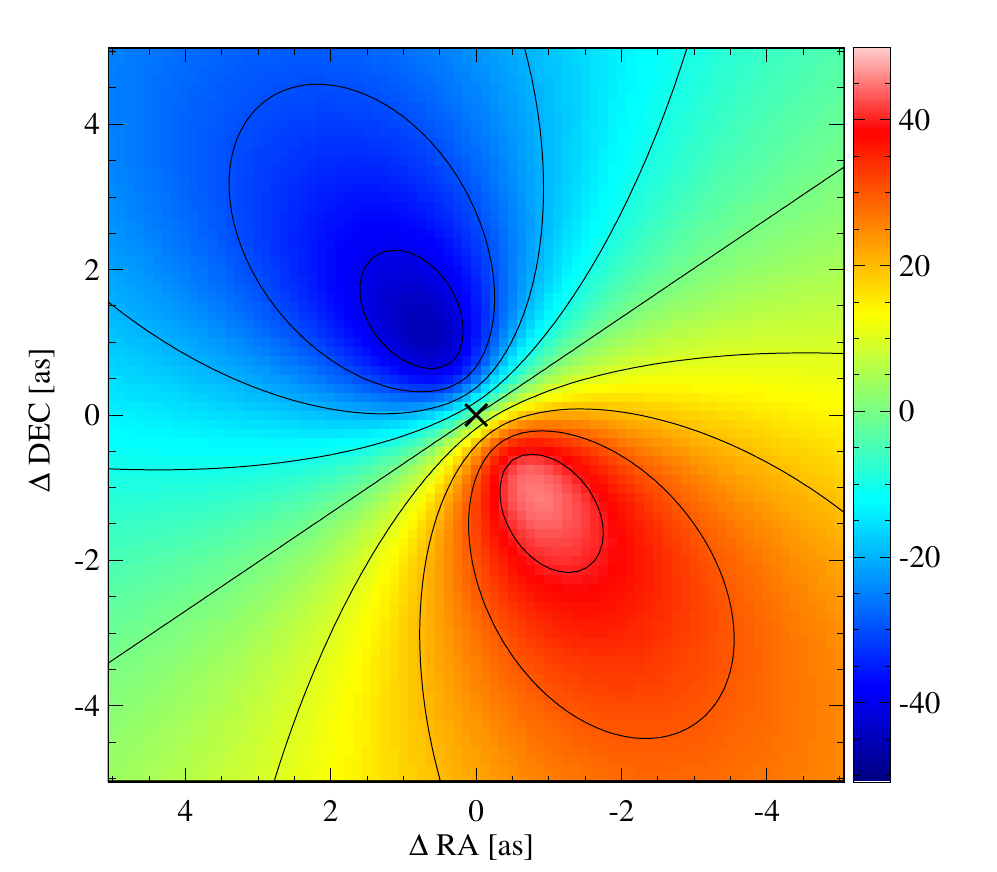}\label{fig:losvco20mvlc}}
\subfigure[Residual LOSV]{\includegraphics[width=0.33\textwidth]{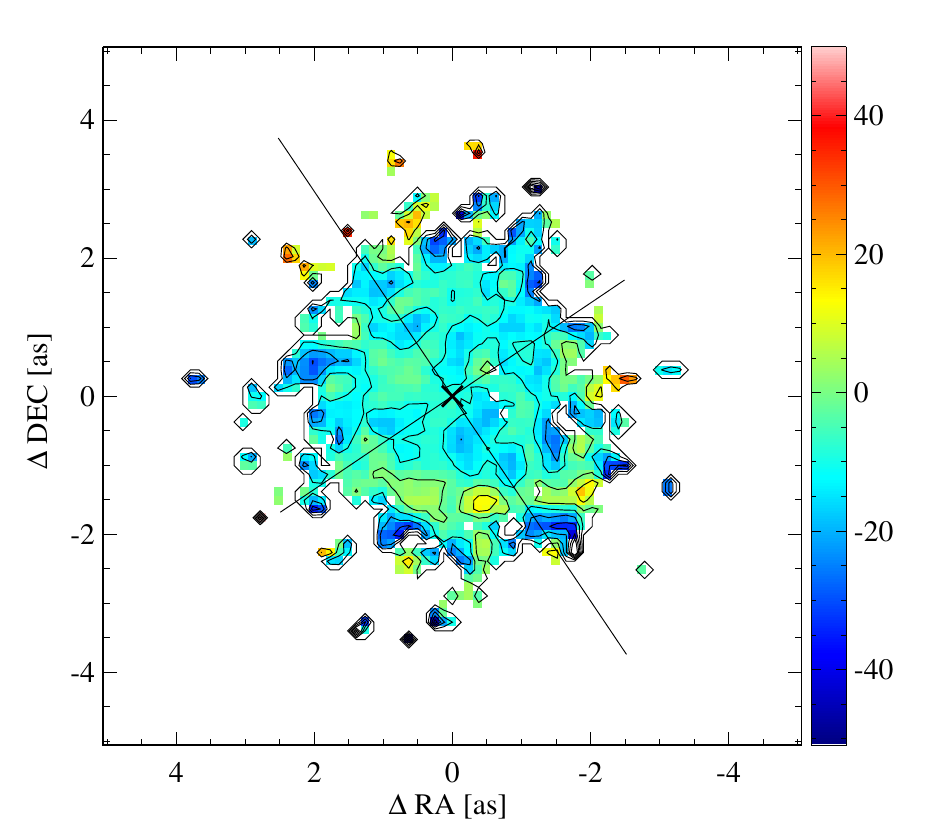}\label{fig:losvco20dmvlc}}
\subfigure[H$_2$(1-0)S(1) LOSV]{\includegraphics[width=0.33\textwidth]{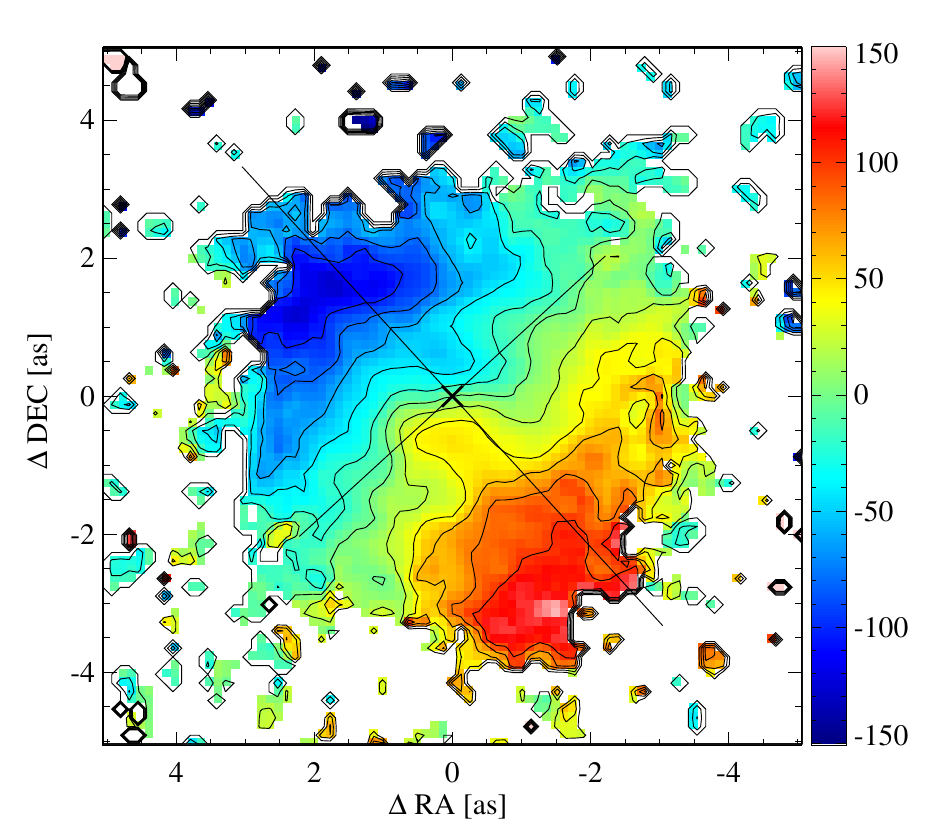}\label{fig:h212losv}}
\subfigure[Model LOSV]{\includegraphics[width=0.33\textwidth]{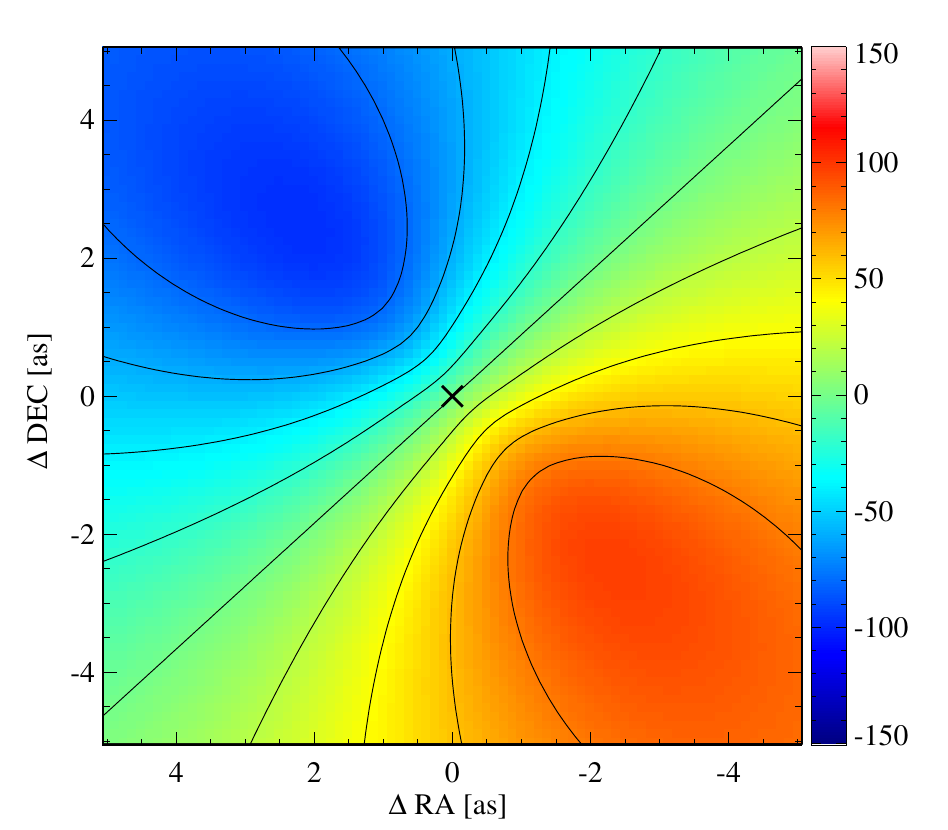}\label{fig:losvmvlc}}
\subfigure[Residual LOSV]{\includegraphics[width=0.33\textwidth]{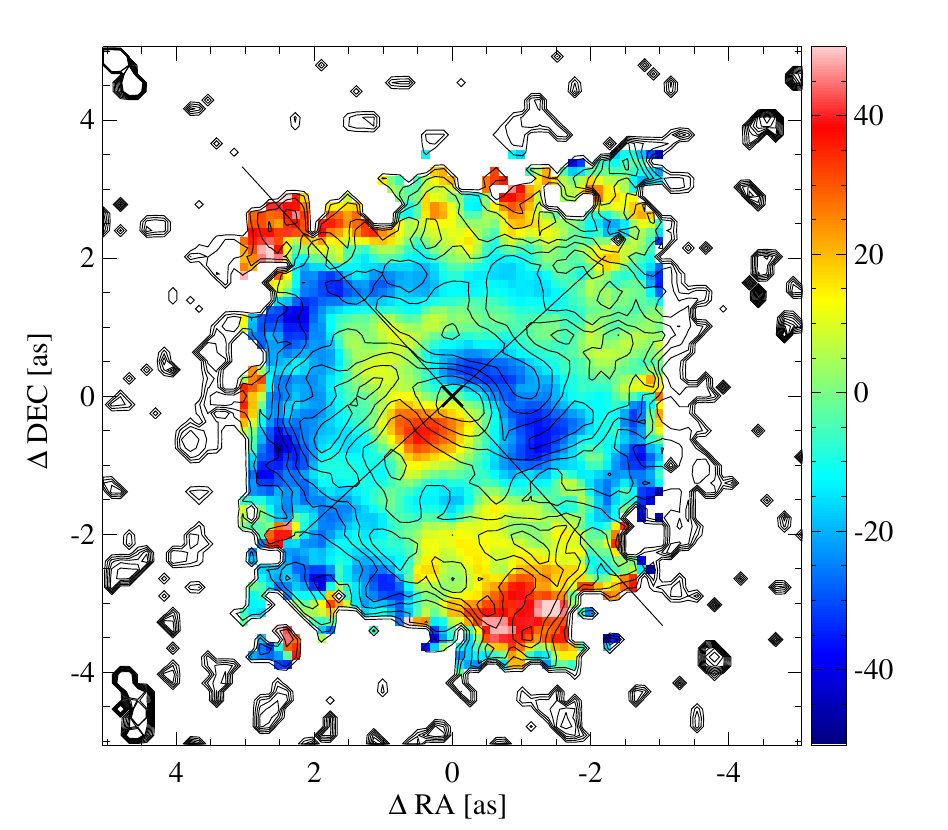}\label{fig:losvdmvlc}}
\caption{Top panel: Stellar LOSV, fitted LOSV model and the subtracted residual in units of [\kms]. Bottom: H$_2$(1-0)S(1) LOSV, fitted LOSV model and the subtracted residual overlayed with contours of the EW map of H$_2$(1-0)S(1) in units of [\kms]. The EW map was overlayed to display the nuclear spiral onto the residual velocity field of the gas. The straight lines denote the orientation of the major and minor rotation axis as determined by the LOSV model. For more details, see Sect. \ref{sec:kin}.
}
\label{fig:vel}
\end{figure*}

The stellar velocity dispersion, LOSV, and intensity distribution are shown in Figs. \ref{fig:conti}, \ref{fig:disp}, and \ref{fig:vel}. The stellar LOSV shows a rather regular rotational field with a PA of $214\degr$ at velocities of $\pm60$~km~s$^{-1}$ with a boxy redshifted side. The stellar velocity dispersion shows velocities of 90 to 125 \kms\ with the very center being at 100 \kms\ and does not show signs of kinematically decoupled regions.

%% file: region5.tex
\begin{table*}[htbp]
\centering
\caption{Emission lines}
\begin{tabular}{c c c c c c c}
\hline\hline
\multicolumn{7}{c}{Flux [$10^{-18}$~W~m$^{-2}$]} \\
 Line & Center $r=1\arcsec$ & Center $r=5\arcsec$ & Center $r=$~PSF & SFr & SF PSF & cPSF \\
\hline
 $[$\ion{Fe}{ii}$]$ & $5.93\pm0.68$ & $...$ & $4.82\pm0.38$ & $0.21\pm0.09$ & $0.06\pm0.02$ & $2.39\pm0.13$ \\
 H$_2$(1-0)S(3) & $3.80\pm0.33$ & $15.15\pm3.18$ & $2.35\pm0.19$ & $0.57\pm0.05$ & $0.14\pm0.01$ & $0.93\pm0.07$ \\
 H$_2$(1-0)S(2) & $1.66\pm0.18$ & $5.46\pm1.35$ & $1.01\pm0.10$ & $0.28\pm0.03$ & $0.08\pm0.01$ & $0.39\pm0.04$ \\
 \ion{He}{i} & $0.51\pm0.14$ & $...$ & $0.42\pm0.09$ & $0.17\pm0.02$ & $0.05\pm0.006$ & $0.23\pm0.04$ \\
 H$_2$(2-1)S(3) & ... & ... & $0.12\pm0.07$ & $0.03\pm0.02$ & $0.012\pm0.006$ & $0.05\pm0.03$ \\
 H$_2$(1-0)S(1) & $3.41\pm0.15$ & $11.21\pm1.2$ & $2.25\pm0.09$ & $0.52\pm0.03$ & $0.13\pm0.008$ & $0.96\pm0.03$ \\
 Br$\gamma$ & $0.71\pm0.16$ & $...$ & $0.60\pm0.10$ & $0.35\pm0.03$ & $0.14\pm0.008$ & $0.31\pm0.04$ \\
 H$_2$(1-0)S(0) & $1.04\pm0.12$ & $3.68\pm1.06$ & $0.64\pm0.07$ & $0.17\pm0.02$ & $0.05\pm0.005$ & $0.25\pm0.03$ \\
 H$_2$(2-1)S(1) & $0.55\pm0.11$ & $...$ & $0.35\pm0.07$ & $0.11\pm0.02$ & $0.03\pm0.005$ & $0.14\pm0.03$ \\
 H$_2$(1-0)Q(1) & $3.68\pm0.35$ & $33.64\pm8.44$ & $2.23\pm0.17$ & $0.74\pm0.11$ & $0.16\pm0.02$ & $0.86\pm0.07$ \\
 H$_2$(1-0)Q(3) & $3.23\pm0.24$ & $49.70\pm5.7$ & $1.80\pm0.12$ & $0.81\pm0.11$ & $0.17\pm0.03$ & $0.69\pm0.05$ \\
\hline

\multicolumn{7}{c}{FWHM [\kms]} \\
\hline
 $[$\ion{Fe}{ii}$]$ & $321\pm48$ & $...$ & $333\pm34$ & $130\pm89$ & $84\pm60$ & $344\pm24$ \\
 H$_2$(1-0)S(3) & $211\pm17$ & $227\pm43$ & $211\pm15$ & $144\pm14$ & $124\pm12$ & $207\pm14$ \\
 H$_2$(1-0)S(2) & $221\pm28$ & $214\pm63$ & $219\pm26$ & $177\pm25$ & $183\pm26$ & $210\pm24$ \\
 \ion{He}{i} & $150\pm54$ & ... & $165\pm43$ & $89\pm19$ & $82\pm16$ & $197\pm37$ \\
 H$_2$(2-1)S(3) & ... & ... & $70\pm46$ & $14\pm10$ & $56\pm32$ & $84\pm44$ \\
 H$_2$(1-0)S(1) & $191\pm10$ & $189\pm25$ & $199\pm9$ & $139\pm10$ & $129\pm10$ & $204\pm8$ \\
 Br$\gamma$ & $165\pm44$ & ... & $199\pm37$ & $96\pm11$ & $103\pm7$ & $225\pm31$ \\
 H$_2$(1-0)S(0) & $173\pm22$ & $210\pm67$ & $178\pm22$ & $116\pm16$ & $112\pm16$ & $179\pm25$ \\
 H$_2$(2-1)S(1) & $172$ & $...$ & $176$ & $115$ & $111$ & $177$ \\
 H$_2$(1-0)Q(1) & $166\pm14$ & $153\pm36$ & $171\pm12$ & $115\pm19$ & $95\pm17$ & $177\pm12$ \\
 H$_2$(1-0)Q(3) & $171\pm14$ & $150\pm20$ & $185\pm14$ & $171\pm26$ & $170\pm30$ & $195\pm14$ \\
\hline

\end{tabular}
\tablefoot{Flux and FWHM table for all regions discussed in this paper (see also Fig.~\ref{fig:lines1} for abbreviations).
The linewidth of H$_2$(2-1)S(1) was tied to the width of emission line H$_2$(1-0)S(0) during the fit, hence no error can be given for the linewidth. The FWHM values are corrected for instrumental broadening.
}
\label{tab:region}
\end{table*}

%% file: discussion.tex

\subsection{The active nucleus}
The AGN of NGC~1566 is a Seyfert 1 nucleus that exhibits broad hydrogen emission lines and is showing variability across the whole wavelength range.

\subsubsection{Mass of the SMBH}

\label{sec:active}

The broad Br$\gamma$ line flux and FWHM are used to estimate the mass of the central SMBH of NGC 1566. A conversion factor of Pa$\alpha$/\brg\ $\sim12$ is used to translate the Br$\gamma$ to Pa$\alpha$ flux and be able to use the equation
\begin{equation}
\mbox{M}_{\bullet}=10^{7.29\pm0.1}\left(\frac{L_{\mbox{\tiny Pa}\alpha}}{10^{42}\mbox{erg s}^{-1}}\right)^{0.43\pm0.03}\left(\frac{FWHM_{\mbox{\tiny Pa}\alpha}}{10^{3}\mbox{km s}^{-1}}\right)^{1.92\pm0.18}\mbox{M}_\odot
\label{eqn:}
\end{equation}
derived by \citet{kim_new_2010}. We estimate a black hole mass of M$_{\bullet}=(3.0\pm0.9)\times10^6$~M$_{\odot}$ using a luminosity of $L_{\mbox{\tiny\brg}}=5\times10^{38}$~erg~s$^{-1}$ and a $FWHM_{\mbox{\tiny\brg}}=2000$~km~s$^{-1}$. This value is less than a factor two smaller than the $5\times10^6$~M${_\odot}$ estimated from optical broad line measurements \citep{kriss_faint_1991}. This difference might be introduced by variations in the activity of the AGN. A flux increase in the broad H$\beta$ line was measured by a factor of four to five within only 24 days \citep{alloin_recent_1985}. Assuming a low activity state during our observation, an increase of a factor four in luminosity will increase the BH mass derived from the \brg\ broad line to $(5.5\pm1.7)\times10^6$~M$_{\odot}$, a factor of two. Hence, the derived value of $(3.0\pm0.9)\times10^6$~M$_{\odot}$ is in good agreement with the literature value.

To use the M$-\sigma_\ast$ relation and estimate the mass of the SMBH the dispersion of the bulge is needed. To determine the bulge dispersion a Gaussian fit was performed on the distribution of the stellar velocity dispersion values (Fig.~\ref{fig:disp}). The fit yields a dispersion of $105\pm10$~\kms. Following \citet{gultekin_m-_2009}
\begin{equation}
M_{\bullet}=10^{8.12\pm0.08}\times\left(\frac{\sigma_{*}}{200\;\mbox{km s}^{-1}}\right)^{4.24\pm0.41}M_{\odot},
\end{equation}
the mass of the SMBH is then estimated to M$_{\bullet}=8.6\pm4.4\times10^6$~M$_{\odot}$. This value is similar to the $8.3\times10^6$~M$_{\odot}$ from \citet{woo_active_2002} but a factor two to three higher than the BH mass estimate from the broad emission lines.

\citet{graham_m_2013} have investigated the M$-\sigma_\ast$ relation for barred galaxies. They find that barred galaxies follow a slightly different M$-\sigma_\ast$ relation than non-barred galaxies. Their best fit
\begin{equation}
M_{\bullet}=10^{7.92\pm0.23}\times\left(\frac{\sigma_{*}}{200\;\mbox{km s}^{-1}}\right)^{5.29\pm1.47}M_{\odot},
\end{equation}
returns 
an upper limit for the BH mass of M$_{\bullet}=6\times10^6$~M$_{\odot}$. When we use their fit with the least root-mean-square scatter
\begin{equation}
M_{\bullet}=10^{7.78\pm0.1}\times\left(\frac{\sigma_{*}}{200\;\mbox{km s}^{-1}}\right)^{4.14\pm0.55}M_{\odot},
\end{equation}
the result is a BH mass of M$_{\bullet}=(4.2\pm2.4)\times10^6$~M$_{\odot}$.
The latter two results agree within the errors with mass estimates using the broad emission lines, i.e., broad \brg\ (see Sect. \ref{sec:ionizedgas}), broad H$\alpha$ \citep{kriss_faint_1991}.

The scatter of the individually estimated BH masses is higher than their uncertainties. Therefore, we derive a mean BH mass from the four relations used in this paper to estimate the BH mass of NGC 1566. The mean BH mass is $(5.3\pm2.9)\times10^6$\msun. All individual BH mass values are within the standard deviation of the mean BH mass. 

\subsubsection{Activity of the SMBH}

NGC~1566 is actively accreting mass as can be inferred from the observed variabilities. The $2-10$~keV X-ray luminosity of the AGN was measured by \citet{levenson_isotropic_2009} to $L_{X,D20}=10^{41.5}$~erg~s$^{-1}$ for a distance of 21.2~Mpc, hence we estimate an $L_X=7.0\times10^{40}$~erg~s$^{-1}$ for 10~Mpc.
From the derived mean BH mass an Eddington luminosity of $L_{Edd}=(6.6\pm3.7)\times10^{44}$~erg~s$^{-1}$ can be determined. Using the relation $L_{\mbox{\tiny bol}}\approx16\times L_X=(1.1\pm0.4)\times10^{42}$~erg~s$^{-1}$ for LLAGN \citep{ho_nuclear_2008,ho_radiatively_2009} the Eddington ratio of the active nucleus in NGC 1566 can be estimated to $\lambda_{Edd}=(2\pm1)\times10^{-3}$. This is a typical value for Seyfert~1 LLAGN \citep{ho_nuclear_2008}. Although \cite{kawamuro_broadband_2013} use different parameters, e.g. BH mass, distance, $L_X$ conversion factor etc., they find a similar value of $3.2\times10^{-3}$.

The mass accretion onto the SMBH can then be estimated with
\begin{equation}
\frac{\mbox{d}M}{\mbox{d}t}=\frac{L_{\mbox{\tiny bol}}}{\eta c^2},
\end{equation}
where $\eta$ is an efficiency factor which is usually of the order of 0.1.
We calculate a mass accretion rate of $(2.0\pm0.7)\times10^{-4}$~M$_{\odot}$~yr$^{-1}$ using $L_{\mbox{\tiny bol}}$ from above.
The cold H$_2$ gas mass in the $10\arcsec\times10\arcsec$ FOV was measured to be $\sim5.4\times10^7$~M$_\odot$. This mass is enclosed in the central $r=3\arcsec$ gas disk (see Fig.~\ref{fig:h212flux}) which then has a column density of $\sim1.9\times10^6$~M$_\odot$~arcsec$^{-2}$ or $\sim7.7\times10^2$~M$_\odot$~pc$^{-2}$. We derive a gas mass density of $3.6\times10^{2}$~\msun~pc$^{-3}$ from the H$_2$ cPSF flux measurement by assuming that the emission stems from a spherical region 29~pc in diameter. Hence, the black hole of NGC 1566 has enough mass enclosed in its central few parcsecs to accrete at the current rate for a few Myrs.

Due to the narrow line ratio of H$\alpha$/H$\beta\sim3.1\pm0.3$, which are typical for Seyfert~2 galaxies, \citet{alloin_recent_1985} conclude that NGC 1566 might be a waking up Seyfert~1, which was in a low ionization state in the past few hundred years as inferred from emission line ratios in the NLR, beginning the harder ionization of the nuclear region. The AGN would have enough fuel to do so and start ionizing hydrogen outside the $\sim13.5$~pc at the center (see Sect. \ref{sec:ionizedgas}). However, on the one hand, the line ratio from \citet{alloin_recent_1985} might have been created by an aperture effect and the \ion{H}{ii} deficiency in the central region of NGC 1566 \citep{comte_galaxy_1982}. On the other hand, the \ion{H}{ii} deficiency might result from the low activity state in the past of NGC 1566 since there is hydrogen in the nuclear region but in molecular form mainly. The activity of NGC 1566 over the last $\sim20$~yr is a variable one \citep[e.g.,][]{alloin_recent_1985,kriss_faint_1991,baribaud_variability_1992,levenson_isotropic_2009} without a specific trend of in- or decrease. Hence, we can confirm that NGC~1566 hosts a Seyfert 1 nucleus at its center and it has a gas reservoir in the central few pc to keep the central engine fueled for several Myrs.

\subsection{Kinematics}
\label{sec:kin}
The kinematics in the center of NGC 1566 show rotations at a PA of $\sim214\degr$ for the stellar kinematics (Fig.~\ref{fig:co20losv}) and at a PA of $\sim222.5\degr$ for the molecular gas (e.g., H$_2$(1-0)S(1), Fig.~\ref{fig:h212losv}). 
The observed velocity field of the molecular gas shows strong deviations from rotation at the center.

The orientation of the galaxy toward the observer can be inferred from considerations of the winding sense of the spiral arms (Fig.~\ref{fig:ngc1566}). These kind of spiral arms are always seen to be trailing. Hence, the near side has to be in the northwest when the arms are trailing because there is blueshifted motion in the northeast and the arms are oriented as seen in Fig.~\ref{fig:ngc1566}.

\subsubsection{Stellar kinematics}
The line of nodes of the stellar rotation is aligned with the stellar continuum major axis at a PA of $34\degr$. We derive a PA from the H$_2$(1-0)S(1) line of $42\fdg5$ in good agreement with \citet{aguero_ngc_2004}. They find a PA of $44\degr\pm8\degr$ from optical emission lines.
The stellar LOSV map shows a smooth rotation field. We fit a model to the observed velocity using the Plummer potential to represent the bulge gravitational potential \citep{barbosa_gemini/gmos_2006}. The model subtracted map shows low residuals (Fig.~\ref{fig:losvco20dmvlc}). The stellar velocity dispersion shows a slight drop at the very center (Sect.~\ref{sec:continuum}). This is often an indication for a stellar disk \citep[e.g.][]{emsellem_dynamics_2001,falcon-barroso_sauron_2006} at scales $<100$~pc. However, there are no features in the stellar residual LOSV map (Fig.~\ref{fig:losvco20dmvlc}) that would support a stellar disk at these scales.

%

The difference in stellar and gaseous PA might be explained by streaming motions of the gas related to a strong spiral wave. But the gaseous PA over the central $8\arcsec\times8\arcsec$ is in agreement with measurements at larger scales, hence the effect has to stem from larger scale spiral density waves rather than the nuclear spiral discussed here. To substantiate any misalignment between gaseous and stellar disk, however, a much better estimate of the PAs is needed. Hence, we find that gaseous and stellar disk are oriented very similar but might be misaligned.


\subsubsection{Gas kinematics}
\label{sec:gaskin}

On scales of $200-300$~pc \citet{combes_ALMA_2014} showed that the gravitational torques are able to transport almost half of the angular momentum of the gas in one rotation period. Here gas is able to move in spiral arms that connect the inner Lindblad resonance (ILR) of the nuclear bar with the central $\le200$~pc.
On smaller scales the angular momentum transport is smaller per rotation period, but the period becomes smaller as well. The inner 200~pc show the molecular nuclear gas disk which exhibits a clear two-arm spiral structure (Figs. \ref{fig:h212flux}, \ref{fig:h212ew}, and \ref{fig:ALMACO32}).

The nuclear spiral disturbs the velocity field of the molecular gas and creates a strong S-shaped feature at the center. To remove the rotational part from the spiral disturbance we again fit a model to the observed velocity using the Plummer potential to represent the bulge gravitational potential. The residual of the fit (Fig.~\ref{fig:losvdmvlc}) highlights the difference in the observed LOSV (Fig.~\ref{fig:h212losv}) and the model (Fig.~\ref{fig:losvmvlc}). The center of the residual map shows blueshifted residuals to the west and redshifted residuals to the east of the nuceleus. The nucleus is situated at $0$~\kms.
Contrary to the case of NGC 1068, discussed by \citet{garcia-burillo_molecular_2014}, where non-virial outflow motions are required to explain the residual velocity field derived for the nuclear $200$~pc gas disk of that Seyfert~2 galaxy, in NGC 1566 residuals shown in Fig.~\ref{fig:losvdmvlc} can be accounted for by streaming motions produced by the spiral/bar structure.
The S-shape shows non-circular motion in one surface as it is the case for bars or warps. Non-circular orbits, e.g. a closed elliptical orbit with axes not parallel to one of the symmetry axis (minor or major), can produce residual velocities as observed in Fig.~\ref{fig:losvdmvlc}.

There are no signs for an outflow from the center. \citet{schmitt_comparison_1996} detect a faint one-sided cone in [\ion{O}{iii}] pointing toward the southeast. The cone is smaller than $0\farcs5$ and can probably be associated with the NLR of NGC 1566. The increase of the FWHM of H$_2$ along the minor axis is in the same direction as the [\ion{O}{iii}] cone but is probably caused by beam smearing along the $0$~\kms\ velocity gradient. At a size of $r\sim0\farcs5$ the [\ion{O}{iii}] cone occupies spatially the nuclear \brg\ and \feii\ emission region, i.e. these lines originate from the same region. However, we do not detect one-sided emission on the nucleus from any emission line (Fig.~\ref{fig:lines1}). The [\ion{Fe}{ii}] emission has a triangular shape but is centered on the nucleus. There is extended flux toward the southeast and also a broadening of the [\ion{Fe}{ii}] line. These are only hints at a possible weak outflow from the nuclear region of NGC 1566 and can as well be associated with the NLR. The LOSV residuals up to $1\arcsec$ from the center in Fig.~\ref{fig:losvdmvlc} fit an outflow along the minor axis, but these residuals are easily explained due to deviations introduced by the density waves of the nuclear spiral. Higher angular resolution observations are needed to be able to compare the [\ion{O}{iii}] emission with shock tracers in the NIR (e.g., H$_2$(1-0)S(1) and \feii) and look for signs of outflowing gas in the LOSV and FWHM.

At $1\farcs5$ east and $1\arcsec$ north the dispersion increases over the nuclear disk average in both investigated molecular species, i.e. H$_2$(1-0)S(1) and $^{12}$CO(3-2), see Fig.~\ref{fig:ALMA}. No increase in line flux can be measured here, but the 0.87~mm continuum shows some substantial emission. \cite{combes_ALMA_2014} find that in their full $18\arcsec\times18\arcsec$ FOV the 0.87~mm continuum is dominated by dust. Hence, it is most probable that heated dust, rather than synchrotron emission from supernovae or free-free emission, is creating the 0.87~mm continuum emission. In combination with the higher dispersion in these regions it must be turbulences in the gas that warm up the dust. What is creating this turbulence is not clear. As mentioned above an outflow from the nucleus is rather improbable as the only signs for an outflow are hinting toward the southeast. Other possibilities are interactions of nuclear disk and spiral arms, e.g. the nuclear spiral. However, these are only speculations since our data is not giving any substantial hints toward the origin of this turbulence.

%% file: con_sum.tex
\section{Conclusion and summary}
\label{sec:consum}

We have analyzed NIR IFS data of the central $10\arcsec\times10\arcsec$ of the Seyfert~1 galaxy NGC 1566. The \ion{H}{ii} deficiency reported by \citet{comte_galaxy_1982} is confirmed. We make a first detection of narrow \brg\ emission at the center and at one region offset to the southwest, in contrast to \citet{reunanen_near-infrared_2002}. They probably were not able to detect the narrow \brg\ emission due to the width of their aperture since the slit positions do coincide with the positions of narrow \brg\ emission.

From the detection of a broad \brg\ component we estimate a BH mass of $(3.0\pm0.9)\times10^6$~\msun\ similar to other BH mass measurements from broad lines \citep{kriss_faint_1991}.
From our continuum decomposition we derive the mass dominating stellar distribution and its velocities. Using the velocity dispersion of the stars in the bulge we estimate the BH mass independently from broad emission lines. We find a BH mass of $(4.2\pm2.4)\times10^6$~\msun, in excellent agreement with the BH estimate from broad emission lines.
We find a BH mass of $(5.3\pm2.9)\times10^6$~\msun\ as a mean value from all methods and relations used in this paper. Furthermore, the Seyfert~1 classification is verified due to hot dust blackbody emission of $\sim1000$~K at the nuclear position, which indicates that the hot inner edge of the torus structure is visible.

\emph{The accretion onto the SMBH is typical for LLAGN} with Eddington ratios of $\lambda_{Edd}=(2\pm1)\times10^{-3}$. The estimated accretion rate is $\sim10^{-4}$ and therefore $10^2$ to $10^4$ times higher than for SgrA* \citep{bower_interferometric_2003,Nayakshin_using_2005}, which has a similar SMBH mass.

Molecular hydrogen is very strong in the observed FOV particularly in an $r=3\arcsec$ disk and in a spiral structure within this disk. Disk and spiral are both detected as well in the warm H$_2$ emission lines with SINFONI as in cold H$_2$ gas from the $^{12}$CO(3-2) emission with ALMA. The shape of the spiral looks similar when comparing the EW of H$_2$(1-0)S(1) with the flux map of $^{12}$CO(3-2). However, H$_2$(1-0)S(1) shows the strongest emission on the nucleus whereas $^{12}$CO(3-2) shows several strong emission spots. 
The southwestern emission spot of $^{12}$CO(3-2) coincides with the \brg\ emission seen in the SF region. From our H$_2$ emission we determine a cold H$_2$ gas mass of $(1.7-9.1)\times10^7$~\msun\ which is in agreement with results based on the ALMA observation by \citet{combes_ALMA_2014}. The molecular gas seems to form a ring-like structure at $r\sim2\arcsec$ best seen in the EW map in Fig.~\ref{fig:h212ew} and the $^{12}$CO(3-2) emission line map (see Fig.~\ref{fig:ALMACO32}).

Along the southern spiral arm a star forming region is detected at a distance of $\sim1\farcs5$ from the center. The measured SFR over a surface area of $\sim1.33$~arcsec$^2$ is $2.6\times10^{-3}$~\msun\ yr$^{-1}$. Comparing this to the global Schmidt law \citep{kennicutt_global_1998} we find that our data point for SFr lies very close to the relation. However, the value for the nuclear $3\arcsec\times3\arcsec$ disk
is situated below the relation with an SFR of $8.0\times10^{-3}$~\msun\ yr$^{-1}$ over a surface area of $\pi\times9$~arcsec$^2$.
The nuclear region of NGC 1566 has a large molecular gas reservoir which it is not using efficiently for star formation.

The excitation mechanism of the molecular gas is partly due to thermal processes (e.g., shocks) and partly due to non-thermal excitation (e.g., UV fluorescence).
The diagnostic diagrams in Figs.~\ref{fig:ddiag1}~\&~\ref{fig:h2diag} show a clear distinction between the nuclear region and the star forming region.
The apertures taken from the nuclear region are situated in the AGN regime in the diagnostic diagram in Fig.~\ref{fig:ddiag1}. The SF region is situated right of the starburst region due to the above mentioned H$_2$ overabundance. The H$_2$ line ratios (Fig.~\ref{fig:h2diag}) imply young star formation at regions SFr and SF~PSF due to a non-thermal excitation fraction of up to 30\%.
From the strong \brg\ but very low \feii\ emission a young starburst of $<9$~Myr is suggested.
\emph{Therefore we find that gas infall is accompanied by young star formation that can be associated with the nuclear spiral arms}.

The central regions are closer to the shock models but show a non-thermal component as well (Fig.~\ref{fig:h2diag}).
The level population diagram (Fig.~\ref{fig:h2lvlpop}) shows that the $v=2$ levels are not thermalized with the $v=1$ levels indicating that the region is excited by a combination of thermal and non-thermal processes in dense gas, i.e. shocked gas and gas ionized by UV-fluorescence from young newly formed bright stars.
Both methods hint on star formation at the nucleus which we are not able to distinguish further from ionizing emission of the AGN.

A region of high dispersion is found in the molecular lines with increased 0.87~mm continuum emission at a distance of $\approx2\arcsec$ from the center. The gas here seems to be shocked, however, what produces the shock is not clear, an outflow or a spiral arm interaction are possible explanations.

\emph{Strong feedback can not be confirmed in the center of NGC~1566}.
Residual velocities that might indicate an outflow and which coincide with higher velocity dispersion in molecular gas. However, these features are aligned with the minor axis which can explain the higher dispersion. The LOSV residuals might be explained by other means, e.g., streaming motions introduced by the nuclear spiral. These might be signs of a \emph{feeding of the SMBH}.
The central $\sim1\arcsec$ harbors possible nuclear star formation as inferred from non-thermal excitation (e.g., UV photons from young bright stars) on the nucleus.
High angular resolution measurements with SINFONI are needed to identify possible feedback and star formation in the central arcsecond of the seemingly waking up Seyfert~1 nucleus. The lack of strong feedback could be a reason for the lack of strong star formation in the nuclear disk if the nuclear spiral is not able to condense gas strong enough to enable star formation.

In the case of NGC~1566 the distribution and velocities of warm and cold molecular gas are very similar. In our first simultaneous observation of NUGA sources in the NIR and sub-mm in NGC~1433 \citep{smajic_ALMA-backed_2014} the similarities were not as high. The distribution was similar but $^{12}$CO(3-2) seemed to be a better tracer of the pseudo-ring in NGC~1433.
In NGC~1566 the $r=3\arcsec$ disk is detected in $^{12}$CO(3-2) and H$_2$(1-0)S(1), which indicates that the central $r\approx200$~pc are warmer than the central region of NGC~1433.
Additionally, in NGC~1433 the LOSV differed in value over the full FOV and in distinct regions in LOSV and dispersion. This is not the case for NGC~1566 where hot and cold gas show similar LOSVs.












%% file: master.bbl
\begin{thebibliography}{90}
\expandafter\ifx\csname natexlab\endcsname\relax\def\natexlab#1{#1}\fi

\bibitem[{Agüero {et~al.}(2004)Agüero, Díaz, \& Bajaja}]{aguero_ngc_2004}
Agüero, E.~L., Díaz, R.~J., \& Bajaja, E. 2004, Astronomy and Astrophysics,
  414, 453

\bibitem[{Alloin {et~al.}(1985)Alloin, Pelat, Phillips, \&
  Whittle}]{alloin_recent_1985}
Alloin, D., Pelat, D., Phillips, M., \& Whittle, M. 1985, The Astrophysical
  Journal, 288, 205

\bibitem[{Alonso-Herrero {et~al.}(2003)Alonso-Herrero, Rieke, Rieke, \&
  Kelly}]{alonso-herrero_[fe_2003}
Alonso-Herrero, A., Rieke, G.~H., Rieke, M.~J., \& Kelly, D.~M. 2003, The
  Astronomical Journal, 125, 1210

\bibitem[{Barbosa {et~al.}(2006)Barbosa, Storchi-Bergmann, Cid~Fernandes,
  Winge, \& Schmitt}]{barbosa_gemini/gmos_2006}
Barbosa, F. K.~B., Storchi-Bergmann, T., Cid~Fernandes, R., Winge, C., \&
  Schmitt, H. 2006, Monthly Notices of the Royal Astronomical Society, 371, 170

\bibitem[{Baribaud {et~al.}(1992)Baribaud, Alloin, Glass, \&
  Pelat}]{baribaud_variability_1992}
Baribaud, T., Alloin, D., Glass, I., \& Pelat, D. 1992, Astronomy and
  Astrophysics, 256, 375

\bibitem[{Bedregal {et~al.}(2009)Bedregal, Colina, Alonso-Herrero, \&
  Arribas}]{bedregal_near-ir_2009}
Bedregal, A.~G., Colina, L., Alonso-Herrero, A., \& Arribas, S. 2009, The
  Astrophysical Journal, 698, 1852

\bibitem[{Black \& van Dishoeck(1987)}]{black_fluorescent_1987}
Black, J.~H. \& van Dishoeck, E.~F. 1987, The Astrophysical Journal, 322, 412

\bibitem[{Bolatto {et~al.}(2013)Bolatto, Wolfire, \&
  Leroy}]{bolatto_co--h2_2013}
Bolatto, A.~D., Wolfire, M., \& Leroy, A.~K. 2013, Annual Review of Astronomy
  and Astrophysics, 51, 207

\bibitem[{Bonnet {et~al.}(2004)Bonnet, Abuter, Baker, Bornemann, Brown,
  Castillo, Conzelmann, Damster, Davies, Delabre, Donaldson, Dumas, Eisenhauer,
  Elswijk, Fedrigo, Finger, Gemperlein, Genzel, Gilbert, Gillet, Goldbrunner,
  Horrobin, Ter~Horst, Huber, Hubin, Iserlohe, Kaufer, Kissler-Patig, Kragt,
  Kroes, Lehnert, Lieb, Liske, Lizon, Lutz, Modigliani, Monnet, Nesvadba,
  Patig, Pragt, Reunanen, Röhrle, Rossi, Schmutzer, Schoenmaker, Schreiber,
  Stroebele, Szeifert, Tacconi, Tecza, Thatte, Tordo, van~der Werf, \&
  Weisz}]{bonnet_first_2004-1}
Bonnet, H., Abuter, R., Baker, A., {et~al.} 2004, The Messenger, 117, 17

\bibitem[{Boone {et~al.}(2007)Boone, Baker, Schinnerer, Combes,
  García-Burillo, Neri, Hunt, Léon, Krips, Tacconi, \&
  Eckart}]{boone_molecular_2007}
Boone, F., Baker, A.~J., Schinnerer, E., {et~al.} 2007, Astronomy and
  Astrophysics, 471, 113

\bibitem[{Bower {et~al.}(2003)Bower, Wright, Falcke, \&
  Backer}]{bower_interferometric_2003}
Bower, G.~C., Wright, M. C.~H., Falcke, H., \& Backer, D.~C. 2003, The
  Astrophysical Journal, 588, 331

\bibitem[{Brand {et~al.}(1989)Brand, Toner, Geballe, Webster, Williams, \&
  Burton}]{brand_constancy_1989}
Brand, P. W. J.~L., Toner, M.~P., Geballe, T.~R., {et~al.} 1989, Monthly
  Notices of the Royal Astronomical Society, 236, 929

\bibitem[{Busch {et~al.}(2015)Busch, Smajić, Scharwächter, Eckart,
  Valencia-S., Moser, Husemann, Krips, \& Zuther}]{busch_low-luminosity_2015}
Busch, G., Smajić, S., Scharwächter, J., {et~al.} 2015, Astronomy and
  Astrophysics, 575, A128

\bibitem[{Busch {et~al.}(2014)Busch, Zuther, Valencia-S., Moser, Fischer,
  Eckart, Scharwächter, Gadotti, \& Wisotzki}]{busch_low-luminosity_2014}
Busch, G., Zuther, J., Valencia-S., M., {et~al.} 2014, Astronomy and
  Astrophysics, 561, 140

\bibitem[{Calzetti(1997)}]{calzetti_reddening_1997}
Calzetti, D. 1997, The Astronomical Journal, 113, 162

\bibitem[{Casasola {et~al.}(2008)Casasola, Combes, García-Burillo, Hunt,
  Léon, \& Baker}]{casasola_molecular_2008}
Casasola, V., Combes, F., García-Burillo, S., {et~al.} 2008, Astronomy and
  Astrophysics, 490, 61

\bibitem[{Casasola {et~al.}(2010)Casasola, Hunt, Combes, García-Burillo,
  Boone, Eckart, Neri, \& Schinnerer}]{casasola_molecular_2010}
Casasola, V., Hunt, L.~K., Combes, F., {et~al.} 2010, Astronomy and
  Astrophysics, 510, 52

\bibitem[{{Cid Fernandes} {et~al.}(2004){Cid Fernandes}, {Gu}, {Melnick},
  {Terlevich}, {Terlevich}, {Kunth}, {Rodrigues Lacerda}, \&
  {Joguet}}]{cidfernandes_star_2004}
{Cid Fernandes}, R., {Gu}, Q., {Melnick}, J., {et~al.} 2004, \mnras, 355, 273

\bibitem[{Combes {et~al.}(2009)Combes, Baker, Schinnerer, García-Burillo,
  Hunt, Boone, Eckart, Neri, \& Tacconi}]{combes_molecular_2009}
Combes, F., Baker, A.~J., Schinnerer, E., {et~al.} 2009, Astronomy and
  Astrophysics, 503, 73

\bibitem[{Combes {et~al.}(2004)Combes, García-Burillo, Boone, Hunt, Baker,
  Eckart, Englmaier, Leon, Neri, Schinnerer, \&
  Tacconi}]{combes_molecular_2004}
Combes, F., García-Burillo, S., Boone, F., {et~al.} 2004, Astronomy and
  Astrophysics, 414, 857

\bibitem[{Combes {et~al.}(2013)Combes, García-Burillo, Casasola, Hunt, Krips,
  Baker, Boone, Eckart, Marquez, Neri, Schinnerer, \&
  Tacconi}]{combes_ALMA_2013}
Combes, F., García-Burillo, S., Casasola, V., {et~al.} 2013, Astronomy and
  Astrophysics, 558, 124

\bibitem[{Combes {et~al.}(2014)Combes, García-Burillo, Casasola, Hunt, Krips,
  Baker, Boone, Eckart, Marquez, Neri, Schinnerer, \&
  Tacconi}]{combes_ALMA_2014}
Combes, F., García-Burillo, S., Casasola, V., {et~al.} 2014, Astronomy and
  Astrophysics, 565, A97

\bibitem[{Comerón {et~al.}(2010)Comerón, Knapen, Beckman, Laurikainen, Salo,
  Martínez-Valpuesta, \& Buta}]{comeron_ainur:_2010}
Comerón, S., Knapen, J.~H., Beckman, J.~E., {et~al.} 2010, Monthly Notices of
  the Royal Astronomical Society, 402, 2462

\bibitem[{Comte \& Duquennoy(1982)}]{comte_galaxy_1982}
Comte, G. \& Duquennoy, A. 1982, Astronomy and Astrophysics, 114, 7

\bibitem[{Dale {et~al.}(2004)Dale, Roussel, Contursi, Helou, Dinerstein,
  Hunter, Hollenbach, Egami, Matthews, Murphy, Lafon, \&
  Rubin}]{dale_near-infrared_2004}
Dale, D.~A., Roussel, H., Contursi, A., {et~al.} 2004, The Astrophysical
  Journal, 601, 813

\bibitem[{Davies {et~al.}(2014)Davies, Maciejewski, Hicks, Emsellem, Erwin,
  Burtscher, Dumas, Lin, Malkan, Müller-Sánchez, Orban~de Xivry, Rosario,
  Schnorr-Müller, \& Tran}]{davies_fueling_2014}
Davies, R.~I., Maciejewski, W., Hicks, E. K.~S., {et~al.} 2014, The
  Astrophysical Journal, 792, 101

\bibitem[{{Davies} {et~al.}(2007){Davies}, {M{\"u}ller S{\'a}nchez}, {Genzel},
  {Tacconi}, {Hicks}, {Friedrich}, \& {Sternberg}}]{davies_close_2007}
{Davies}, R.~I., {M{\"u}ller S{\'a}nchez}, F., {Genzel}, R., {et~al.} 2007,
  \apj, 671, 1388

\bibitem[{Davies {et~al.}(2003)Davies, Sternberg, Lehnert, \&
  Tacconi-Garman}]{davies_molecular_2003}
Davies, R.~I., Sternberg, A., Lehnert, M., \& Tacconi-Garman, L.~E. 2003, The
  Astrophysical Journal, 597, 907

\bibitem[{Davies {et~al.}(2005)Davies, Sternberg, Lehnert, \&
  Tacconi-Garman}]{davies_molecular_2005}
Davies, R.~I., Sternberg, A., Lehnert, M.~D., \& Tacconi-Garman, L.~E. 2005,
  The Astrophysical Journal, 633, 105

\bibitem[{Davies {et~al.}(2006)Davies, Thomas, Genzel, Müller~Sánchez,
  Tacconi, Sternberg, Eisenhauer, Abuter, Saglia, \&
  Bender}]{davies_star-forming_2006}
Davies, R.~I., Thomas, J., Genzel, R., {et~al.} 2006, The Astrophysical
  Journal, 646, 754

\bibitem[{Draine \& McKee(1993)}]{draine_theory_1993}
Draine, B.~T. \& McKee, C.~F. 1993, Annual Review of Astronomy and
  Astrophysics, 31, 373

\bibitem[{Draine \& Woods(1990)}]{draine_h2_1990}
Draine, B.~T. \& Woods, D.~T. 1990, The Astrophysical Journal, 363, 464

\bibitem[{{Eisenhauer} {et~al.}(2003){Eisenhauer}, {Abuter}, {Bickert},
  {Biancat-Marchet}, {Bonnet}, {Brynnel}, {Conzelmann}, {Delabre}, {Donaldson},
  {Farinato}, {Fedrigo}, {Genzel}, {Hubin}, {Iserlohe}, {Kasper},
  {Kissler-Patig}, {Monnet}, {Roehrle}, {Schreiber}, {Stroebele}, {Tecza},
  {Thatte}, \& {Weisz}}]{eisenhauer_sinfoni_2003}
{Eisenhauer}, F., {Abuter}, R., {Bickert}, K., {et~al.} 2003, in Society of
  Photo-Optical Instrumentation Engineers (SPIE) Conference Series, Vol. 4841,
  Society of Photo-Optical Instrumentation Engineers (SPIE) Conference Series,
  ed. {M.~Iye \& A.~F.~M.~Moorwood}, 1548--1561

\bibitem[{Emsellem {et~al.}(2001)Emsellem, Greusard, Combes, Friedli, Leon,
  Pécontal, \& Wozniak}]{emsellem_dynamics_2001}
Emsellem, E., Greusard, D., Combes, F., {et~al.} 2001, Astronomy and
  Astrophysics, 368, 52

\bibitem[{Falcón-Barroso {et~al.}(2006)Falcón-Barroso, Bacon, Bureau,
  Cappellari, Davies, de~Zeeuw, Emsellem, Fathi, Krajnović, Kuntschner,
  McDermid, Peletier, \& Sarzi}]{falcon-barroso_sauron_2006}
Falcón-Barroso, J., Bacon, R., Bureau, M., {et~al.} 2006, Monthly Notices of
  the Royal Astronomical Society, 369, 529

\bibitem[{Ferrarese \& Merritt(2000)}]{ferrarese_fundamental_2000}
Ferrarese, L. \& Merritt, D. 2000, The Astrophysical Journal Letters, 539, L9

\bibitem[{Fischer {et~al.}(2006)Fischer, Iserlohe, Zuther, Bertram,
  Straubmeier, Schödel, \& Eckart}]{fischer_nearby_2006}
Fischer, S., Iserlohe, C., Zuther, J., {et~al.} 2006, Astronomy and
  Astrophysics, 452, 827

\bibitem[{García-Burillo {et~al.}(2003)García-Burillo, Combes, Hunt, Boone,
  Baker, Tacconi, Eckart, Neri, Leon, Schinnerer, \&
  Englmaier}]{garcia-burillo_molecular_2003}
García-Burillo, S., Combes, F., Hunt, L.~K., {et~al.} 2003, Astronomy and
  Astrophysics, 407, 485

\bibitem[{García-Burillo {et~al.}(2005)García-Burillo, Combes, Schinnerer,
  Boone, \& Hunt}]{garcia-burillo_molecular_2005}
García-Burillo, S., Combes, F., Schinnerer, E., Boone, F., \& Hunt, L.~K.
  2005, Astronomy and Astrophysics, 441, 1011

\bibitem[{García-Burillo {et~al.}(2014)García-Burillo, Combes, Usero, Aalto,
  Krips, Viti, Alonso-Herrero, Hunt, Schinnerer, Baker, Boone, Casasola,
  Colina, Costagliola, Eckart, Fuente, Henkel, Labiano, Martín, Márquez,
  Muller, Planesas, Ramos~Almeida, Spaans, Tacconi, \& van~der
  Werf}]{garcia-burillo_molecular_2014}
García-Burillo, S., Combes, F., Usero, A., {et~al.} 2014, Astronomy and
  Astrophysics, 567, 125

\bibitem[{García-Burillo {et~al.}(2009)García-Burillo, Fernández-García,
  Combes, Hunt, Haan, Schinnerer, Boone, Krips, \&
  Márquez}]{garcia-burillo_molecular_2009}
García-Burillo, S., Fernández-García, S., Combes, F., {et~al.} 2009,
  Astronomy and Astrophysics, 496, 85

\bibitem[{Graham \& Scott(2013)}]{graham_m_2013}
Graham, A.~W. \& Scott, N. 2013, The Astrophysical Journal, 764, 151

\bibitem[{Gültekin {et~al.}(2009)Gültekin, Richstone, Gebhardt, Lauer,
  Tremaine, Aller, Bender, Dressler, Faber, Filippenko, Green, Ho, Kormendy,
  Magorrian, Pinkney, \& Siopis}]{gultekin_m-_2009}
Gültekin, K., Richstone, D.~O., Gebhardt, K., {et~al.} 2009, The Astrophysical
  Journal, 698, 198

\bibitem[{Ho(2008)}]{ho_nuclear_2008}
Ho, L.~C. 2008, Annual Review of Astronomy and Astrophysics, 46, 475

\bibitem[{Ho(2009)}]{ho_radiatively_2009}
Ho, L.~C. 2009, The Astrophysical Journal, 699, 626

\bibitem[{Howell(2000)}]{howell_handbook_2000}
Howell, S.~B. 2000, Handbook of {CCD} {Astronomy}

\bibitem[{Hunt {et~al.}(2008)Hunt, Combes, García-Burillo, Schinnerer, Krips,
  Baker, Boone, Eckart, Léon, Neri, \& Tacconi}]{hunt_molecular_2008}
Hunt, L.~K., Combes, F., García-Burillo, S., {et~al.} 2008, Astronomy and
  Astrophysics, 482, 133

\bibitem[{Kawamuro {et~al.}(2013)Kawamuro, Ueda, Tazaki, \&
  Terashima}]{kawamuro_broadband_2013}
Kawamuro, T., Ueda, Y., Tazaki, F., \& Terashima, Y. 2013, The Astrophysical
  Journal, 770, 157

\bibitem[{Kennicutt(1998)}]{kennicutt_global_1998}
Kennicutt, Jr., R.~C. 1998, The Astrophysical Journal, 498, 541

\bibitem[{Kim {et~al.}(2010)Kim, Im, \& Kim}]{kim_new_2010}
Kim, D., Im, M., \& Kim, M. 2010, The Astrophysical Journal, 724, 386

\bibitem[{Koribalski {et~al.}(2004)Koribalski, Staveley-Smith, Kilborn, Ryder,
  Kraan-Korteweg, Ryan-Weber, Ekers, Jerjen, Henning, Putman, Zwaan, de~Blok,
  Calabretta, Disney, Minchin, Bhathal, Boyce, Drinkwater, Freeman, Gibson,
  Green, Haynes, Juraszek, Kesteven, Knezek, Mader, Marquarding, Meyer, Mould,
  Oosterloo, O'Brien, Price, Sadler, Schröder, Stewart, Stootman, Waugh,
  Warren, Webster, \& Wright}]{koribalski_1000_2004}
Koribalski, B.~S., Staveley-Smith, L., Kilborn, V.~A., {et~al.} 2004, The
  Astronomical Journal, 128, 16

\bibitem[{Krips {et~al.}(2005)Krips, Eckart, Neri, Pott, Leon, Combes,
  García-Burillo, Hunt, Baker, Tacconi, Englmaier, Schinnerer, \&
  Boone}]{krips_molecular_2005}
Krips, M., Eckart, A., Neri, R., {et~al.} 2005, Astronomy and Astrophysics,
  442, 479

\bibitem[{Kriss {et~al.}(1991)Kriss, Hartig, Armus, Blair, Caganoff, \&
  Dressel}]{kriss_faint_1991}
Kriss, G.~A., Hartig, G.~F., Armus, L., {et~al.} 1991, The Astrophysical
  Journal Letters, 377, L13

\bibitem[{Larkin {et~al.}(1998)Larkin, Armus, Knop, Soifer, \&
  Matthews}]{larkin_near-infrared_1998}
Larkin, J.~E., Armus, L., Knop, R.~A., Soifer, B.~T., \& Matthews, K. 1998, The
  Astrophysical Journal Supplement Series, 114, 59

\bibitem[{Lester {et~al.}(1988)Lester, Harvey, \&
  Carr}]{lester_properties_1988}
Lester, D.~F., Harvey, P.~M., \& Carr, J. 1988, The Astrophysical Journal, 329,
  641

\bibitem[{Levenson {et~al.}(2009)Levenson, Radomski, Packham, Mason, Schaefer,
  \& Telesco}]{levenson_isotropic_2009}
Levenson, N.~A., Radomski, J.~T., Packham, C., {et~al.} 2009, The Astrophysical
  Journal, 703, 390

\bibitem[{Lindt-Krieg {et~al.}(2008)Lindt-Krieg, Eckart, Neri, Krips, Pott,
  García-Burillo, \& Combes}]{lindt-krieg_molecular_2008}
Lindt-Krieg, E., Eckart, A., Neri, R., {et~al.} 2008, Astronomy and
  Astrophysics, 479, 377

\bibitem[{Läsker {et~al.}(2014)Läsker, Ferrarese, van~de Ven, \&
  Shankar}]{lasker_supermassive_2014}
Läsker, R., Ferrarese, L., van~de Ven, G., \& Shankar, F. 2014, The
  Astrophysical Journal, 780, 70

\bibitem[{{Magorrian} {et~al.}(1998){Magorrian}, {Tremaine}, {Richstone},
  {Bender}, {Bower}, {Dressler}, {Faber}, {Gebhardt}, {Green}, {Grillmair},
  {Kormendy}, \& {Lauer}}]{magorrian_demography_1998}
{Magorrian}, J., {Tremaine}, S., {Richstone}, D., {et~al.} 1998, \aj, 115, 2285

\bibitem[{Maiolino {et~al.}(1996)Maiolino, Rieke, \&
  Rieke}]{maiolino_correction_1996}
Maiolino, R., Rieke, G.~H., \& Rieke, M.~J. 1996, The Astronomical Journal,
  111, 537

\bibitem[{Marconi \& Hunt(2003)}]{marconi_relation_2003}
Marconi, A. \& Hunt, L.~K. 2003, The Astrophysical Journal Letters, 589, L21

\bibitem[{Mazzalay {et~al.}(2013)Mazzalay, Saglia, Erwin, Fabricius, Rusli,
  Thomas, Bender, Opitsch, Nowak, \& Williams}]{mazzalay_molecular_2013}
Mazzalay, X., Saglia, R.~P., Erwin, P., {et~al.} 2013, Monthly Notices of the
  Royal Astronomical Society, 428, 2389

\bibitem[{McMullin {et~al.}(2007)McMullin, Waters, Schiebel, Young, \&
  Golap}]{mcmullin_casa_2007}
McMullin, J.~P., Waters, B., Schiebel, D., Young, W., \& Golap, K. 2007, in ,
  127

\bibitem[{Mouri(1994)}]{mouri_molecular_1994}
Mouri, H. 1994, The Astrophysical Journal, 427, 777

\bibitem[{Mulchaey \& Regan(1997)}]{mulchaey_fueling_1997}
Mulchaey, J.~S. \& Regan, M.~W. 1997, The Astrophysical Journal Letters, 482,
  L135

\bibitem[{Nayakshin(2005)}]{Nayakshin_using_2005}
Nayakshin, S. 2005, Astronomy and Astrophysics, 429, L33

\bibitem[{Netzer(1977)}]{netzer_profiles_1977}
Netzer, H. 1977, Monthly Notices of the Royal Astronomical Society, 181, 89P

\bibitem[{Nussbaumer \& Storey(1988)}]{nussbaumer_transition_1988}
Nussbaumer, H. \& Storey, P.~J. 1988, Astronomy and Astrophysics, 193, 327

\bibitem[{Panuzzo {et~al.}(2003)Panuzzo, Bressan, Granato, Silva, \&
  Danese}]{panuzzo_dust_2003}
Panuzzo, P., Bressan, A., Granato, G.~L., Silva, L., \& Danese, L. 2003,
  Astronomy and Astrophysics, 409, 99

\bibitem[{Reunanen {et~al.}(2002)Reunanen, Kotilainen, \&
  Prieto}]{reunanen_near-infrared_2002}
Reunanen, J., Kotilainen, J.~K., \& Prieto, M.~A. 2002, Monthly Notices of the
  Royal Astronomical Society, 331, 154

\bibitem[{Riffel {et~al.}(2013)Riffel, Storchi-Bergmann, \&
  Winge}]{riffel_feeding_2013}
Riffel, R.~A., Storchi-Bergmann, T., \& Winge, C. 2013, Monthly Notices of the
  Royal Astronomical Society, 430, 2249

\bibitem[{Riffel {et~al.}(2008)Riffel, Storchi-Bergmann, Winge, McGregor, Beck,
  \& Schmitt}]{riffel_mapping_2008}
Riffel, R.~A., Storchi-Bergmann, T., Winge, C., {et~al.} 2008, Monthly Notices
  of the Royal Astronomical Society, 385, 1129

\bibitem[{Riffel {et~al.}(2014)Riffel, Vale, Storchi-Bergmann, \&
  McGregor}]{riffel_feeding_2014}
Riffel, R.~A., Vale, T.~B., Storchi-Bergmann, T., \& McGregor, P.~J. 2014,
  Monthly Notices of the Royal Astronomical Society, 442, 656

\bibitem[{{Rodr{\'{\i}}guez-Ardila} {et~al.}(2004){Rodr{\'{\i}}guez-Ardila},
  {Pastoriza}, {Viegas}, {Sigut}, \&
  {Pradhan}}]{rodriguez-ardila_molecular_2004}
{Rodr{\'{\i}}guez-Ardila}, A., {Pastoriza}, M.~G., {Viegas}, S., {Sigut},
  T.~A.~A., \& {Pradhan}, A.~K. 2004, \aap, 425, 457

\bibitem[{{Rodr{\'{\i}}guez-Ardila} {et~al.}(2005){Rodr{\'{\i}}guez-Ardila},
  {Riffel}, \& {Pastoriza}}]{rodriguez-ardila_molecular_2005}
{Rodr{\'{\i}}guez-Ardila}, A., {Riffel}, R., \& {Pastoriza}, M.~G. 2005,
  \mnras, 364, 1041

\bibitem[{Schmitt \& Kinney(1996)}]{schmitt_comparison_1996}
Schmitt, H.~R. \& Kinney, A.~L. 1996, The Astrophysical Journal, 463, 498

\bibitem[{Scoville {et~al.}(1982)Scoville, Hall, Ridgway, \&
  Kleinmann}]{scoville_velocity_1982}
Scoville, N.~Z., Hall, D. N.~B., Ridgway, S.~T., \& Kleinmann, S.~G. 1982, The
  Astrophysical Journal, 253, 136

\bibitem[{Sheth {et~al.}(2005)Sheth, Vogel, Regan, Thornley, \&
  Teuben}]{sheth_secular_2005}
Sheth, K., Vogel, S.~N., Regan, M.~W., Thornley, M.~D., \& Teuben, P.~J. 2005,
  The Astrophysical Journal, 632, 217

\bibitem[{Smajic {et~al.}(2014)Smajic, Moser, Eckart, Valencia-S., Combes,
  Horrobin, García-Burillo, García-Marín, Fischer, \&
  Zuther}]{smajic_ALMA-backed_2014}
Smajic, S., Moser, L., Eckart, A., {et~al.} 2014, Astronomy and Astrophysics,
  567, 119

\bibitem[{Smajić {et~al.}(2012)Smajić, Fischer, Zuther, \&
  Eckart}]{smajic_unveiling_2012}
Smajić, S., Fischer, S., Zuther, J., \& Eckart, A. 2012, Astronomy and
  Astrophysics, 544, 105

\bibitem[{Sternberg \& Dalgarno(1989)}]{sternberg_infrared_1989}
Sternberg, A. \& Dalgarno, A. 1989, The Astrophysical Journal, 338, 197

\bibitem[{Sternberg \& Dalgarno(1995)}]{sternberg_chemistry_1995}
Sternberg, A. \& Dalgarno, A. 1995, The Astrophysical Journal Supplement
  Series, 99, 565

\bibitem[{Sternberg \& Neufeld(1999)}]{sternberg_ratio_1999}
Sternberg, A. \& Neufeld, D.~A. 1999, The Astrophysical Journal, 516, 371

\bibitem[{Turner \& Ostriker(1977)}]{turner_mass--light_1977}
Turner, E.~L. \& Ostriker, J.~P. 1977, The Astrophysical Journal, 217, 24

\bibitem[{Valencia-S. {et~al.}(2012)Valencia-S., Zuther, Eckart,
  García-Marín, Iserlohe, \& Wright}]{valencia-s._is_2012}
Valencia-S., M., Zuther, J., Eckart, A., {et~al.} 2012, Astronomy and
  Astrophysics, 544, 129

\bibitem[{van~der Laan {et~al.}(2011)van~der Laan, Schinnerer, Boone,
  García-Burillo, Combes, Haan, Leon, Hunt, \&
  Baker}]{van_der_laan_molecular_2011}
van~der Laan, T. P.~R., Schinnerer, E., Boone, F., {et~al.} 2011, Astronomy and
  Astrophysics, 529, 45

\bibitem[{Winge {et~al.}(2009)Winge, Riffel, \&
  Storchi-Bergmann}]{winge_gemini_2009}
Winge, C., Riffel, R.~A., \& Storchi-Bergmann, T. 2009, The Astrophysical
  Journal Supplement Series, 185, 186

\bibitem[{Wolniewicz {et~al.}(1998)Wolniewicz, Simbotin, \&
  Dalgarno}]{wolniewicz_quadrupole_1998}
Wolniewicz, L., Simbotin, I., \& Dalgarno, A. 1998, The Astrophysical Journal
  Supplement Series, 115, 293

\bibitem[{Woo \& Urry(2002)}]{woo_active_2002}
Woo, J.-H. \& Urry, C.~M. 2002, The Astrophysical Journal, 579, 530

\bibitem[{Zuther {et~al.}(2007)Zuther, Iserlohe, Pott, Bertram, Fischer, Voges,
  Hasinger, \& Eckart}]{zuther_mrk_2007}
Zuther, J., Iserlohe, C., Pott, J.-U., {et~al.} 2007, Astronomy and
  Astrophysics, 466, 451

\end{thebibliography}
